\documentclass[submission, Phys]{SciPost}

\hypersetup{
    colorlinks,
    linkcolor={red!50!black},
    citecolor={blue!50!black},
    urlcolor={blue!80!black}
}

\usepackage[bitstream-charter]{mathdesign}
\DeclareSymbolFont{usualmathcal}{OMS}{cmsy}{m}{n}
\DeclareSymbolFontAlphabet{\mathcal}{usualmathcal}

\usepackage[utf8]{inputenc}
\usepackage{todonotes}

\newcommand{\pdag}{{\phantom{\dagger}}}

\usepackage{amsmath}        
\usepackage{amsthm}         
\usepackage{bm}             
\usepackage{graphicx}       
\usepackage{booktabs}       
\usepackage{subcaption}     
\usepackage{multirow}
\usepackage{bm}             

\usepackage{xcolor}
\usepackage{slashed,physics,braket} 
\usepackage[version=4]{mhchem}
\usepackage[normalem]{ulem} 
\usepackage{placeins}

\begin{document}

\begin{center}{\Large \textbf{
Correlated States in Quantum Dot Clusters Coupled to a Common Superconductor
}}\end{center}

\begin{center}
Martin \v{Z}onda\textsuperscript{1$\star$},
Jakub R\k{e}kas\textsuperscript{2},
Tobi\'a\v{s} Pol\'a\v{c}ek\textsuperscript{3,4},
Jana Kodrlov\'a\textsuperscript{1,3},
Vladislav Pokorn\'y\textsuperscript{5},
Martin Fri\'ak\textsuperscript{3,4}
\end{center}

\begin{center}
{\bf 1} Department of Condensed Matter Physics, Faculty of Mathematics and Physics, Charles University, Ke Karlovu 5, CZ-121~16 Praha 2, Czech Republic \\
{\bf 2} Department of Theoretical Physics, Faculty of Fundamental Problems of Technology, Wroc\l aw University of Science and Technology, 50-370 Wroc\l aw, Poland \\
{\bf 3}  Institute of Physics of Materials, Czech Academy of Sciences, v.v.i., \v{Z}i\v{z}kova 22, CZ-616~00 Brno, Czech Republic \\
{\bf 4} Department of Condensed Matter Physics,  Faculty of Science, Masaryk University, \\ Kotlářská 2, CZ-611 37 Brno, Czech Republic \\
{\bf 5} Institute of Physics (FZU), Czech Academy of Sciences, Na Slovance 2, CZ-182~00 Praha 8, Czech Republic \\
%
${}^\star$ {\small \sf martin.zonda@matfyz.cuni.cz}
\end{center}

\begin{center}
\today
\end{center}


\section*{Abstract}
{
We study an effective model of regular quantum dot clusters coupled to a common superconductor. By applying a canonical transformation, we map the system onto a particle-number-conserving representation, enabling an efficient treatment with standard fermionic neural-network quantum-state variational Monte Carlo methods. We show that the excitation gap can close under a particular high-symmetry condition, which, in finite non-interacting systems, corresponds to crossings between singlet ground states of different character. At this high-symmetry condition the model can be mapped to a standard Hubbard model in grand-canonical ensemble with fixed chemical potential away from half-filling. Combining exact methods, density matrix renormalization group, and neural quantum-state variational Monte Carlo calculations, we identify three distinct interacting regimes: a trivial superconducting singlet phase, a strongly correlated regime related to an effective Heisenberg model, and a critical intermediate regime with qualitatively different behavior in one and two dimensions. In one-dimensional systems, the intermediate regime exhibits a sequence of singlet-doublet transitions and becomes gapless in the thermodynamic limit even for finite Coulomb interaction. In two-dimensional clusters, we find robust triplet ground states. These results connect the physics of superconducting nanostructures with correlated lattice models and show that standard fermionic neural quantum states can capture the relevant interacting regimes.
}

\vspace{10pt}
\noindent\rule{\textwidth}{1pt}
\tableofcontents\thispagestyle{fancy}
\noindent\rule{\textwidth}{1pt}
\vspace{10pt}

\section{Introduction
\label{sec:intro}}
Magnetic adatoms, molecules, and various realizations of quantum dots (QDs) coupled to superconducting (SC) leads or deposited on SC surfaces have emerged as a versatile platform for studying the interplay between superconductivity, strong electron correlations, magnetism, and related phenomena~\cite{Wernsdorfer-2010,Heinrich-2018}. At the same time, these systems provide a promising platform for engineering new quantum matter with desired functionalities. Modern experimental setups provide the means to study and tune the in-gap Yu–Shiba–Rusinov excitations, Andreev spectra and gate-controlled supercurrents, advancing toward scalable nanoscopic devices for applications in SC electronics, classical and quantum computing, and sensor technologies~\cite{Balatsky2006impurity,Krantz-2019-qubits,Natarajan-2012-detectors,Persky-2022-squid}. For example, molecular assemblies on SC surfaces allow for construction of switchable information units~\cite{Li2025individual} and there are proposals to use them as building blocks for SC quantum dot cellular automata~\cite{Li2025negative}. Furthermore, the minimal Kitaev chains, realized with magnetic adatoms or QD arrays, provide a flexible testbed for exploring Majorana physics, parity control, and nonlocal correlations~\cite{dvir2023realization,bordin2025enhanced,SeoaneSouto-2024}.

Systems with one or two impurities were treated using a wide set of methods, including perturbation expansion techniques~\cite{Martin-Rodero-2012,Zonda2015perturbation}, numerical/functional renormalization group~\cite{Pillet2013tunneling,Karrasch-2008-fRG}, continuous-time quantum Monte Carlo (QMC)~\cite{Luitz2010weak,Luitz-2012,Pokorny-2018} and various effective models~\cite{Vecino-2003-ZBW,Meng-2009self,Pokorny-2023,Bobok2025che} (for a review, see~\cite{Meden2019the}). However, as these systems grow in complexity, from single impurities to coupled arrays and molecular layers, the theoretical modeling becomes increasingly demanding. At some point, even the simplest effective low-energy descriptions require advanced numerical techniques capable of treating large Hilbert spaces and long coherence lengths, such as the density matrix renormalization group (DMRG)~\cite{White1992DMRG,White1993DMRG,Catarina2023DMRG} and various QMC methods~\cite{Gubernatis2016QMC}.

DMRG has proven to be a particularly powerful tool for SC nanohybrids in quasi-one-dimensional geometries. It enables highly accurate simulations of interacting quantum impurity chains and proximitized nanowires, including regimes with strong correlations and induced pairing~\cite{Bacsi2023Exchange,Bobok2025che}. Formulations in terms of matrix product states (MPS) naturally accommodate local SC pairing terms. However, these advantages come with important limitations. The efficiency of DMRG rapidly deteriorates in higher spatial dimensions due to the growth of entanglement entropy, which leads to an exponential increase in the required bond dimension~\cite{Cirac2021MPS,Banuls2023TN}. In addition, particle-number-nonconserving Hamiltonians typically break the global $U(1)$ symmetry that is often exploited to block-diagonalize the Hilbert space, further increasing the computational cost. Although extensions to two-dimensional systems and SC states~\cite{verstraete2004PEPS,Orus2014TN,Banuls2023TN,Cirac2021MPS} are possible, they are, in general,
computationally demanding and require careful handling of fermionic statistics.

A promising complementary direction is provided by neural quantum state (NQS) variational Monte Carlo (VMC) methods~\cite{carleo2017solving,carleo2018_constructing}. These machine-learning–based many-body methods have recently shown remarkable success in modeling strongly correlated lattice systems~\cite{Medvidovic2024NQS,chooSymmetriesManyBodyExcitations2018,dawidModernApplicationsMachine,Padila2025autoregressive,mezera2023NNQS,gu2026solving}, providing compact and systematically improvable representations of complex quantum states. However, their application to SC systems remains comparatively underexplored~\cite{carrasquillaNeuralNetworksQuantum2021,kim2024ultra}. The central challenge originates from the absence of particle-number conservation: SC pairing mixes different particle-number sectors, whereas many standard NQS architectures are tailored to fixed particle-number Hilbert spaces. Several strategies have been proposed to overcome this limitation. A conceptually simple approach is to employ a Jordan–Wigner or related fermion-to-spin transformation, mapping the fermionic degrees of freedom to a spin-$1/2$ lattice of distinguishable sites~\cite{jordan1928paulische,Barrett2022autoregressive}. Such mappings are fully compatible with particle-number-nonconserving Hamiltonians, and thereby enable the use of established NQS techniques developed for spin systems. However, in more than one spatial dimension, these transformations generally introduce long-range, highly nonlocal spin interactions, which can severely degrade the efficiency and trainability of the resulting models. 

An alternative route consists of augmenting NQS with Pfaffian wave functions~\cite{Bajdich2006pfaffian,gao2023generalizing,gao2024neural}. 
Pfaffian states provide the natural extension of Slater determinants to fermionic wave functions with pairing correlations. 
Recent works have demonstrated that Pfaffian-based variational ans\"atze provide an accurate description of paired and SC fermionic phases~\cite{kim2024ultra,Chen2024empowering,chen2025neural}. From a computational perspective, however, Pfaffian evaluations scale cubically with system size, similarly to determinant evaluations, but typically involve larger numerical prefactors and less efficient update schemes in VMC calculations~\cite{Xu2022optimized}. As a result, large-scale optimization of Pfaffian-based states is generally more demanding than for determinant-based approaches.

As we discuss in this work, a much simpler alternative can be applied to a broad class of SC nanosystems. In particular, a standard canonical transformation can map the SC problem onto a particle-number-conserving one, allowing the use of standard fermionic NQS methods. We study an effective lattice model that can be interpreted either as a cluster of atoms or molecules with a single active orbital deposited on an SC substrate, or as an array of QDs coupled through SC bridges. Our goals are twofold: first, to map out and characterize the ground-state structure over a broad parameter range, including its evolution between qualitatively distinct many-body regimes; and second, to establish whether relatively standard fermionic NQS-VMC approaches can provide a common and reliable description across these regimes, with DMRG and exact methods serving as complementary benchmarks where applicable.

To address the first of these goals, we analyze the model in detail across both noninteracting and interacting regimes. We begin with the noninteracting limit, where we derive analytical conditions for the closing of the effective spectral gap and characterize transitions between distinct singlet ground states. We then turn to interacting clusters accessible by exact diagonalization and determine their ground-state phase diagrams, focusing on how the competition between local interactions, SC pairing, and interdot coupling gives rise to qualitatively different correlated regimes. We show that one-dimensional systems exhibit a sequence of singlet and doublet ground states with different local occupations, local moments, and entanglement properties. A particularly useful simplification occurs at a high-symmetry condition (HSC), defined explicitly below, where the model maps exactly onto the repulsive Hubbard model. This provides a direct interpretation of the different regimes: the strongly correlated region corresponds to the half-filled Mott regime, while the intermediate region maps onto the doped, gapless Tomonaga--Luttinger-liquid regime. Away from the HSC, this exact correspondence is lost, and even a relatively small finite hopping can qualitatively modify the phase structure and open a gap in the intermediate regime. In two-dimensional square clusters, additional higher-spin ground states appear, including robust triplet phases that can persist away from the HSC.

For larger interacting systems, we employ both tensor network and neural network approaches. MPS simulations optimized using the DMRG method provide highly accurate results for long one-dimensional chains. We show that three distinct regions emerge with increasing system size, characterized by the energy gap and bond entanglement entropy. For two-dimensional clusters, MPS calculations serve as a controlled benchmark for moderate lattice sizes.

In parallel, we apply NQS-VMC to two-dimensional clusters. Here, the transformation to a particle-number conserving representation allows avoiding explicit Pfaffian evaluations. We benchmark different variational architectures, including Slater and Jastrow--Slater wave functions, restricted Boltzmann machine (RBM) constructions~\cite{carleo2018_constructing,netket2:2019,netket3:2022,Nomura2017restricted,stokes2020phases}, and neural backflow states~\cite{luo2019backflow,Liu2024Unifying}. While several of these ans\"atze perform well in specific parameter regimes, we find that the neural backflow ansatz provides the most consistent accuracy across different interaction strengths and phases. For this reason, neural backflow states constitute the primary NQS approach used throughout this work.

\section{Model and methods
\label{sec:model}}
We now introduce the effective model describing hybrid QD clusters coupled to SC leads or surfaces in the SC atomic limit (SC-AL), where the SC gap $\Delta$ is assumed to be the largest energy scale ($\Delta \to \infty$). 
In this limit, the system is governed by the competition between local electronic correlations and induced pairing. We also outline the methodological framework employed throughout this work, which forms the basis for both the analytical treatment and the numerical simulations presented below.
\subsection{Effective model of quantum dot cluster on superconducting surface}
We study finite regular clusters of single-level impurities, hereafter referred to as QDs, on an SC surface. As discussed in Appendix~\ref{app:SCAL}, the model follows from the superconducting Anderson impurity model in the SC-AL together with several additional simplifying assumptions. The resulting effective Hamiltonian reads 
\begin{equation}
\begin{aligned}
\mathcal{H} &= \sum_{j=1}^{L} \sum_{\sigma} \epsilon_j n_{j \sigma} -\sum_{i=1}^{L}\sum_{j\neq i}^{L}\sum_{\sigma}t_{ij} d_{i\sigma}^\dagger d_{j\sigma}^\pdag
+ \sum_{j=1}^{L}U_j n_{j\uparrow}n_{j\downarrow} \\
 &- \sum_{i=1}^{L}\sum_{j\neq i}^{L}\Gamma_{ji}  \left(d_{i\uparrow}^\dagger d_{j\downarrow}^\dagger + d_{i\downarrow}^\pdag d_{j\uparrow}^\pdag\right) - \sum_{j=1}^{L} \Gamma_{jj}\left( d_{j\uparrow}^\dagger d_{j\downarrow}^\dagger + d_{j\downarrow} d_{j\uparrow} \right),
 \label{eq:Ham1}
\end{aligned}
\end{equation}
where $d_{j\sigma}^\dag$ ($d_{j\sigma}^\pdag$) creates (annihilates) an electron with spin $\sigma$ on the QD with index $j$ and on-site energy $\epsilon_j$. Here $n_{j\sigma}=d_{j\sigma}^\dagger d_{j\sigma}^\pdag$ and $L$ is the number of QDs, i.e., the total cluster/lattice size of the system. The second term describes direct hopping between the dots with $t_{jj}=0$ for all $j$. The third term describes the local Coulomb interaction with strength (charging energy) $U_j$ on the dot $j$. 
The last two terms describe effective SC coupling where $\Gamma_{jj}\equiv\Gamma_j$ represents the local induced SC pairing on the dots, and the cross terms $\Gamma_{i \neq j}$ describe processes where a Cooper pair splits onto two different QDs.

The geometry of the ensemble is encoded in the inter-dot matrix elements $t_{ij}$ and $\Gamma_{i\neq j}$. In the following, we restrict both terms to nearest-neighbor bonds. We assume uniform onsite parameters $U_j\equiv U$ and $\Gamma_{jj}\equiv\Gamma$, and take the same hopping amplitude $t_{ij}\equiv t$ and nonlocal pairing amplitude $\Gamma_{ij}\equiv\zeta\Gamma$ ($\zeta\in[-1,1]$) for all nearest-neighbor pairs $\langle i,j\rangle$; all other inter-dot matrix elements are set to zero. 
Here, $\zeta$ should be understood as an effective measure of the nonlocal correlations mediated by the common SC environment, rather than as a direct parametrization of the inter-dot distance; its microscopic interpretation and the limitations associated with the SC-AL are discussed in Appendix~\ref{app:SCAL}.
We define shifted energy levels $\varepsilon=\epsilon_j+U/2$ such that $\varepsilon=0$ corresponds to the half-filled case on a bipartite lattice whenever at least one of $t$, $\Gamma$, or $\zeta$ is zero. Unless stated otherwise, we assume open boundary conditions.

Throughout the paper we use $\Gamma$ as the energy unit and choose the interaction range such that it covers both sides of the isolated-dot
singlet--doublet transition, which occurs at $U=2\Gamma$ for $\varepsilon=0$. We focus mainly on small direct hopping $t\ll\Gamma$ and the vicinity of the half-filling, as this is a typical experimental regime~\cite{Li2025individual,Li2025negative}, while selected larger values are used to illustrate the evolution away from this regime. 

It is convenient to rewrite the Hamiltonian into the form
\begin{equation}
\begin{aligned}
\mathcal{H} &= \, \varepsilon \sum_{j=1}^{L} \sum_{\sigma} \left( n_{j \sigma} - \frac{1}{2} \right) 
-t\sum_{\langle i,j\rangle}\sum_{\sigma} \left(d_{i\sigma}^\dagger d_{j\sigma}^\pdag + d_{j\sigma}^\dagger d_{i\sigma}^\pdag\right)
+ \frac{U}{2} \sum_{j=1}^{L} \left( n_{j \uparrow} + n_{j \downarrow} - 1 \right)^2 \\
& - \Gamma \sum_{j=1}^{L} \left( d_{j\uparrow}^\dagger d_{j\downarrow}^\dagger + d_{j\downarrow} d_{j\uparrow} \right)
 - \zeta \Gamma \sum_{\langle i,j \rangle} \left( d_{j\uparrow}^\dagger d_{i\downarrow}^\dagger + d_{i\uparrow}^\dagger d_{j\downarrow}^\dagger + d_{j\downarrow} d_{i\uparrow} + d_{i\downarrow} d_{j\uparrow} \right),
 \label{eq:Ham2}
\end{aligned}
\end{equation}
where we omitted an irrelevant constant energy shift. Because the SC pairing terms change the particle number by $\pm 2$, the Hamiltonian does not conserve the total particle number $N$. It does, however, conserve fermion parity, thereby separating the Hilbert space into even- and odd-parity sectors. Since we assume spin-independent hopping and do not include spin-orbit coupling, Zeeman terms, or triplet SC pairing, the Hamiltonian also conserves the total spin, i.e.,
\begin{equation}
\left[\mathcal{H},\bm{S}_T^2\right]=0,\;\;\;\;\left[\mathcal{H},S_z\right]=0.
\end{equation}
This allows us to analyze the properties of the ground-state by classifying phases according to the total spin $S$, defined by $\bm{S}_T^2\ket{\psi}=S(S+1)\ket{\psi}$, together with other expectation values such as double occupancy $\mathcal{D}=\sum_i \langle n_{i\uparrow} n_{i\downarrow}\rangle$, which measures the number of sites doubly occupied. The latter quantity is related to the local spin moment via
\begin{equation}
\mathcal{S}^{(2)}_z=\sum_{i=1}^L \langle S_{zi}^2 \rangle = \frac{1}{4} \langle N\rangle - \frac{1}{2} \mathcal{D},
\end{equation}
where $S_{zi}=(n_{i\uparrow}-n_{i\downarrow})/2$. Note that even under the above assumptions, namely in the absence of explicit spin-orbit coupling or spin-dependent hopping, the system can still be tuned into topologically nontrivial regimes. This, however, requires a finite phase difference between the lattice sites, leading to uneven and complex pairing amplitudes $\Gamma_{ij}$~\cite{Antonelli2025exploring,Ebert2025sextets}. Here, for simplicity, we assume that no SC phase difference exists between the quantum-dot regions, and restrict our analysis to topologically trivial regimes. 


\subsection{Transformation to particle conserving basis}
The absence of particle-number conservation in Hamiltonian~\eqref{eq:Ham2} poses a technical challenge for standard fermionic NQS and their associated Monte Carlo update schemes, as well as for conventional implementations of DMRG and continuous-time QMC methods. In addition, many numerical frameworks, such as NetKet~\cite{netket3:2022} or TeNPy~\cite{tenpy2024}, provide highly optimized Hilbert-space constructions for fermionic systems with fixed particle number, which are substantially more efficient than their nonconserving counterparts.

One possible strategy is to employ generalized variational ans\"atze tailored to particle-number–nonconserving states. However, these constructions are typically more involved and computationally demanding. 
Instead, we employ a standard partial particle--hole transformation, often referred to as the Shiba  transformation~\cite{Shiba-1972}, which maps between repulsive and attractive Hubbard-type representations and has been extensively discussed in the literature~\cite{lieb-1989,Scalettar-1989,yang-1990,essl-2024}. The same type of transformation was previously used in SC impurity problems~\cite{Luitz2010weak,Pokorny2021footprints}. In the present
context, its practical advantage is that it restores particle-number conservation:
\begin{equation}
\begin{aligned}
   & d_{j\uparrow}^\dagger \rightarrow \tilde{d}_{j\uparrow}^\dagger,\enspace
    d_{j\uparrow} \rightarrow \tilde{d}_{j\uparrow},\enspace
    d_{j\downarrow}^\dagger \rightarrow \tilde{d}_{j\downarrow}, \enspace
    d_{j\downarrow} \rightarrow \tilde{d}_{j\downarrow}^\dagger, \\
   & n_{j \uparrow}=d_{j\uparrow}^\dagger d_{j\uparrow} \rightarrow \tilde{n}_{j \uparrow}, \\
   & n_{j \downarrow}=d_{j\downarrow}^\dagger d_{j\downarrow} \rightarrow \tilde{d}_{j\downarrow}\tilde{d}_{j\downarrow}^\dagger=1-\tilde{d}_{j\downarrow}^\dagger\tilde{d}_{j\downarrow}=1-\tilde{n}_{j \downarrow}.
   \label{eq:trans}
\end{aligned}
\end{equation}
This transformation maps the SC system to a model with attractive local interaction $-U$ between electrons with different spins that is described as
\begin{equation}
\begin{aligned}
\tilde{\mathcal{H}} &= \, \varepsilon \sum_{j=1}^{L} \left( \tilde{n}_{j \uparrow} -  \tilde{n}_{j \downarrow}\right) 
-t\sum_{\langle i,j\rangle}\left(\tilde{d}_{i\uparrow}^\dagger \tilde{d}_{j\uparrow}^\pdag - \tilde{d}_{i\downarrow}^\dagger \tilde{d}_{j\downarrow}^\pdag + \text{H.c.}\right)
+ \frac{U}{2} \sum_{j=1}^{L} \left( \tilde{n}_{j \uparrow} - \tilde{n}_{j \downarrow} \right)^2 \\
& - \Gamma \sum_{j=1}^{L} \left( \tilde{d}_{j\uparrow}^\dagger \tilde{d}_{j\downarrow}^\pdag + \tilde{d}_{j\downarrow}^\dagger \tilde{d}_{j\uparrow}^\pdag \right)
 - \zeta \Gamma \sum_{\langle i,j \rangle} \left( \tilde{d}_{j\uparrow}^\dagger \tilde{d}_{i\downarrow}^\pdag + 
 \tilde{d}_{i\uparrow}^\dagger \tilde{d}_{j\downarrow}^\pdag+
  \tilde{d}_{j\downarrow}^\dagger \tilde{d}_{i\uparrow}^\pdag+
 \tilde{d}_{i\downarrow}^\dagger \tilde{d}_{j\uparrow}^\pdag \right).
 \label{eq:Ham3}
\end{aligned}
\end{equation}
This introduces effective spin-flip terms; therefore, the transformed Hamiltonian no longer conserves the total spin. As used here, the transformation is useful when it leads to conservation of the total number of particles $\tilde{N}_f$ in the transformed system. According to the last line of Eq.~\eqref{eq:trans}, this requires that the original model must conserve the total spin projection $S_z=\sum_i (n_{i\uparrow}-n_{i\downarrow})/2$. This requirement immediately excludes models containing spin-dependent hopping or spin-orbit coupling. For example, terms of the form $d^\dagger_{i\uparrow} d_{j\downarrow}$ are transformed into particle-number-nonconserving contributions such as $\tilde{d}^\dagger_{i\uparrow} \tilde{d}^\dagger_{i\downarrow}$. In such situations, alternative analytical or numerical approaches are more suitable.

Nevertheless, it should be emphasized that the class of systems for which this transformation is advantageous remains broad, encompassing both effective and microscopic descriptions of complex assemblies on SC surfaces as well as multidot and/or multiterminal SC junctions.

In NQS implementations, we typically employ spinless fermions in order to avoid technical complications arising from the fact that the transformed particle conserving system does not conserve the total spin. This is achieved by a simple relabeling of the lattice degrees of freedom, ${j,\uparrow}\rightarrow 2j-1$ and ${j,\downarrow}\rightarrow 2j$. Since in the Monte Carlo sampling, we typically use an update scheme that moves a randomly selected particle to a different randomly chosen empty mode, this procedure explicitly preserves the spin-parity constraint $(-1)^{\tilde{N}_\uparrow-\tilde{N}_\downarrow}$, which follows from the conservation of particle-number parity in the original Hamiltonian.

To distinguish expectation values in the original SC representation from those in the transformed particle conserving representation, we denote quantities in the latter by a tilde. For example, the total spin ${\cal S}$ refers to the SC representation, whereas the total particle number $\tilde{N}_f$ refers to the transformed representation.

As a final remark, let us identify the physical processes in Hamiltonian~\eqref{eq:Ham3} by applying the standard fermion-(pseudo)spin mapping 
\begin{equation}
\begin{aligned}
\tilde{\mathcal{H}} =& 
-t\sum_{\langle i,j\rangle}\left(\tilde{d}_{i}^\dagger \sigma_z \tilde{d}_{j}^\pdag + \text{H.c.}\right)- \zeta 2 \Gamma \sum_{\langle i,j \rangle} \left( \tilde{d}_{j\uparrow}^\dagger \tilde{d}_{i\downarrow}^\pdag +  \tilde{d}_{i\downarrow}^\dagger \tilde{d}_{j\uparrow}^\pdag \right)\\
&+2\varepsilon\sum_{j=1}^{L} \tilde{S}_j^z
-2\Gamma\sum_{j=1}^{L} \tilde{S}_j^x + 2U\sum_{j=1}^{L} \left(\tilde{S}_j^z\right)^2.
\end{aligned}
\end{equation}
The first term, which involves the Pauli matrix $\sigma_z$, represents spin-resolved hopping. The second term corresponds to a nonlocal spin-flip hopping process. The third term describes an effective magnetic field $2\varepsilon$ oriented along the $z$ axis, while the fourth term accounts for the on-site transverse magnetic field $-2\Gamma$ in the $x$ direction. The fifth term represents a uniaxial anisotropy.

\subsection{Methods}
\subsubsection{Exact methods} 

As shown in Sect.~\ref{sec:NoInt}, the noninteracting limit can be treated analytically. Analytical solutions can also be obtained for small interacting systems, see Appendix~\ref{app:SQD} for $L=1$ and Appendix~\ref{app:L2} for $L=2$ at $\varepsilon=0$ and $t=0$. Small-size systems ($L \leq 12$) with finite $U$  were investigated by exact diagonalization (ED), either via full diagonalization of the Hamiltonian ($L \leq 6$) or using the Lanczos algorithm as implemented in standard numerical libraries. Both the original and the canonically transformed formulations were benchmarked. For larger system sizes, we employ NQS-VMC implemented within the open-source framework NetKet~\cite{netket3:2022} and MPS-DMRG implemented within the open-source framework TeNPy~\cite{tenpy2024}.
    
\subsubsection{Variational Monte Carlo and neural quantum states} 

NQS provide a flexible class of variational wave functions that can efficiently approximate many-body quantum states within the VMC framework~\cite{Gubernatis2016QMC}. 
Their use is motivated by the universal approximation theorem, which ensures that sufficiently expressive neural networks can approximate continuous functions to arbitrary accuracy~\cite{hornik1989multilayer,devore2021neural}. 
Over the past decade, a broad spectrum of NQS architectures has been developed~\cite{Carleo2017_notes,lange2024architectures}, tailored to different symmetry constraints and correlation structures.

\paragraph{Jastrow--Slater architecture:}
We first consider a hierarchy of variational ans\"atze of increasing expressivity. 
As a baseline, we use a simple Slater determinant
\begin{equation}
\Psi_{\mathrm{S}}(n) = \det A(n), 
\qquad 
A(n) = M_{R(n)},
\end{equation}
where \(n\in\{0,1\}^{N_{\mathrm{orb}}}\) is an occupation configuration with $\tilde{N}_f$ particles, and 
\(M\in\mathbb{C}^{N_{\mathrm{orb}}\times \tilde{N}_f}\) is a variational orbital matrix. 
Denoting by \(R(n)=\{r_1,\dots,r_{\tilde{N}_f}\}\) the occupied orbitals in configuration \(n\), the matrix
$A(n)_{ij}=M_{r_i,j}$
is the \(\tilde{N}_f\times \tilde{N}_f\) submatrix of \(M\) obtained by selecting the occupied rows. This construction enforces fermionic antisymmetry exactly, but the orbitals are fixed and do not depend on the configuration, which limits the ability to describe many-body correlations.

A neural correlator can be introduced via a Jastrow factor
\begin{equation}
\Psi_{\mathrm{J}}(n) = J(n;\theta),
\end{equation}
where $J(n;\theta)$ is parametrized by a neural network. 
In conventional variational wave functions, a Jastrow factor is a configuration-dependent multiplicative weight, often written in an exponential form involving density-density correlators, which is used to incorporate correlations beyond an uncorrelated reference state. The neural-network parametrization generalizes this idea by allowing $J(n;\theta)$ to represent more complex correlations directly from the occupation configuration $n$.
While such ans\"atze are formally expressive, they do not impose antisymmetry explicitly, and the fermionic sign structure must be learned implicitly during optimization.

Combining both ingredients yields the Jastrow--Slater ansatz
\begin{equation}
\Psi_{\mathrm{JS}}(n) = J(n;\theta)\,\det A(n).
\end{equation}
Here, the determinant fixes the antisymmetric structure and defines the nodal surface, while the Jastrow factor introduces correlations through a multiplicative reweighting of configurations. 
For standard choices of $J(n;\theta)$ (e.g.\ positive-definite exponentials), the nodal structure is entirely determined by $M$, and the Jastrow factor modifies only the amplitude of the wave function. 
Thus, correlations enter outside the determinant and do not affect its internal structure. In practice, we parametrize $J(n;\theta)$ using two standard architectures: a feed-forward neural network acting on $\log|\Psi|$ (JS-NQS) and an RBM, leading to an RBM--Slater (RBMS-NQS) ansatz. 

\paragraph{Neural backflow ansatz:}
As the main NQS ansatz in this work, we employ a neural-network backflow (BF) construction~\cite{luo2019backflow}. 
In contrast to the Jastrow--Slater form, where correlations enter multiplicatively, the BF ansatz introduces correlations directly into the determinant by making the orbital matrix configuration-dependent.

The wave function is written as
\begin{equation}
\Psi_{\theta}(n) = \det A_{\theta}(n), 
\qquad 
A_{\theta}(n) = \bigl[M + F_\theta(n)\bigr]_{R(n)},
\end{equation}
where $M \in \mathbb{C}^{N_{\mathrm{orb}} \times \tilde{N}_f}$ is a static orbital matrix and 
$
F_\theta(n) \in \mathbb{C}^{N_{\mathrm{orb}} \times \tilde{N}_f}
$
is a configuration-dependent correction generated by a neural network. 

In our implementation, $F_\theta(n)$ is produced by a feed-forward neural network acting on the binary occupation vector $n$, mapping
$
n \;\longmapsto\; F_\theta(n),
$
where the network output is reshaped into an $N_{\mathrm{orb}} \times \tilde{N}_f$ matrix. 
For each configuration, the rows corresponding to occupied orbitals are selected to form $A_\theta(n)$, and the logarithm of the wave function is evaluated via $\log \det A_\theta(n)$.

This construction can be viewed as a configuration-dependent deformation of the orbital matrix,
$
M \;\longrightarrow\; M + F_\theta(n),
$
which promotes the single-particle orbitals to effective many-body objects. 
As a result, correlations are incorporated directly into the antisymmetric structure of the wave function. 
In particular, unlike the Jastrow--Slater ansatz, the nodal surface is no longer fixed by $M$, but can adapt dynamically through $F_\theta(n)$. 
This added flexibility proved to be crucial for accurately describing strongly correlated fermionic systems.

From a physical perspective, the BF transformation can be interpreted as dressing the fermions into effective quasiparticles whose orbitals depend on the surrounding configuration. 
This leads to a nonlinear reparameterization of both amplitudes and nodal structure, significantly enhancing the expressive power of the ansatz.

Recent work by Liu and Clark~\cite{Liu2024Unifying} places neural BF within a broader class of fermionic neural-network constructions, showing that several ans\"atze can be recast in terms of configuration-dependent orbital matrices. 
Within this framework, direct BF parameterizations of the form $M + F_\theta(n)$ are found to be particularly effective, typically yielding improved variational energies compared to more indirect or low-rank constructions. 
Moreover, once a sufficiently expressive BF transformation is included, additional multiplicative factors such as Jastrow or determinant corrections provide only marginal improvements.

For completeness, we also investigated several alternative NQS architectures, including pure Jastrow and RBM states, and MultiSlater expansions, which improve the wave function through linear combinations of determinants, enhancing the flexibility of the nodal structure in a controlled manner. 
Within the parameter regimes considered here, however, these approaches typically capture only subsets of the relevant phases, often those characterized by low entanglement. 
A detailed comparison is provided in the Appendix~\ref{app:NQS}.
 
\subsubsection{Density matrix renormalization group and matrix product states} 

To obtain independent tensor-network reference results, we employ MPS-DMRG as implemented in the TeNPy library~\cite{tenpy2024}. All calculations are performed in the particle-number-conserving representation, with fermionic statistics handled internally through a Jordan--Wigner transformation.

For one-dimensional chains, DMRG provides highly accurate results whose convergence can be systematically checked by increasing the bond dimension $\chi$. We therefore use it as the main reference method for long one-dimensional systems.

For two-dimensional clusters, the lattice is mapped onto a one-dimensional chain using a snake-like ordering. This introduces effective long-range couplings and increases the entanglement cost, so the required bond dimension grows rapidly with the system width. In this case, MPS-DMRG is used primarily as an independent benchmark for the NQS calculations on moderate system sizes.

Ground states are obtained using one- and two-site DMRG algorithms. 
We employ bond dimensions up to $\chi \approx 2000$. 
For one-dimensional chains, convergence can be systematically checked by increasing $\chi$,
and DMRG therefore provides highly accurate reference results. For the
two-dimensional clusters, the increased entanglement associated with the
one-dimensional MPS mapping limits the accessible system sizes and bond dimensions. In this case, we use MPS-DMRG primarily as an independent
benchmark for the NQS calculations.

Tensor-network approaches designed directly for two-dimensional systems, such as projected entangled pair states, provide a more natural representation of two-dimensional area-law states, but require approximate contraction and more involved optimization schemes. A systematic comparison with such approaches is beyond the scope of the present work, and we therefore restrict the tensor-network calculations to MPS-DMRG.

The Jupyter notebooks containing the implementations used in this work, together with illustrative examples and benchmark scripts, are available in the repository Ref.~\cite{gitlabrepo}.

\section{Results}
\subsection{Noninteracting case}
\label{sec:NoInt}
The core of the work focuses on the interacting finite clusters in different geometries. Nevertheless, to gain some important insight into the physical properties of model~\eqref{eq:Ham2}, it is useful to first investigate the non-interacting case ($U=0$) for both finite systems with open boundary conditions as well as the one- and two-dimensional systems in the $L\rightarrow\infty$ limit with periodic boundary conditions. The latter case is readily solved by using Bogoliubov-de-Gennes (BdG) and Fourier transformations. The solution for a lattice with spacing constants $a=1$ yields the spectrum
\begin{figure}[h]
    \centering
    \includegraphics[width=\linewidth]{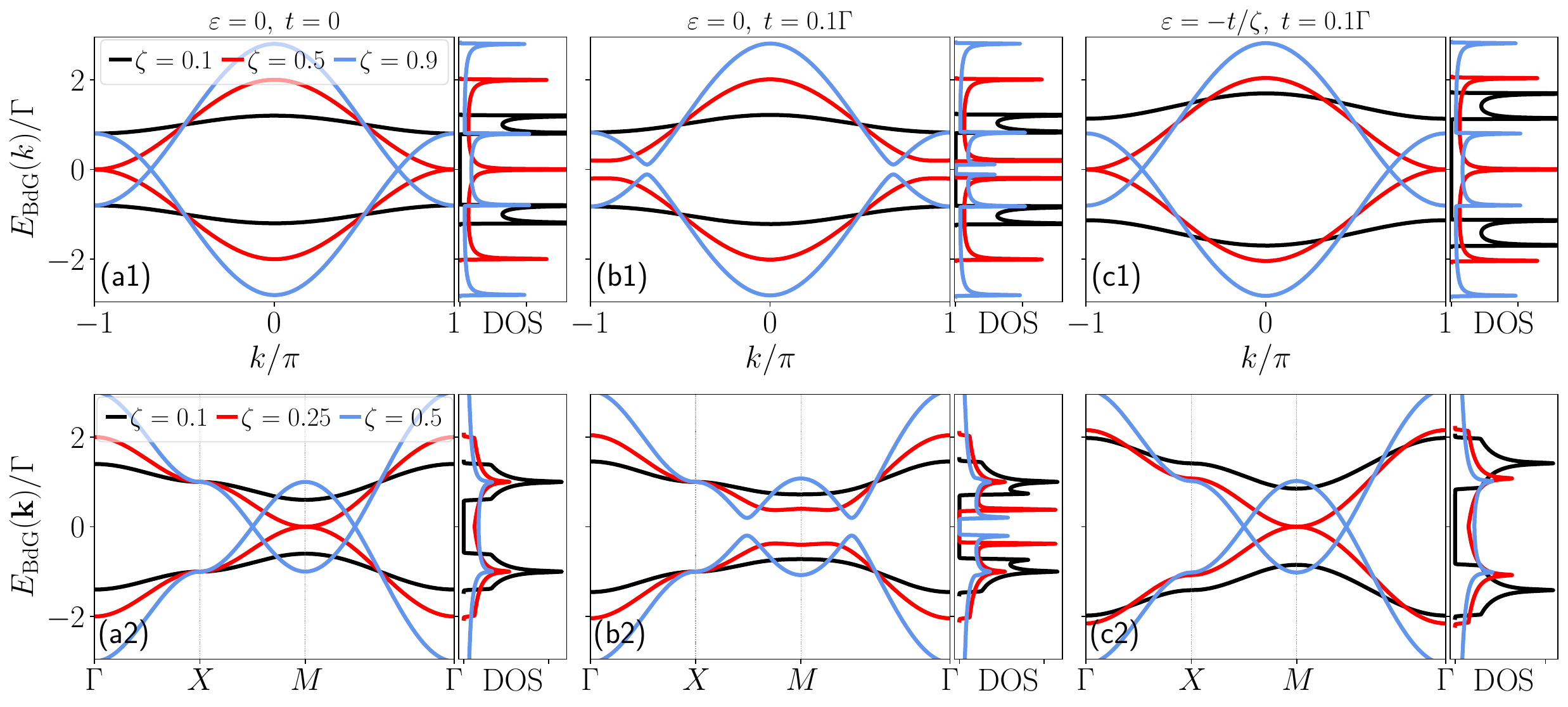}
    \caption{Examples of the band structure and the corresponding single-particle DOS in one (first row) and two (second row) spatial dimensions for infinite noninteracting system with periodic boundary conditions. 
Panels (a1), (a2), (c1), and (c2) display parameter regimes in which HSC is satisfied, leading to a closing of the gap. 
In contrast, panels (b1) and (b2) illustrate generic situations where this condition is not fulfilled and the spectral gap remains finite.
\label{fig:SpecInf}}
\end{figure}

\begin{equation}
        E_{\mathrm{BdG}}(\mathbf{k})=\pm \sqrt{(\epsilon-t\gamma_{\mathbf{k}})^2+\Gamma^2(1+\zeta \gamma_{\mathbf{k}})^2},
        \label{eq:disp_rel}
\end{equation}
where $\gamma_{\mathbf{k}}=2\cos(k)$ for one-dimensional chain and $\gamma_{\mathbf{k}}=2[\cos(k_x)+\cos(k_y)]$ for a two-dimensional square lattice. For a general case, there is a gap in the spectrum at the Fermi energy, which can, however, close at some special points.  The conditions for these special points are quite restrictive. A real-valued solution of $E(\mathbf{k})=0$ for $\gamma_{\mathbf{k}}$ exists, apart from the trivial case $\Gamma=0$, only when $t=-\varepsilon \zeta$, which we call the high-symmetry condition (HSC). For HSC the energies $ E^{\mathrm{HSC}}_{\mathrm{BdG}}(\mathbf{k})=\pm \left|1+\zeta \gamma_{\bm{k}}\right|\sqrt{\epsilon^2+\Gamma^2}$ are linear in $\zeta$. The gap then closes at $\gamma_{\mathbf{k}}^{E=0}=-1/\zeta$, which is, however, realizable only for $\abs{\zeta}\geq 1/2$ in one-dimensional chain and $\abs{\zeta}\geq 1/4$ on a two-dimensional square lattice. Note that at HSC, all spectra for a given dimension and $\zeta$ are identical up to a factor of $\sqrt{\epsilon^2+\Gamma^2}$. 

Figure~\ref{fig:SpecInf} shows the evolution of the band structure and the single-particle density of states (DOS). The top (bottom) row corresponds to the one-dimensional (two-dimensional) case. The first and third columns represent HSC scenarios with $\varepsilon=0$ and $t=0$ in panels (a1) and (a2), and $\varepsilon=-t/\zeta$ with $t=0.1\Gamma$ in panels (c1) and (c2). The central column illustrates a general parameter regime in which the gap does not close for any $\zeta$.

The values $\zeta=0.1$, $0.5$, and $0.9$ in one dimension and $\zeta=0.1$, $0.25$, and $0.5$ in two dimensions are chosen to lie below, at, and above the critical value $\zeta_c$ at which the gap closes, provided the remaining parameters satisfy the necessary conditions.

For columns (a) and (c), the interpretation is straightforward. When $\zeta<\zeta_c$, two well-separated bands are present, each exhibiting a tight-binding-like DOS with van Hove singularities at the band edges in one dimension and at the band center in two dimensions. For $\zeta\geq\zeta_c$, the bands overlap and the gap closes. By contrast, in column (b) the HSC is not fulfilled, and therefore the gap remains open for all values of $\zeta$. 

\begin{figure}[h]
    \centering
    \includegraphics[width=\linewidth]{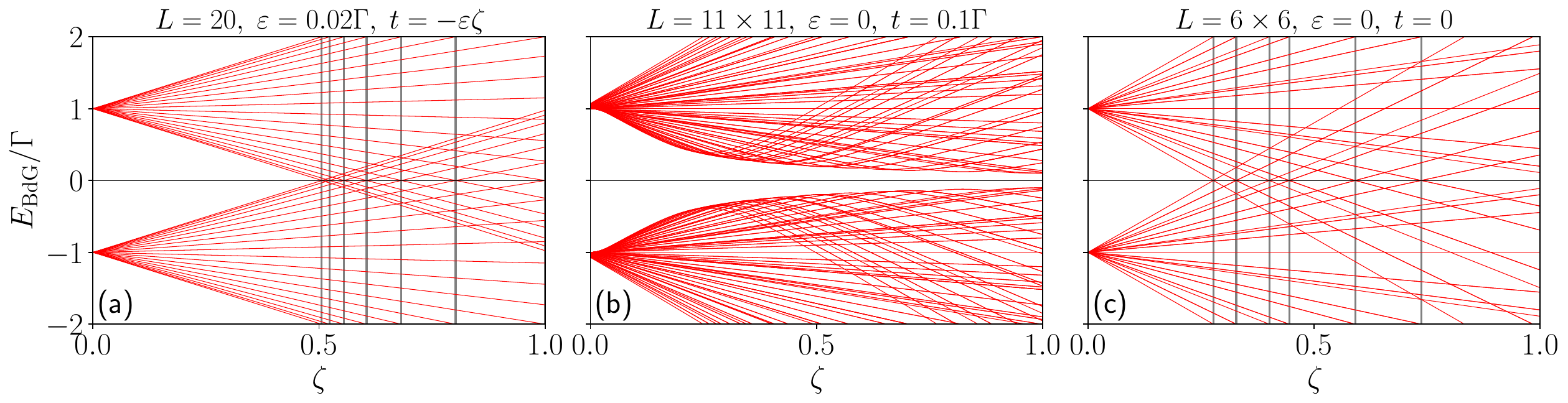}
    \caption{Evolution of the BdG single-particle spectrum for a finite noninteracting system with $L$ sites as a function of $\zeta$. Panel (a) shows a chain with $L=20$ in the HSC regime, where the gap closes at the critical values indicated by the vertical lines. Panel (b) displays an $11\times11$ two-dimensional cluster in a generic parameter regime, where the gap remains finite for all $\zeta$. Panel (c) illustrates a $6\times6$ cluster with trivially fulfilled HSC condition ($\varepsilon=0$, $t=0$). In panels (a) and (c), the vertical lines mark the analytical predictions from Eq.~\eqref{eq:crzeta} for the zero-energy crossings, signaling sharp transitions between distinct singlet ground states.
\label{fig:SpecFin}}
\end{figure}

These results extend directly to finite clusters. The single-particle energies of a regular rectangular cluster with open boundary conditions read
\begin{equation}
        E^{d}_{\mathrm{BdG}}(L;\bm{n})=\pm \sqrt{\left[\epsilon-t\gamma_{d}(L;\bm{n})\right]^2+\Gamma^2\left[1+\zeta \gamma_{d}(L;\bm{n})\right]^2},
        \label{eq:disp_rel}
\end{equation}
where 
\begin{equation}
\gamma_{1D}(L;n) = 2\cos{\left(\frac{n \pi}{L+1}\right)},
\end{equation}
for one-dimensional chain with integer parameter $1\leq n \leq L$ and 
\begin{equation}
\gamma_{2D}(L_x,L_y;n_x,n_y)= 2\left[\cos{\left(\frac{n_x \pi}{L_x+1}\right)} + \cos{\left(\frac{n_y \pi}{L_y+1}\right)}\right]
\end{equation}
for rectangular cluster of size $L_x\times L_y=L$ with integer parameters $1\leq n_x\leq L_x$, and $1\leq n_y \leq L_y$. 
If the necessary condition $t=-\varepsilon \zeta$ is satisfied, the single-particle energies reach zero at discrete values of $\zeta$ given by
\begin{equation}
\zeta_{1D}(L;n) = \frac{-1}{\gamma_{1D}(L;n)},\quad\text{or}\quad
\zeta_{2D}(L_x,L_y;n_x,n_y) = \frac{-1}{\gamma_{2D}(L_x,L_y;n_x,n_y)},
\label{eq:crzeta}
\end{equation}
which is illustrated in Fig.~\ref{fig:SpecFin}. As we discuss in the next section, these critical values of $\zeta$ play an important role in the phase diagram in the interacting case.   

\begin{figure}[h]
    \centering
    \includegraphics[width=\linewidth]{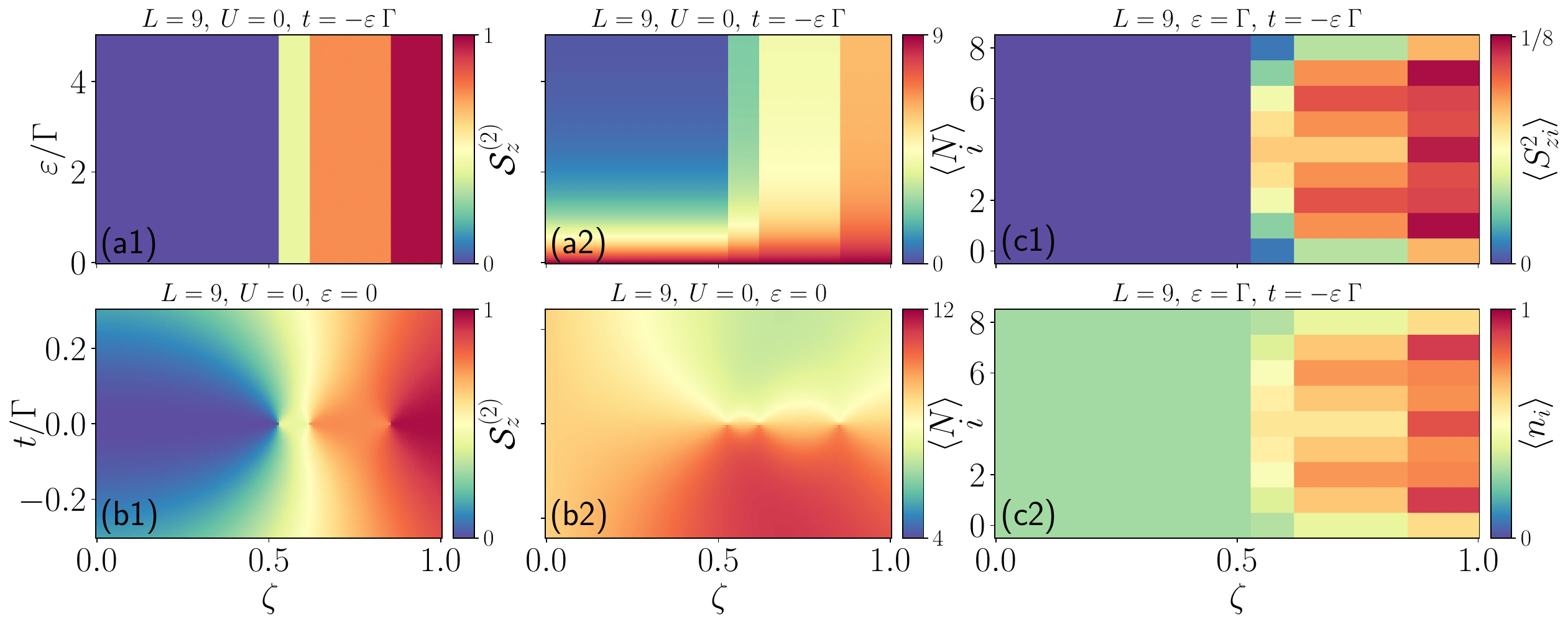}
    \caption{Local $z$-moment $\mathcal{S}^{(2)}_z$ (a1) and mean total electron occupation $\langle N \rangle$ (a2) in the $\zeta$-$\varepsilon$ plane within the HSC regime for a noninteracting chain of length $L=9$. 
Panels (b1) and (b2) show the corresponding quantities away from HSC at $\varepsilon=0$, where the evolution is smooth up to special points $\zeta_d(L,\bm{n})$ at $t=0$. 
Panels (c1) and (c2) illustrate the spatial distribution of the local moment and particle occupation for representative singlet ground states at $\varepsilon=\Gamma$, highlighting the qualitative difference between the homogeneous trivial singlet at $\abs{\zeta}<0.5$ and the inhomogeneous singlets realized at larger $|\zeta|$.
\label{fig:U0maps}}
\end{figure}

In the noninteracting limit, the many-body ground state is always a singlet; however, its structure changes at each critical value $\zeta_d(L,\bm{n})$. Within the HSC regime, these transitions are discontinuous. This behavior is illustrated in Fig.~\ref{fig:U0maps}(a1), where we show the local moment $\mathcal{S}^{(2)}_z$ in the $\zeta$–$\varepsilon$ plane for a chain of length $L=9$. The local moment remains constant within each interval bounded by consecutive $\zeta_d(L,\bm{n})$, but increases stepwise at the critical points, starting from zero at small $|\zeta|$ (negative $\varepsilon$ and $\zeta$ are omitted from the plot due to symmetry). As shown in Fig.~\ref{fig:U0maps}(a2), away from half filling this discontinuous change in the local moment is accompanied by a sudden jump in the mean electron occupation. Importantly, although not shown explicitly, the HSC ensures that, except for the trivial singlet realized at $|\zeta|<\zeta_1(L,1)$, the mean filling never reaches zero ($2L$), even in the limit $\varepsilon \gg \Gamma$ ($\varepsilon \ll -\Gamma$).

Away from the HSC regime, the evolution becomes smooth, as demonstrated in Fig.~\ref{fig:U0maps}(b1) and (b2). Discontinuities in the $\zeta$–$t$ plane occur only directly at the HSC condition, i.e., at $t=0$ and $\varepsilon=0$.

The qualitative distinction between the singlet states in the HSC regime is most clearly seen in their real-space structure. Figures~\ref{fig:U0maps}(c1) and (c2), shown for $\varepsilon=\Gamma$ at HSC, reveal that both the local moment and the mean occupation are spatially homogeneous for $|\zeta|<\zeta_{1D}(L,L)$, while the other singlets exhibit a characteristic inhomogeneous distribution. The corresponding results in two dimensions are analogous.

\subsection{High-symmetry condition and Nambu rotation}

To reveal the physical meaning of the HSC, it is useful to revisit the Hamiltonian transformation. Although the same result can be obtained by applying an additional transformation to Hamiltonian~\eqref{eq:Ham3}, it is more transparent to derive it directly from Hamiltonian~\eqref{eq:Ham2} using a standard Bogoliubov transformation.

By introducing the Nambu spinor $\Psi_j^\dagger=(d_{j\uparrow}^\dagger,d_{j\downarrow})$, we can rewrite the non-interacting part of Hamiltonian~\eqref{eq:Ham2} as
\begin{equation}
\begin{aligned}
\mathcal{H}(U=0) &= \sum_{j=1}^{L} \Psi_j^\dagger\left(\varepsilon\sigma_z - \Gamma \sigma_x\right)\Psi_j 
-\sum_{\langle i,j\rangle}\left( \Psi_i^\dagger\left(
t\sigma_z + \zeta \Gamma\sigma_x
\right) \Psi_j + \mathrm{H.c.}\right).
\label{eq:HamNambu}
\end{aligned}
\end{equation}

We can now perform a Nambu rotation
\begin{equation}
\Psi_j=
\begin{pmatrix}
d_{j\uparrow}\\
d_{j\downarrow}^{\dagger}
\end{pmatrix}
=
\begin{pmatrix}
\cos\frac{\theta}{2} & -\sin\frac{\theta}{2}\\
\sin\frac{\theta}{2} & \cos\frac{\theta}{2}
\end{pmatrix}
\begin{pmatrix}
c_{j\uparrow}\\
c_{j\downarrow}^{\dagger}
\end{pmatrix},
\end{equation}
where, assuming $\Gamma>0$,
\begin{equation}
{\cal E}=\sqrt{\varepsilon^2+\Gamma^2},
\qquad
\cos\frac{\theta}{2}=\sqrt{\frac{{\cal E}+\varepsilon}{2{\cal E}}},
\qquad
\sin\frac{\theta}{2}=-\sqrt{\frac{{\cal E}-\varepsilon}{2{\cal E}}}.
\end{equation}

Applying this transformation to Hamiltonian~\eqref{eq:Ham2}, and noting that the interaction term is invariant under the rotation, we obtain
\begin{equation}
\begin{aligned}
\mathcal{H}^c
={}&
{\cal E}\sum_{j=1}^{L}
\left(
n^{c}_{j\uparrow}
+
n^{c}_{j\downarrow}
-1
\right)
-
{\cal T}\sum_{\langle i,j\rangle}\sum_{\sigma}
\left(
c_{i\sigma}^{\dagger}c_{j\sigma}
+
c_{j\sigma}^{\dagger}c_{i\sigma}
\right)
+
\frac{U}{2}
\sum_{j=1}^{L}
\left(
n^{c}_{j\uparrow}
+
n^{c}_{j\downarrow}
-1
\right)^2
\\
&-
{\cal D}\sum_{\langle i,j\rangle}
\Big(
c_{i\uparrow}^{\dagger}c_{j\downarrow}^{\dagger}
+
c_{j\uparrow}^{\dagger}c_{i\downarrow}^{\dagger}
+
c_{j\downarrow}c_{i\uparrow}
+
c_{i\downarrow}c_{j\uparrow}
\Big),
\end{aligned}
\label{eq:HBog}
\end{equation}
with
\begin{equation}
{\cal T}=
\left(\varepsilon t-\zeta\Gamma^2\right)/
{\cal E},
\qquad
{\cal D}=
\Gamma(t+\varepsilon\zeta)/
{\cal E}.
\label{eq:hubTD}
\end{equation}

In general, this Hamiltonian is not particle conserving because of the last anomalous term. However, under the high-symmetry condition
$
(t+\varepsilon\zeta=0),
$ one has ${\cal D}=0$ and ${\cal T}=-\zeta{\cal E}$. The anomalous term therefore vanishes and the original model maps onto the standard Hubbard model
\begin{equation}
\mathcal{H}^c_{\mathrm{HSC}}
=
-t_{\mathrm H}
\sum_{\langle i,j\rangle}\sum_{\sigma}
\left(
c_{i\sigma}^{\dagger}c_{j\sigma}
+
c_{j\sigma}^{\dagger}c_{i\sigma}
\right)
+
U\sum_{j=1}^{L}
n^c_{j\uparrow}n^c_{j\downarrow}
-
\mu\sum_{j=1}^{L}\sum_{\sigma}
n^c_{j\sigma},
\label{eq:HBog_Hubbard}
\end{equation}
where
\begin{equation}
t_{\mathrm H}=-\zeta{\cal E},
\qquad
\mu=\frac{U}{2}-{\cal E},
\end{equation}
and we left out a constant energy shift
$L\left(\frac{U}{2}-{\cal E}\right)$. 

The mapped model is thus a standard repulsive-$U$ Hubbard model with the chemical potential shifted from the particle-hole-symmetric value $\mu=U/2$ by $-{\cal E}$. Changes of the parameters of the original model can therefore modify both the effective hopping and the filling of the Hubbard model.

At the HSC, conservation of the Hubbard particle number $N^c=\sum_{j\sigma}n^c_{j\sigma}$ provides a $U(1)$ symmetry, with fermion parity contained as its subgroup. In the original SC representation this corresponds to an additional quasispin symmetry that is hidden by the anomalous terms. The Bogoliubov transformation is spin singlet and leaves the complete spin algebra invariant, so that the total-spin quantum number is identical in the two representations.

Finally, in the absence of a magnetic field the Hamiltonian retains full spin-rotation symmetry. For a fixed $N^c$, it is therefore sufficient to diagonalize the sector with the minimal possible spin polarization: $N^c_\uparrow=N^c_\downarrow$ for even $N^c$, and $|N^c_\uparrow-N^c_\downarrow|=1$ for odd $N_c$. This sector contains a representative of every allowed total-spin multiplet, while the corresponding value of $S$ can be identified from $\langle\mathbf{S}^2\rangle$.

We use the exact correspondence between the original model at the HSC and the grand-canonical Hubbard model at fixed chemical potential for additional checks and for interpreting the results. However, to keep the numerical treatment uniform across the full parameter space, all DMRG and NQS calculations, including those at the HSC, are performed using the particle-number-conserving representation of Eq.~\eqref{eq:Ham3} obtained from the original canonical transformation.

\subsection{Interacting case}
For both one dimensional chains and two-dimensional clusters we begin with relatively small systems that remain accessible to ED, enabling the construction of detailed phase diagrams in the presence of Coulomb repulsion. Note that analytical solutions for the general $L=1$ case and the special $L=2$ case used in analysis of the numerical results are provided in Appendix~\ref{app:SQD} and Appendix~\ref{app:L2}.

\subsubsection{Open chains}  

\begin{figure}[h]
    \centering
    \includegraphics[width=\linewidth]{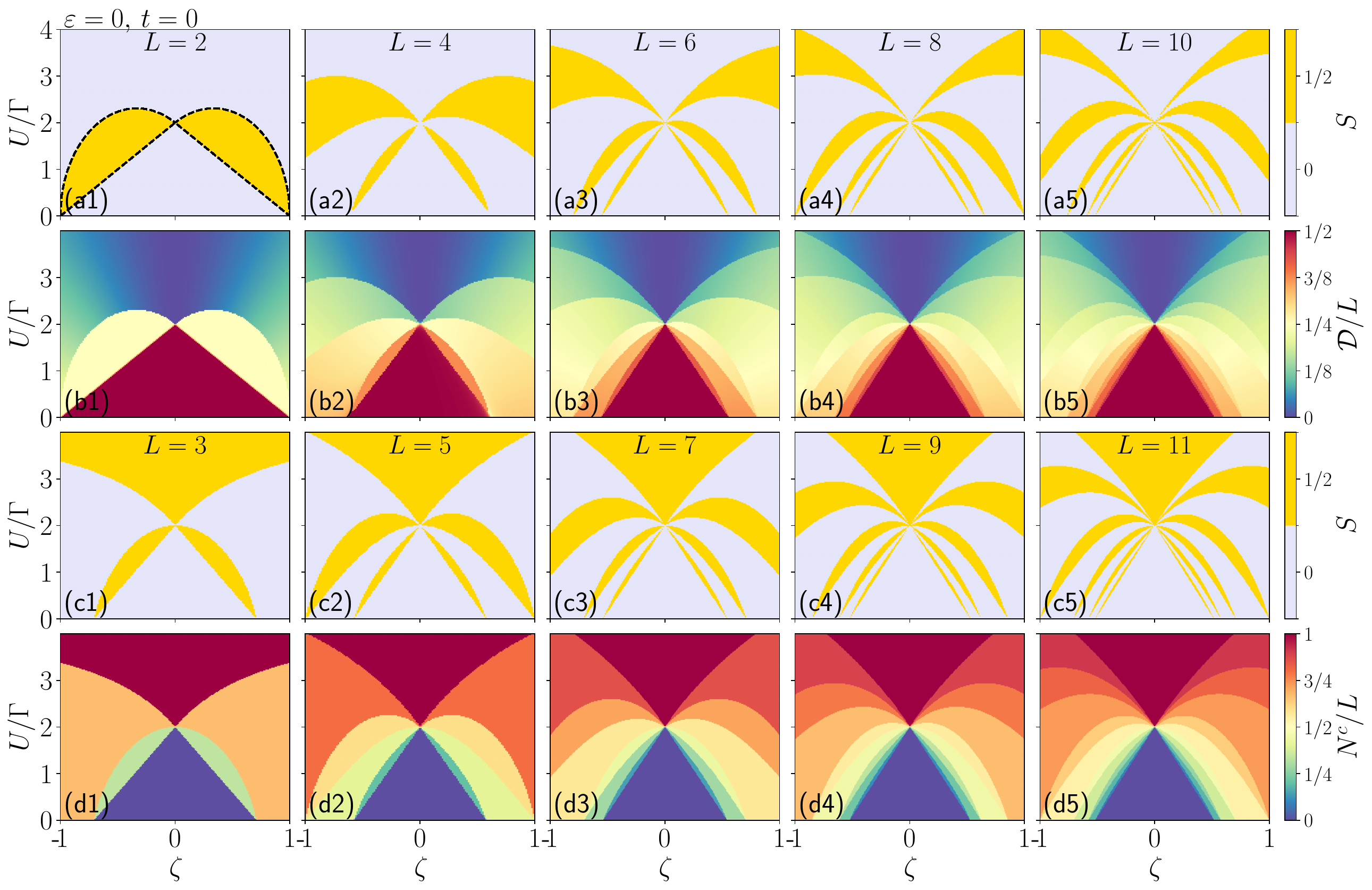}
    \caption{Phase diagrams in the $\zeta$–$U$ plane for a one-dimensional chain with an even (first and second row)  and odd (third and fourth row) number of sites,  $\varepsilon=0$ and $t=0$. 
    (a) and (c): Total spin of the many-body ground state. 
    (b): Average double occupancy $\mathcal{D}/L$. (d): Average occupancy of equivalent Hubbard model~\eqref{eq:HBog_Hubbard}.  Dashed black lines in (a1) are analytical phase boundaries calculated in Appendix~\ref{app:L2}.}
    \label{fig:1DUzetaEO}
\end{figure}

To isolate the competition between the on-site Coulomb interaction and the induced SC pairing, we first focus on the simplest HSC case, defined by $\varepsilon=0$ and $t=0$. Figure~\ref{fig:1DUzetaEO} shows the resulting phase diagrams in the $\zeta$--$U$ plane for chains of even and odd length. The first and third rows display the total spin of the ground state. For even chains, the second row shows the average double occupancy $\mathcal{D}/L$, while for odd chains, the fourth row shows the occupation number $N^c$ of the equivalent Hubbard model defined by Eq.~\eqref{eq:HBog_Hubbard}.

Only singlet and doublet ground states occur. This is consistent with the Lieb--Mattis ordering for the one-dimensional Hubbard model with open boundaries, for which the ground state has the minimum allowed total spin, $S=0$ for even particle number and $S=1/2$ for odd particle
number~\cite{LiebMattis1962,lieb-1989,Tasaki1998} as well as with the
exact Bethe-ansatz phase structure of the repulsive one-dimensional
Hubbard model at zero magnetic field~\cite{Luo-2024}.

Within each sector, distinct phases can be identified through the behavior of $\mathcal{D}/L$ respective $N^c/L$, indicating qualitatively different correlation patterns. 
The total-spin diagrams exhibit characteristic leaf-like structures. 
The doublet “leaves” originate at $(\zeta=0,\, U=2\Gamma)$ and terminate at $(\zeta_{1\mathrm{D}}(L,n), U=0)$, where $\zeta_{1\mathrm{D}}(L,n)$ is given by Eq.~\eqref{eq:crzeta}.

At $\zeta=0$ the dots are decoupled, and the onset of the doublet phase coincides with the quantum phase transition of an isolated SC Anderson impurity as given by SC-AL, which at the particle-hole symmetry point occurs at $U=2\Gamma$~\cite{Meden2019the} (see also Appendix~\ref{app:SQD}). 
At the opposite end of the leaves, the critical values $\zeta_{1\mathrm{D}}(L,n)$ correspond to level crossings in the single-particle BdG spectrum. 
At these points, the many-body ground state changes its character because adding or removing a particle does not cost energy, and even a small finite $U$ can stabilize the doublet configuration.
\begin{figure}[h]
    \centering
    \includegraphics[width=1.0\linewidth]{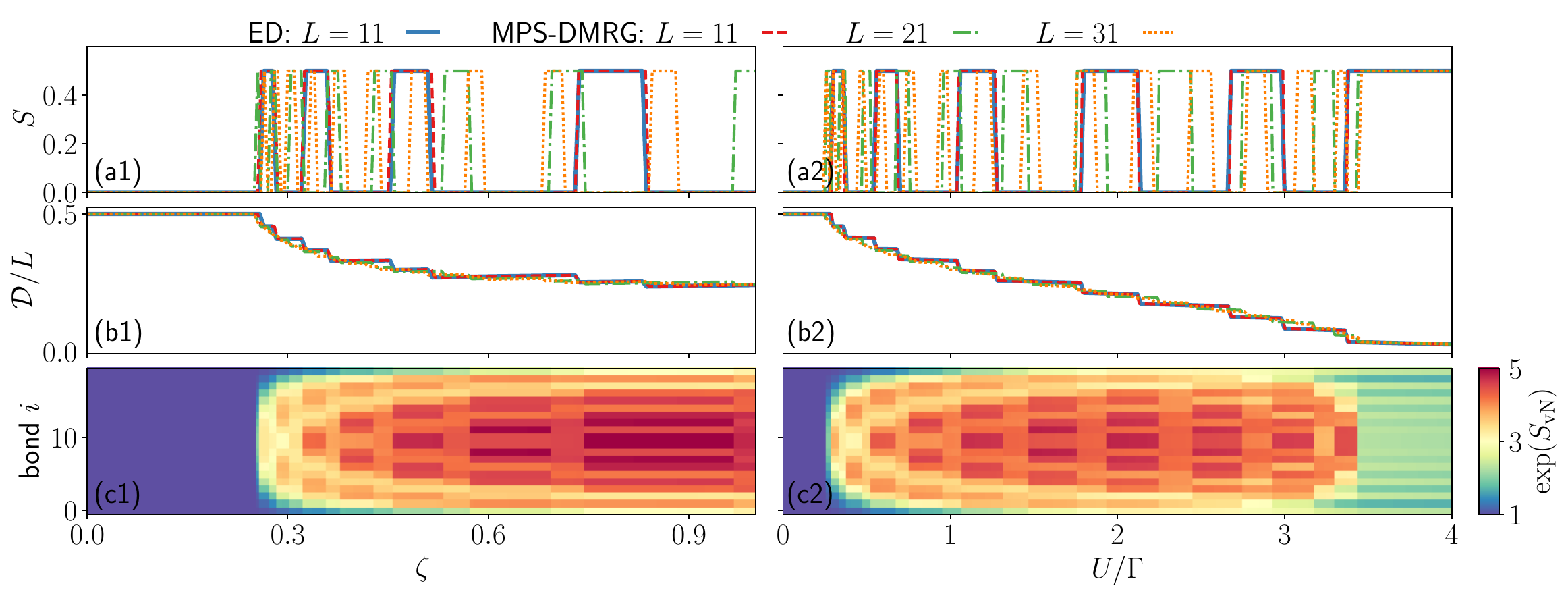}
    \caption{Evolution of the ground-state total spin (first row) and double occupancy per site (second row) with $\zeta$ (first column) and $U$ (second column) for (a1,b1) $U=\Gamma$ and $t=0$ and (a2,b2) $\zeta=0.44$ at $t=0$ and $\varepsilon=0$ for open chain with different lengths. Third row shows examples of entanglement entropy at each bond of $L=21$ chain as function of $\zeta$ and $U$.}
    \label{fig:Cuts1D}
\end{figure}

A qualitative difference emerges between even and odd chain lengths in the strong-interaction regime $U>2\Gamma$.  
For even $L$, the phase diagram is dominated by singlet ground states, whereas for odd $L$ the doublet sector prevails.
With respect to $\zeta_{1\mathrm{D}}(L,n)$ this can be attributed to the fact that for odd $L$ there is a divergent solution at $n=(L+1)/2$. For even $L$, no zero mode $\gamma_{1D}(L;n)=0$ exists.
Physically, this parity effect can be understood from the $\zeta\rightarrow 0$, $U>2\Gamma$ case, where singly occupied sites are energetically favored. 
At half filling, an even number of sites allows complete compensation of local moments leading to a global singlet stabilized by the $\zeta\Gamma$ term. In contrast, an odd number of sites necessarily leaves one unpaired electron, resulting in a many-body doublet ground state.       

Figure~\ref{fig:Cuts1D} shows the evolution of the total spin $S$ (first row) and the scaled double occupancy $\mathcal{D}/L$ (second row) along two cuts of the $\zeta$–$U$ plane. Panels (a1) and (b1) display the dependence on $\zeta$ at fixed $U=1$, while panels (a2) and (b2) show the dependence on $U$ at $\zeta=0.44$, that is, slightly below the first critical point $\zeta_{1\mathrm{D}}(L,L)$. The thick blue lines correspond to ED results obtained with the Lanczos method, whereas the remaining curves are obtained using DMRG in the MPS framework. Panels (c1) and (c2) of Fig.~\ref{fig:Cuts1D} illustrate the spatial distribution of the bond entanglement entropy $S_{vN}$, more precisely, $\exp(S_{vN})$, as a function of $\zeta$ and $U$. 
For an MPS with bond dimension $\chi$, $S_{\rm vN}\leq\ln\chi$, therefore $e^{S_{\rm vN}}$ provides lower-bound scale required by the entanglement across the corresponding cut.

The data reveal three distinct regions. The first region  corresponds to a robust trivial singlet phase, which extends over the parameter space bounded by the points $(\zeta_{1\mathrm{D}}(L,1), U=0)$, $(\zeta_{1\mathrm{D}}(L,L), U=0)$, and $(\zeta=0, U=2\Gamma)$. In this regime, the ground state is a simple product of local BCS singlets on individual dots,
\begin{equation} 
\ket{\Psi_{\mathrm{BCS}}} = \prod_{j=1}^{L} \frac{1}{\sqrt{2}} \left( 1 + d_{j\uparrow}^\dagger d_{j\downarrow}^\dagger \right) \ket{0},\quad 
\ket{\tilde{\Psi}_{\mathrm{BCS}}} = \prod_{j=1}^{L} \frac{1}{\sqrt{2}} \left( \tilde{d}_{j\uparrow}^\dagger + \tilde{d}_{j\downarrow}^\dagger \right) \ket{\tilde{0}},
\label{eq:BCS}
\end{equation}
written here in both the direct SC and transformed bases. The corresponding eigenenergy is $E_{\mathrm{BCS}} = -L\Gamma$ (for $\varepsilon=0$, $t=0$, and after including the constant energy shift). As expected for a product state, the bond entanglement entropy vanishes, see $S_{vN}$ Fig.~\ref{fig:Cuts1D}(c1),(c2) in this region.

The second region is the phase spreading above the critical point $(\zeta=0, U=2\Gamma)$. In the particle conserving representation, the ground state at $\zeta=0$ is any combination of doubly occupied and empty dots (see Appendix~\ref{app:SQD}) with fixed $\tilde{N}_f$. However, as we argue in detail in Appendix~\ref{app:2nd} finite $\zeta$ stabilizes a ground state that can be understood as a perturbation (due to the hopping term) of a checkerboard ordering of these doubly occupied (doublons) and empty sites. Note that a perfect checkerboard pattern of this type requires $\tilde{N}_f=L$ particles for even $L$, while for odd $L$ it requires $\tilde{N}_f=L-1$ or $L+1$. This also explains the singlet/doublet phases in the original representation. As shown in Appendix~\ref{app:2nd}, in the small $\zeta$ regime of this region, the model can be mapped onto an antiferromagnetic Heisenberg chain with exchange coupling $J_H=2J=4\zeta^2\Gamma^2/U$ where $J$ is defined in Eq.~\eqref{eq:Heff}. The DMRG energies of longer chains in the full model agree well with the Bethe ansatz result~\cite{bethe1931theorie} for the effective model (see Fig.~\ref{fig:app:reg2}). 
The corresponding entanglement entropy shows a nonuniform spatial profile similar to that in Fig.~\ref{fig:Cuts1D}(c2), but remains smaller than in the intermediate regime for all
system sizes considered in Appendix~\ref{app:MPS1D}. Accordingly, this regime is less demanding for MPS-DMRG than the intermediate one.

The third region, located between the two regimes discussed above (see Fig.~\ref{fig:1DUzetaEO}), presents a significantly more complex behavior. As the chain length increases, the number of singlet–doublet crossings grows while their width decreases. For longer chains, identifying the true ground state becomes increasingly challenging, as small parameter variations can lead to rapid alternations between competing states. At the same time, the entanglement entropy increases and develops a highly structured profile. The finite-size analysis in Appendix~\ref{app:MPS1D} shows that, at the HSC, its maximum value
grows approximately as $\log L$, while both the charge and sector gaps decrease with increasing $L$.

The exact mapping to the Hubbard model at the HSC provides a complementary interpretation of these three regions. The trivial BCS product state maps onto the empty lattice of the equivalent repulsive Hubbard model as shown in bottom panels of Fig.~\ref{fig:1DUzetaEO}. The second region corresponds to the half-filled Mott regime; in the strong-coupling limit its low-energy spin sector reduces to the antiferromagnetic Heisenberg chain, consistent with the effective description discussed above. The intermediate region corresponds instead to the Hubbard model away from half filling. For a finite chain in the grand-canonical ensemble, the Hubbard particle number $N^c$ changes discretely as the model parameters are varied. The observed sequence of singlet--doublet crossings can thus be understood as successive parity changes of the Hubbard particle-number sector. In the thermodynamic limit, these discrete filling steps evolve into the continuously doped, gapless Tomonaga--Luttinger-liquid regime~\cite{Luo-2024}.
\begin{figure}[h]
    \centering
    \includegraphics[width=\linewidth]{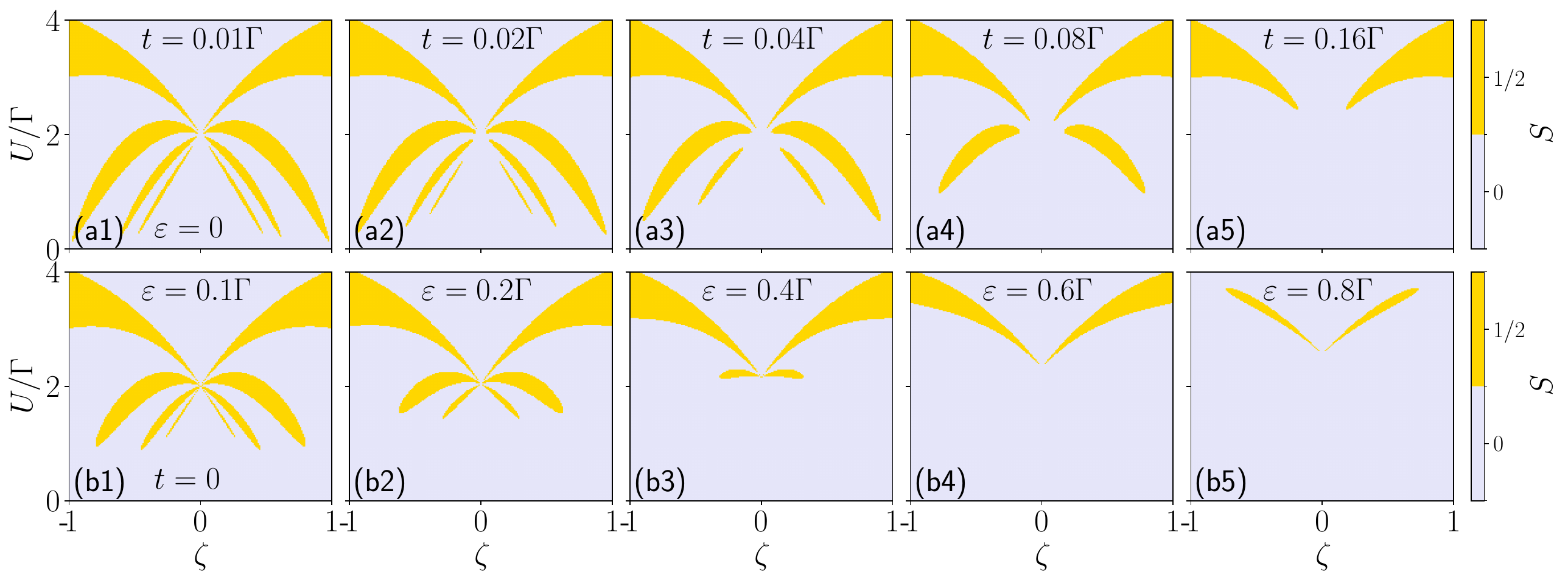}
    \caption{Evolution of the phase diagrams in the $\zeta$--$U$ plane for a one-dimensional chain with $L=8$ as the system is driven away from the HSC. The first row shows increasing hopping $t$ at $\varepsilon=0$, while the second row shows increasing onsite energy $\varepsilon$ at $t=0$.}
    \label{fig:tepsfin}
\end{figure}

It is worth stressing that this correspondence is exact only at the HSC. When the HSC is broken, e.g., by introducing finite hopping $t$ or shifting the original model away from particle-hole symmetry through finite $\varepsilon$, the doublet regions are progressively suppressed, as shown in Fig.~\ref{fig:tepsfin}. This effect is particularly pronounced for $U<U_c= 2\sqrt{\varepsilon^2+\Gamma^2}$. The influence of $\varepsilon$ is weaker at small $\zeta$, which can be understood from the second relation in Eq.~\eqref{eq:hubTD}, where $\varepsilon$ enters multiplied by $\zeta$. Consequently, for small $\zeta$ its main effect is to shift the critical value $U_c$ upward (see Appendix~\ref{app:SQD}). In contrast, even a small finite $t$ suppresses the doublet regions around the critical point $(\zeta=0, U=U_c)$. As discussed in detail in Appendix~\ref{app:MPS1D}, breaking the HSC also changes the finite-size scaling: with increasing lattice size a small but finite entanglement entropy appears already in the first regime, while the third regime becomes gapped, with saturated entanglement entropy and a finite charge gap.
\begin{figure}[ht]
    \centering
    \includegraphics[width=1.0\linewidth]{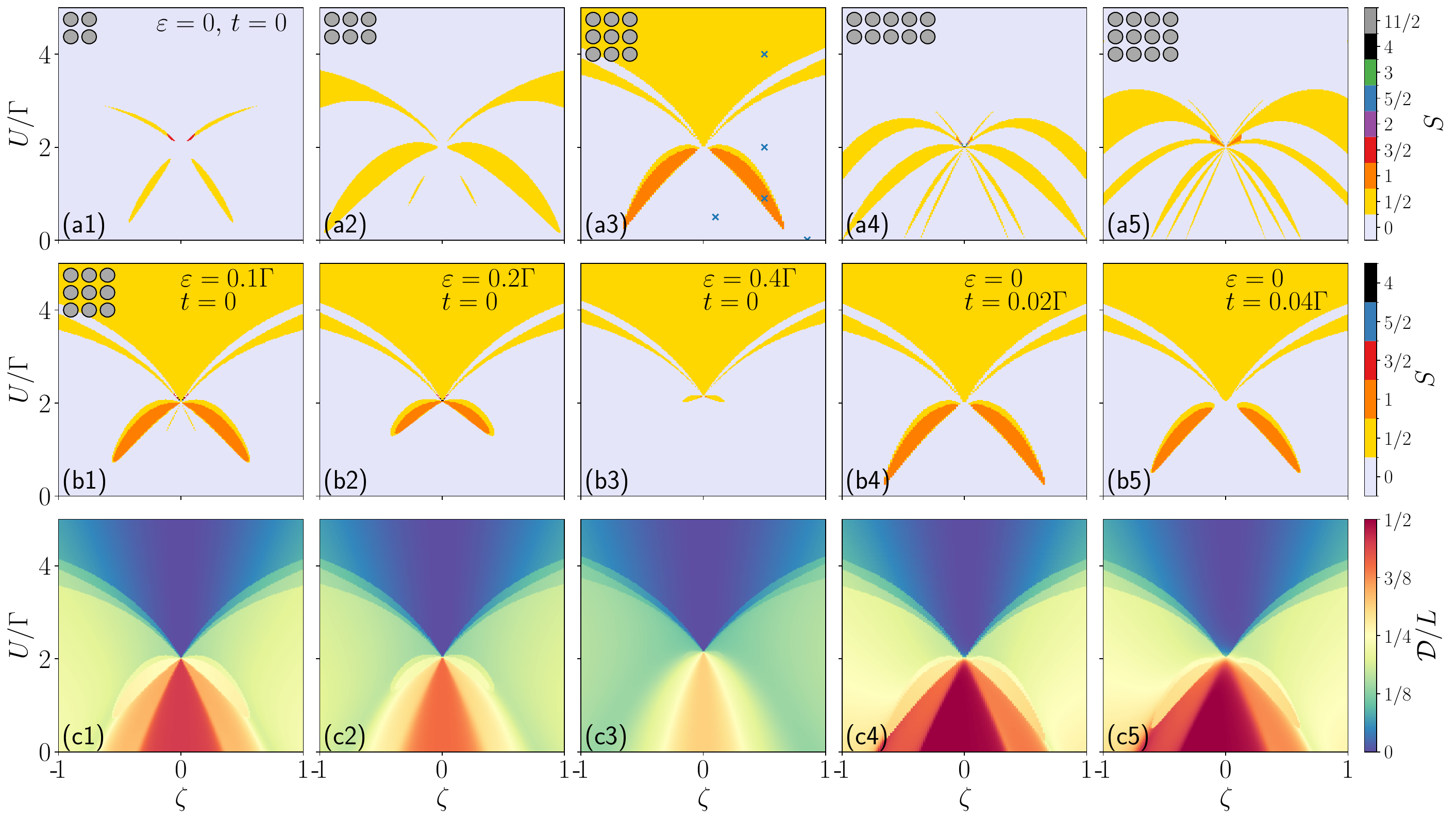}
    \caption{Phase diagrams of the total spin of the many-body ground state in the $\zeta$--$U$ plane for several two-dimensional systems. (a1)--(a5) The first row shows the HSC case $\varepsilon=0$, $t=0$, with the corresponding geometries illustrated in the upper-left corner of each column. Blue crosses in panel (a3) mark the benchmark points listed in Tab.~\ref{tab01:benchmarks} (Appendix~\ref{app:NQS}). (b1)--(b5) The second row shows examples away from the HSC for the $L=3\times3$ lattice. The color bars include only the spin sectors present in the corresponding row of panels. Regions with $S>1$ typically occur only close to the critical point $U=U_c$ at $\zeta=0$. (c1)--(c5) The average double
    occupancy for the same parameters as in second row. }
    \label{fig:1DUzetat0}
\end{figure}

\subsubsection{Two-dimensional clusters}  

Examples of the two-dimensional clusters shown in Fig.~\ref{fig:1DUzetat0} exhibit analogous leaf-like structures that for HSC again originate at $(\zeta=0, U=U_c)$ and typically terminate at $(\zeta_{2\mathrm{D}}(L,\bm{n}), U=0)$. However, in contrast to one-dimensional clusters, $\zeta_{2\mathrm{D}}(\bm{L},\bm{n})$ can diverge at various combinations of $\bm{n}=(n_x,n_y)$, e.g. $(1,2)$ and $(2,1)$ for $L=2\times 2$ or $(1,3)$, $(2,2)$, $(3,1)$ for $L=3\times 3
$, leading to multiple leaves that do not terminate at $U=0$ (not even for $|\zeta| > 1$). As in the one-dimensional case, this also leads to a clear distinction between lattices with an odd and even number of sites showing as a large doublet leaf above the $(\zeta=0, U=2\Gamma)$ point for odd systems.

\begin{figure}[h]
    \centering
    \includegraphics[width=1.0\linewidth]{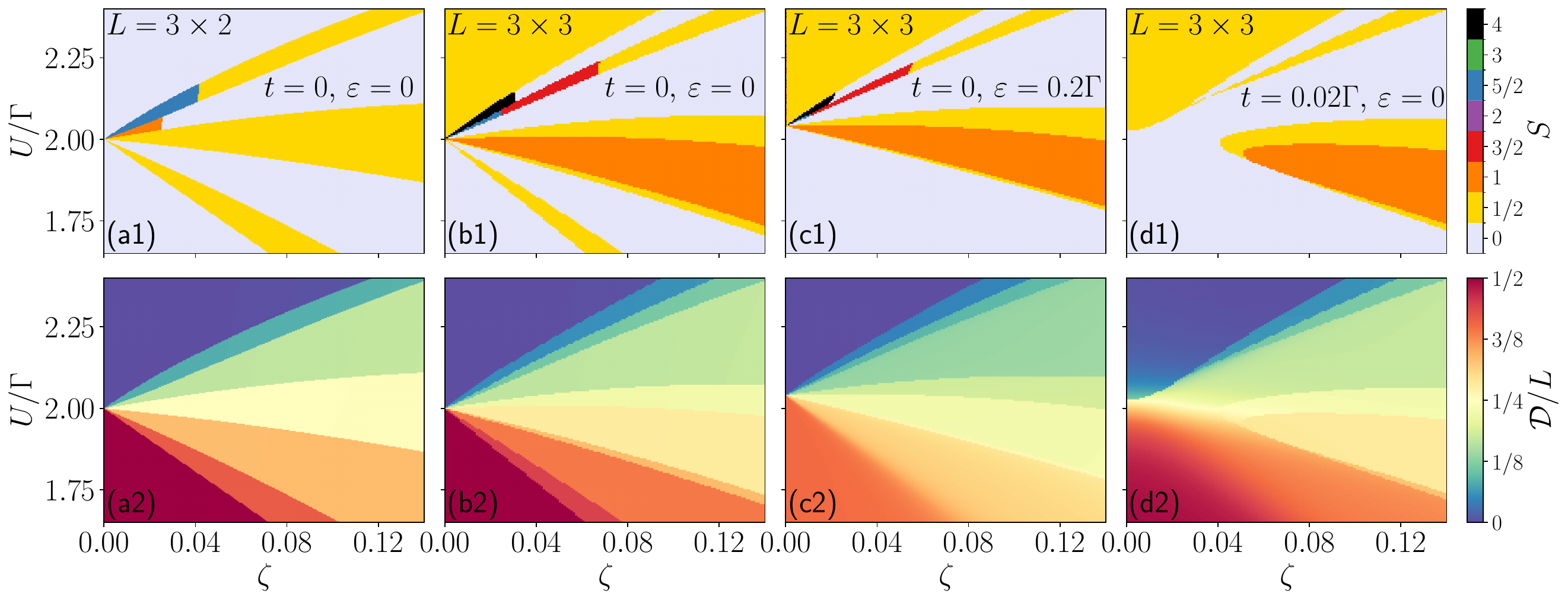}
    \caption{Details of the phase diagrams in the $\zeta$--$U$ plane near the critical point $(\zeta=0,U=U_c)$ for several two-dimensional systems: at the HSC $\varepsilon=0$, $t=0$ in panels (a)--(b), and away from HSC for $\varepsilon=0.2\Gamma$, $t=0$ in panel (c) and $\varepsilon=0$, $t=0.02\Gamma$ in panel (d). First row: total spin of the many-body ground state, with color bars showing only the spin sectors present in the corresponding panels. Second row: average double occupancy $\mathcal{D}/L$.}
    \label{fig:1DUzetatDet}
\end{figure}

Another qualitative difference of the two-dimensional clusters is the occurrence of ground states with total spin larger than $1/2$. A prominent example is the triplet phase forming the second pair of leaves for the $L=3\times3$ cluster in panel (a3), which persists even away from the HSC, as shown in the second row of Fig.~\ref{fig:1DUzetat0}. Higher-spin phases also appear in the vicinity of the point $(\zeta=0,U=U_c=2\sqrt{\varepsilon^2+\Gamma^2})$. Since these phases occupy only narrow regions of parameter space, we show this regime in greater detail in Fig.~\ref{fig:1DUzetatDet}, where stable ground states with total spin up to $S=4$ are resolved.

Although some of the higher-spin phases occupy relatively narrow regions in the $(\zeta,U)$ parametrization, this should not be interpreted directly as requiring equally fine experimental tuning. The parameter $\zeta$ represents an effective nonlocal SC coupling whose relation to microscopic quantities such as intermolecular distance, molecular orientation, and contact geometry is generally nontrivial. The occurrence of these phases near the singlet--doublet transition of isolated dots, i.e., close to $(\zeta=0,U=U_c)$, is therefore particularly interesting for experimentally studied systems of individual molecules on SC surfaces (see, e.g., Refs.~\cite{Li2025individual,Li2025negative}). In such systems, variations of intermolecular geometry or of the local coupling $\Gamma$, for instance due to the proximity of a scanning tip, may shift the system between different total-spin sectors. It is therefore important to examine how robust these phases remain once the HSC is broken. 

As illustrated in the second row of Fig.~\ref{fig:1DUzetat0} and in more detail in Fig.~\ref{fig:1DUzetatDet}(c1),(d1), finite $t$ or $\varepsilon$ modifies the leaf structure in a qualitatively similar way to the one-dimensional case. In both cases, the endpoints of the leaves are shifted from $U=0$ to finite $U$. Their effect on the higher-spin regions near the critical point $(\zeta=0,U=U_c)$, however, is markedly different. Finite $t$ suppresses these regions already for the small hopping values, e.g. $t=0.02\Gamma$ in Fig.~\ref{fig:1DUzetatDet}(d1), whereas several higher-spin regions remain visible even for $\varepsilon=0.2\Gamma$ in panel (c1). As in one dimension, this difference can be understood from Eq.~\eqref{eq:hubTD}: the contribution of $\varepsilon$ to the term breaking the Hubbard correspondence is proportional to $\varepsilon\zeta$ and therefore vanishes as $\zeta\rightarrow0$, while the contribution from $t$ remains finite. Thus, several of the higher-spin phases persist away from the HSC, in particular for finite $\varepsilon$, i.e., away from half-filling.

The corresponding changes are also visible in the double occupancy, see the third row of Fig.~\ref{fig:1DUzetat0}. Even when finite $t$ or $\varepsilon$ replaces sharp transitions between different spin sectors by avoided crossings between singlet states, $\mathcal{D}/L$ still shows clear variations and therefore remains a useful indicator of changes in the ground-state structure.

\begin{figure}[ht]
    \centering
    \includegraphics[width=1.0\linewidth]{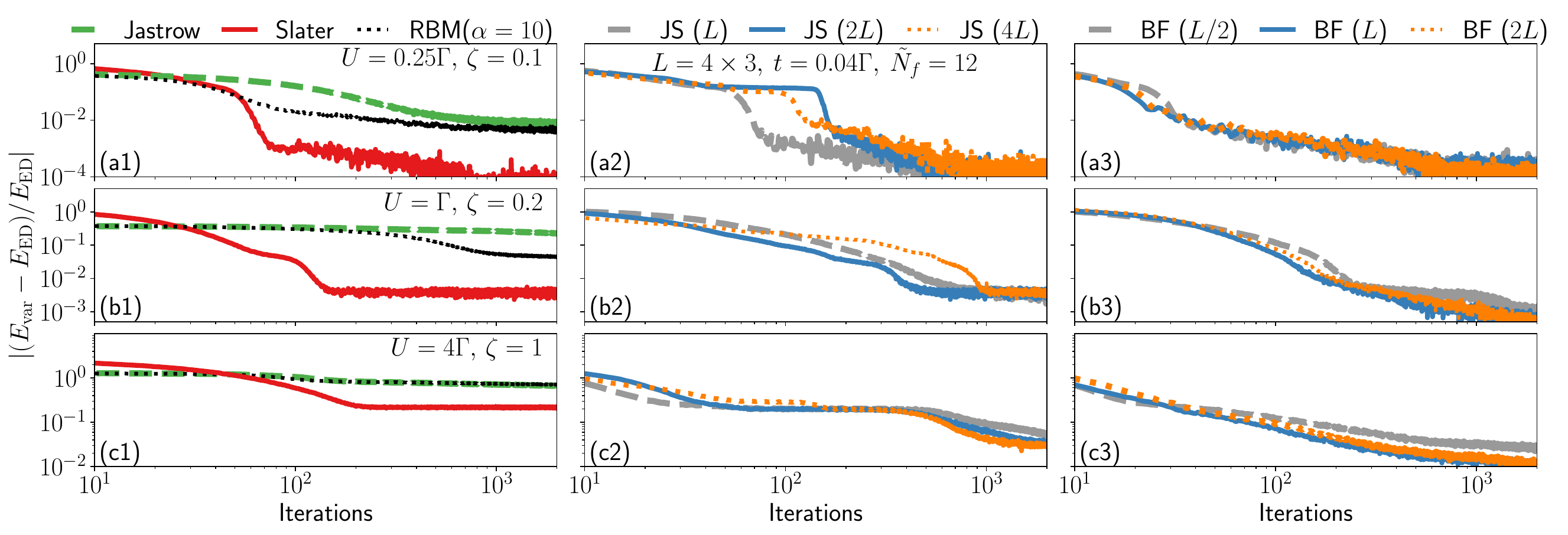}
    \caption{Examples of typical learning curves for different NQS architectures in the weakly correlated (first row), intermediate (second row), and strongly correlated (third row) regimes away from HSC. The calculations are performed for a $L=4\times 3$ system with $t=0.04\Gamma$ and $\varepsilon=0$. The remaining parameters are $U=0.25\Gamma$, $\zeta=0.1$ in panels (a1)--(a3); $U=\Gamma$, $\zeta=0.2$ in panels (b1)--(b3); and $U=4\Gamma$, $\zeta=1$ in panels (c1)--(c3) and the relative error in energy is calculated against the exact diagonalization resuls. The left column compares the Jastrow, Slater, and RBM ans\"atze, with the RBM using $\alpha=10$ to achieve a parameter count comparable to the more expressive architectures shown in the other columns. The middle and right columns show Jastrow--Slater and BF results, respectively, for networks of different sizes. In all cases, we use the Adam optimizer with a learning rate decreased from $0.002$ to $0.0002$ and $2^{13}$ samples. Parameter counts: Jastrow $276$, Slater $288$, RBM $6024$, JS $(588,\,888,\,3288)$, and BF $(2454,\,4332,\,8088)$.}
    \label{fig:lc}
\end{figure}
\begin{figure}[ht]
    \centering
    \includegraphics[width=1.0\linewidth]{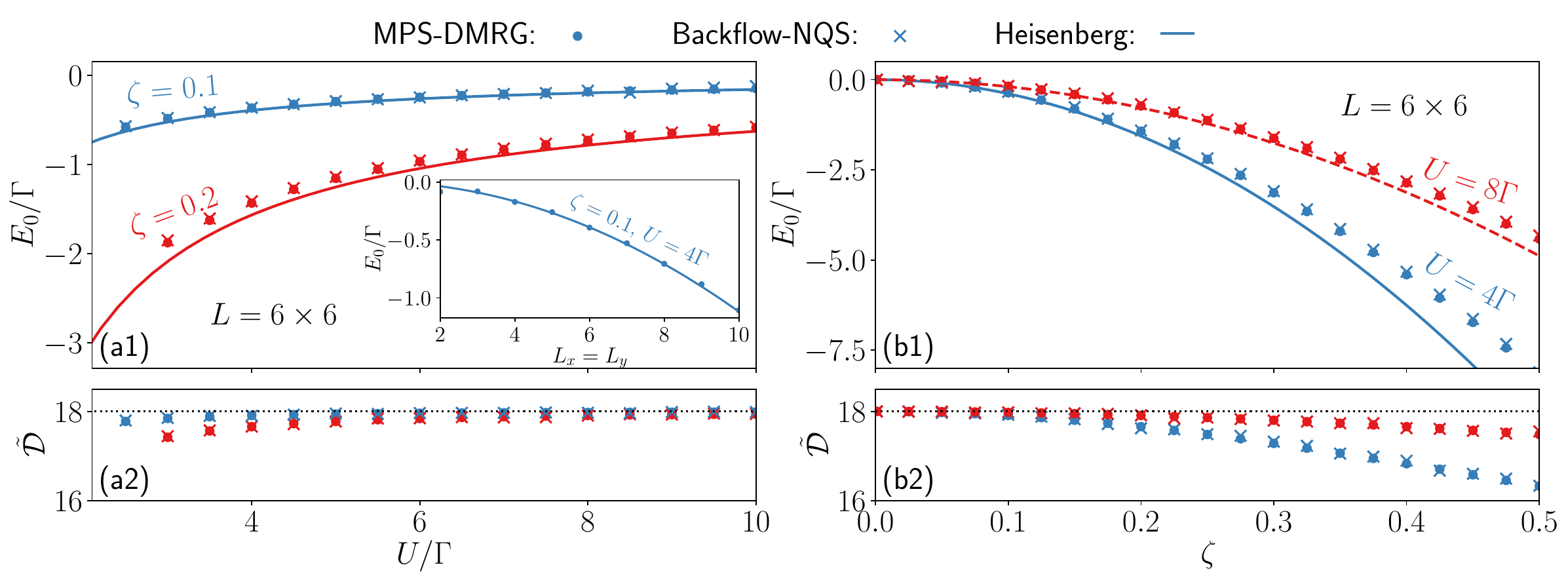}
   \caption{
Analysis of the second regime at HSC point (\(\varepsilon=0\), \(t=0\)) for regular two-dimensional clusters. Circles denote MPS-DMRG results, crosses BF-NQS calculations, and solid lines the semi-analytical prediction of Eq.~\eqref{eq:Heis2D}. 
(a1) Ground-state energy versus \(U\) for \(\zeta=0.1\) (blue) and \(\zeta=0.2\) (red) and $L=6\times 6$. Inset: finite-size scaling of the ground-state energy comparing the semi-analytical prediction with DMRG calculations. 
(a2) Double occupancy
$
{\tilde {\cal D}}=\sum_i \tilde n_{i\uparrow}\tilde n_{i\downarrow}
$
as a function of $U$. 
(b1) Ground-state energy versus $\zeta$ for $U=4\Gamma$ (blue) and $U=8\Gamma$ (red). 
(b2) Corresponding double occupancy $\tilde {\cal D}$ as a function of $\zeta$.
}
    \label{fig:L2D}
\end{figure}
\begin{figure}[ht]
    \centering
    \includegraphics[width=1.0\linewidth]{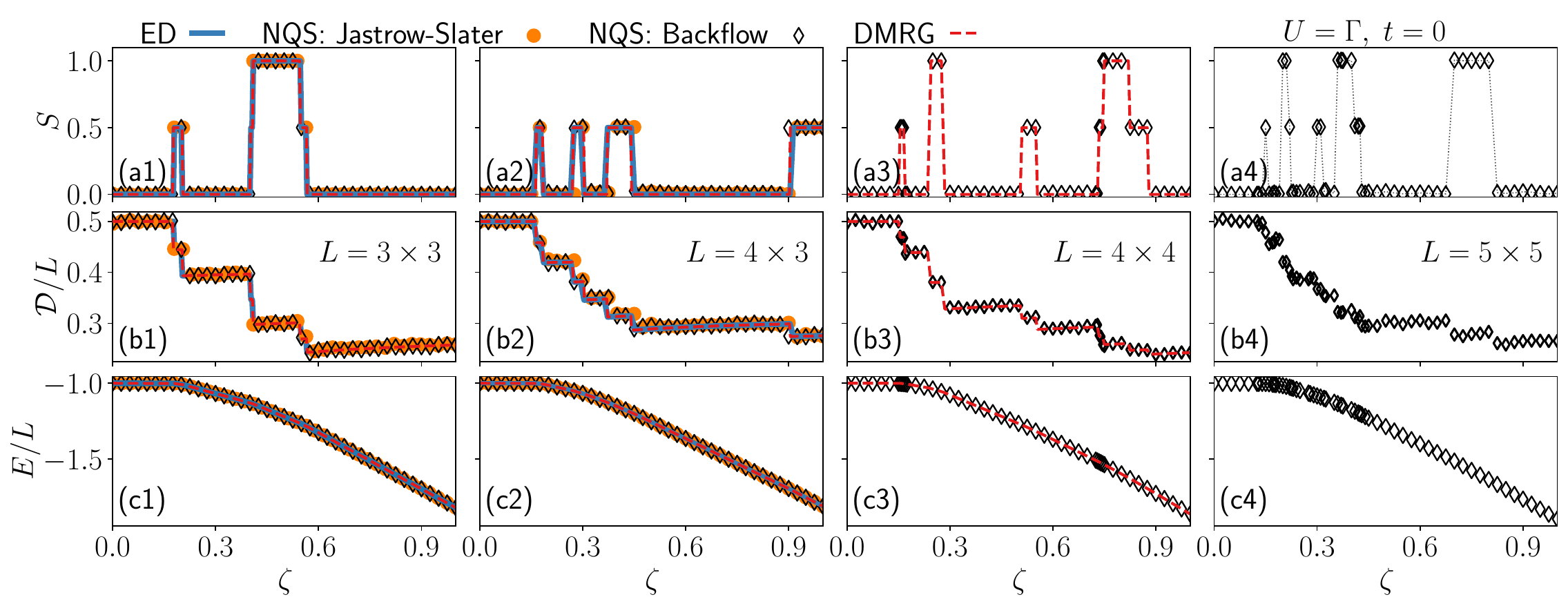}
    \caption{
    Evolution of the ground-state total spin (first row), double occupancy per
site (second row), and energy per site (third row) as a function of $\zeta$
for $U=\Gamma$, $t=0$, and $\varepsilon=0$, for different lattice sizes. Comparison of Lanzos ED results (only for $L=3\times 3$ and $L=4\times 3$) with Jastrow-Slater (orange circles), neural BF NQS (black diamonds) VMC, and MPS based DMRG (dashed red lines).}
    \label{fig:Cuts2D}
\end{figure}
\begin{figure}[ht]
    \centering
    \includegraphics[width=1.0\linewidth]{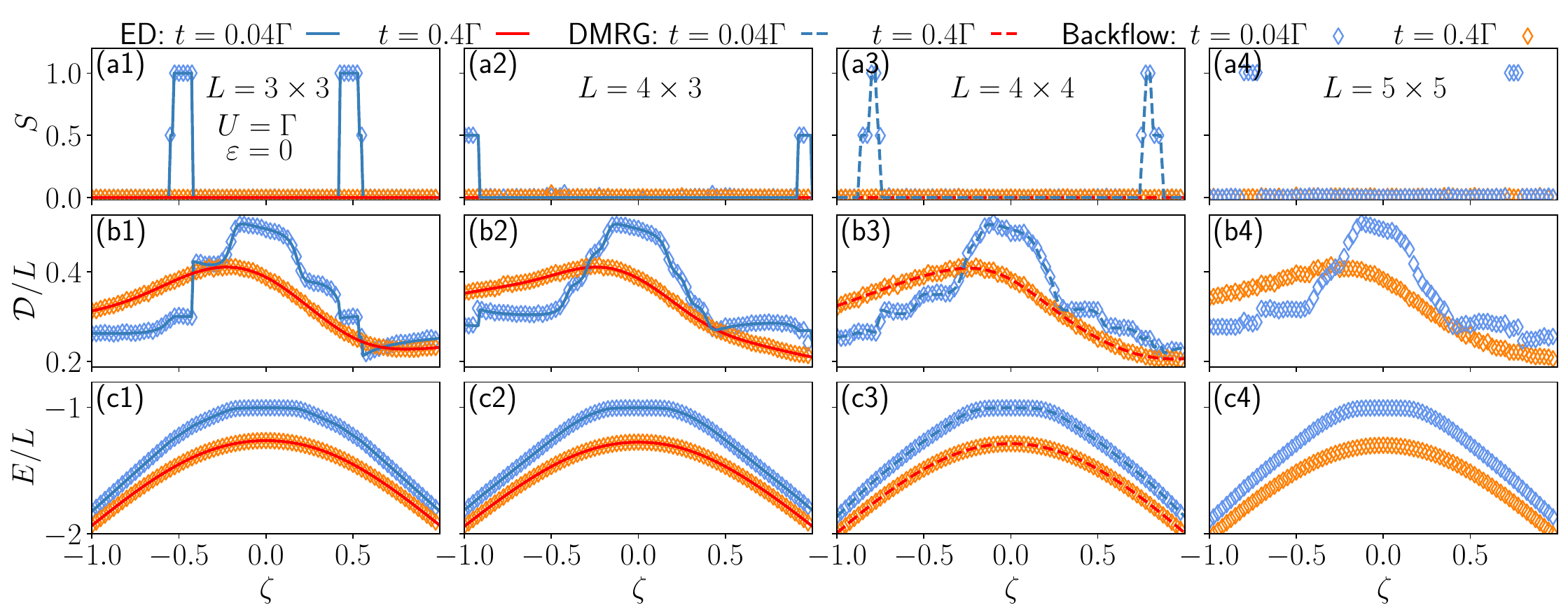}
    \caption{Evolution of the ground-state total spin (first row), double occupancy per site (second row), and energy per site (third row) as a function of $\zeta$ for $U=\Gamma$ and two hopping amplitudes, $t=0.04\Gamma$ (blue lines and symbols) and $t=0.4\Gamma$ (red lines and orange symbols), for several lattice sizes. We compare Lanczos exact-diagonalization results, available for $L=3\times3$ and $L=4\times3$, MPS results for $L=4\times4$, and neural BF NQS results.}
    \label{fig:Cuts2Dft}
\end{figure}

In analogy to the one-dimensional case, the two-dimensional phase diagrams can again be divided into three characteristic regimes. Before discussing their physical properties in detail, we first compare the performance of the different NQS architectures at representative points from each regime. This allows us to identify an ansatz suitable for the subsequent calculations on larger two-dimensional clusters.

Figure~\ref{fig:lc} shows representative learning curves away from HSC for $L=4\times3$, $\varepsilon=0$, and $t=0.04\Gamma$. The three rows correspond to parameter points in the weakly correlated ($U=0.25\Gamma$, $\zeta=0.1$), intermediate ($U=\Gamma$, $\zeta=0.2$), and strongly correlated ($U=4\Gamma$, $\zeta=1$) regimes, respectively. Simple architectures such as a Jastrow and RBM provide a reasonable approximation to the ground state only in the weakly correlated regime. We use an RBM with $\alpha=10$, giving a number of variational parameters comparable to that of the Jastrow--Slater and BF networks. The differences in accuracy can therefore not be attributed simply to the network size.

Including the fermionic structure explicitly improves the results: the Slater ansatz performs better in the weakly correlated regime and considerably better in the intermediate regime. As the correlations become stronger, however, the Jastrow--Slater and especially the neural BF ans\"atze provide substantially more accurate results. Among the architectures considered here, BF gives the most consistent performance across all three regimes. We consequently use it as the principal NQS ansatz for the larger-system calculations discussed below. Additional HSC benchmarks for the different architectures are summarized in Appendix~\ref{app:NQS}.

It is important to distinguish the formal expressivity of a variational ansatz from its practical trainability. The results in
Fig.~\ref{fig:lc} do not provide a strict separation of these two effects: the architecture determines both the variational manifold and the optimization landscape. Nevertheless, all ans\"atze were optimized using the same sampling and optimization protocol, and the RBM with $\alpha=10$ was chosen such that its number of variational parameters is comparable to that of the Jastrow--Slater and BF states. We also tested Jastrow--Slater and BF architectures with larger parameter counts, up to approximately $8\times10^3$, but observed only modest further improvement. The pronounced differences in convergence behavior and final relative variational-energy errors therefore cannot be attributed simply to the number of parameters or to the optimizer. Rather, they indicate that explicitly incorporating fermionic structure and configuration-dependent correlations into the ansatz is important for achieving high accuracy and favorable optimization behavior in the more strongly correlated regimes.


We now turn back to the physical properties of the three regimes. At HSC, the first corresponds to the same trivial singlet phase as in one dimension, i.e., a product of isolated QDs in the BCS singlet state described by Eq.~\eqref{eq:BCS}. Owing to its product-state character, this regime is straightforward to treat using both MPS-DMRG and NQS (see Appendix~\ref{app:NQS}).

The second regime is more challenging and can be further divided into low- and high-$\zeta$ regions. As in one dimension, the low-$\zeta$ part is connected to a weakly perturbed doublon checkerboard state and can be mapped onto an antiferromagnetic Heisenberg model on a two-dimensional lattice, as discussed in detail in Appendix~\ref{app:2nd}. This limit is well understood from previous studies of the two-dimensional Heisenberg model~\cite{sandvik1997finite}.


In Fig.~\ref{fig:L2D}(a1),(b1), we compare the ground-state energy obtained from BF-NQS, MPS-DMRG with the semi-analytical prediction for the corresponding Heisenberg model (see Appendix~\ref{app:2nd}),
\begin{equation}
E_{2D}^{o}\simeq \frac{\zeta^2\Gamma^2}{U}\left(
-4.6777L + L_x+L_y\right),
\label{eq:Heis2D}
\end{equation}
where we omit the trivial constant energy shift arising from the transformation. Panel (a1) shows the ground-state energy as a function of $U$ for $\zeta=0.1$ and $0.2$, while the inset presents the finite-size scaling
for $U=4\Gamma$ and $\zeta=0.1$. The semi-analytical expression already provides a good description for clusters as small as $L_x=L_y=4$. The BF NQS results (crosses) closely reproduce the DMRG energies
(circles) throughout this regime.

Even for relatively small lattices, converged DMRG simulations require sizeable bond dimensions. We attribute this to the combined effect of the Jordan--Wigner mapping, which generates effectively long-range couplings
in two dimensions, and the near degeneracy of low-energy states within the $N_f=L$ sector (or $N_f=L-1$ for odd lattices). The MPS bond dimension was restricted to $1024$, which is sufficient to converge the ground-state
energy but can be inadequate for more demanding observables.

For $L_x=L_y=6$, the agreement with the effective Heisenberg description systematically improves with increasing $U$ and decreasing $\zeta$, as
shown in panels (a1)--(b2). Increasing $\zeta$ enhances corrections beyond the strong-coupling approximation, whereas larger $U$ stabilizes the doublon checkerboard state and extends the parameter range over which the effective description remains accurate. The same trend is reflected in the double occupancy, shown in panels (a2) and (b2).

Note that BF-NQS convergence was improved by transfer learning between neighboring parameter points. Rather than optimizing each point independently, we reuse converged variational parameters to initialize nearby calculations. Starting from larger values of $\zeta$, where the optimization is more robust, we gradually decrease $\zeta$ while carrying over the optimized parameters. The states obtained at $U=4\Gamma$ are subsequently used to initialize calculations at $U=8\Gamma$. This procedure considerably accelerates convergence, particularly at small $\zeta$.

The third regime differs qualitatively from the one-dimensional case. ED for a $3\times3$ cluster already shows a stable triplet ground state, although such a small system alone does not establish whether finite-spin phases survive for larger lattices. Figure~\ref{fig:Cuts2D} demonstrates that the triplet regions remain robust also for larger two-dimensional systems. At HSC, we compare ED, DMRG, Jastrow--Slater NQS, and BF NQS along a cut in $\zeta$ at $U=\Gamma$ and $\varepsilon=0$. The figure shows the total spin, double occupancy, and energy for systems up to $L=5\times5$.

These quantities correspond to the original SC representation but are evaluated using the particle conserving basis. Consequently, the total spin provides a useful consistency check for the NQS calculations, since it should remain close to the symmetry-allowed values ($0$, $1/2$, $1$, \dots). For the larger lattices, we primarily use the BF ansatz, whose performance across the different regimes was established in Fig.~\ref{fig:lc}. The triplet phase appears for both even and odd lattices, although along this particular cut we do not observe stable phases with even larger total spin. We benchmark these results against MPS-DMRG calculations with bond dimensions truncated to $768$--$2048$. This is sufficient for the level of precision required in the present comparison, although not always for full convergence as the third regime is considerably more demanding for MPS-DMRG.


To test whether the agreement between the different numerical approaches persists away from HSC, Fig.~\ref{fig:Cuts2Dft} shows analogous cuts for $U=\Gamma$, $\varepsilon=0$, and two finite hopping amplitudes, $t=0.04\Gamma$ and $t=0.4\Gamma$. We include both positive and negative $\zeta$. It can be shown that spectra and gauge-invariant observables are invariant under simultaneous $(t,\zeta)\rightarrow(-t,-\zeta)$. At $t=0$, the relation implies a symmetry under $\zeta\rightarrow-\zeta$, whereas at fixed finite
$t$ this symmetry is lost. The resulting asymmetry therefore reflects the physical relevance of the relative sign of the direct hopping and nonlocal SC coupling once both are finite.

More importantly, the figure shows that the correlated structure identified at HSC is not immediately destroyed when this condition is broken. For $t=0.04\Gamma$, the triplet region survives with increasing lattice size, including for the $5\times5$ cluster, demonstrating that it is neither an artifact of the exact Hubbard correspondence nor restricted to the smallest clusters. The double occupancy retains a pronounced step-like structure, which separates distinct ground-state regimes even where the total spin remains singlet. Increasing the hopping to $t=0.4\Gamma$ smooths these features and suppresses the finite-spin region, but the double occupancy still exhibits a strong and asymmetric dependence on $\zeta$. Thus, breaking HSC produces a crossover from the sharply differentiated weak-hopping regimes toward a smoother, hopping-dominated state rather than simply removing all correlation effects.


Taken together, these benchmarks demonstrate that standard fermionic NQS can provide a common and reliable description of qualitatively different two-dimensional regimes, both at and away from HSC. Here our aim is not to optimize a single parameter point to the highest possible accuracy, but to test whether the same NQS framework can reproduce the relevant energies and observables over an extended parameter range and across distinct many-body states. The agreement with ED and DMRG shows that this is indeed the case, while the comparison of architectures identifies the BF NQS as the most consistently accurate ansatz in the more strongly correlated regimes and enables its application to larger clusters beyond the reach of ED.

\FloatBarrier
\section{Summary and discussion}
In this work, we studied interacting SC nanostructures within the SC-AL description using a combination of exact analytical and numerical methods, DMRG, and VMC based on fermionic NQS. By combining these complementary approaches, we were able to analyze systems ranging from small exactly solvable clusters to larger one- and two-dimensional structures, while simultaneously benchmarking the accuracy and convergence behavior of the variational ans\"atze. A crucial ingredient in this respect was the performed transformation of the Hamiltonian from the SC to the particle-number-conserving basis. This transformation allowed us to formulate the problem in a representation suitable for both tensor-network and neural-network methods, while preserving direct access to physically relevant observables.

The analysis of the non-interacting case revealed the existence of a high-symmetry condition necessary for closing the spectral gap. At the HSC, the interacting model can furthermore be mapped exactly onto a standard repulsive Hubbard model in the grand-canonical ensemble, providing a complementary interpretation of the different regimes discussed below. In the interacting case, we identified three distinct regions, whose boundaries can be partially understood analytically. The first stable region corresponds to a simple product state of local SC singlets with negligible entanglement entropy. This state survives up to $\zeta=1/2$ in one-dimensional chains and up to $\zeta=1/4$ in two-dimensional clusters for small $U$, or alternatively up to $U=2\Gamma$ for negligible $\zeta$.

The second regime, realized at large $U$, can for small $\zeta$ be understood as a perturbed checkerboard arrangement of doubly occupied (doublon) and empty sites in the particle conserving representation. At the HSC, this regime corresponds to the Mott phase of the equivalent half-filled Hubbard model and, in the strong-coupling limit, maps onto an effective antiferromagnetic Heisenberg model. This correspondence allowed us to benchmark both MPS-DMRG and NQS-VMC calculations against previously known results.

The third regime, located between the first and second regions, is the most complex and differs significantly between one- and two-dimensional systems. For chains of QDs at HSC, we showed that the ground state in this region can only be either a doublet or a singlet; however, these states alternate rapidly with changing $\zeta$ or $U$, particularly for longer chains. In addition, each stable singlet (or doublet) has a different character. The Hubbard mapping identifies these transitions with successive changes of the particle-number sector, while our finite-size scaling is consistent with a gapless Tomonaga--Luttinger-liquid regime in the thermodynamic limit. Away from HSC, this behavior changes qualitatively: even a relatively small finite hopping can suppress the sequence of doublet regions and open a finite gap.

The two-dimensional case exhibits substantially richer behavior. Already for small clusters, ED reveals stable triplet ground states in a broad parameter regime near HSC. Our DMRG and NQS calculations show that these triplet phases remain robust also for larger lattices, for both even and odd system sizes, and can persist away from HSC. In addition, narrow regions with still higher total spin appear close to the critical point of isolated QD. The persistence of the triplet state demonstrates that the competition between SC pairing and strong correlations can generate nontrivial magnetic structures even in relatively small SC arrays.

From the methodological perspective, one of the main results of this work is the demonstrated applicability of relatively standard fermionic NQS to SC nanostructures, although optimization becomes increasingly challenging in the more strongly correlated regimes. The learning curves for representative points from all three regimes show that relatively simple Jastrow and RBM states can already describe the weakly correlated regime, whereas Jastrow--Slater and particularly BF ans\"atze provide substantially more accurate and robust results as the correlations increase. Since the different architectures were trained using the same optimization protocol, and an RBM with a comparable number of variational parameters was included in the comparison, the observed differences cannot be attributed simply to the optimizer or to the network size. At the same time, the architecture affects both the variational manifold and the optimization landscape, and therefore, the present comparison does not provide a strict separation between expressivity and trainability.

The advantage of the neural-network parametrization should not be understood as a replacement for conventional VMC, which also relies on compact parametrized wave functions and stochastic optimization. Rather, neural networks provide a flexible way of representing configuration-dependent correlations without prescribing a specific analytical form. This flexibility can be combined with established fermionic structures, such as Slater determinants, Jastrow factors, and BF transformations. The neural BF ansatz used here follows this strategy: fermionic antisymmetry is built into the determinant, while the neural network parametrizes the configuration-dependent modification of the single-particle orbitals. Its good performance in the present calculations should therefore be attributed to the combination of an appropriate fermionic structure with a flexible neural parametrization, rather than to the use of neural networks alone.

The final question is to what extent the present predictions remain valid beyond the SC-AL approximation. The model considered here inherits several important limitations from this description. In realistic systems, the SC gap $\Delta$ is often comparable to or smaller than $U$ and $\Gamma$, and the effective model therefore cannot be expected to reproduce quantitatively the full SC Anderson impurity problem. Generalized atomic-limit approaches can partially improve this situation through a renormalization of the effective parameters~\cite{Zonda2023generalized,Pokorny-2023}, although important qualitative limitations remain~\cite{zalom2024double,Bobok2025che}.  Nevertheless, because realistic SC nanostructures are often difficult to model in full detail, atomic-limit effective models are widely used to interpret experiments and identify the relevant low-energy physics. They have been employed, for example, in studies of minimal Kitaev chains~\cite{dvir2023realization,bordin2025enhanced}, topological phases in multiterminal SC nanostructures~\cite{Antonelli2025exploring}, multidot systems~\cite{annabi2026fermion}, magnetic chains and more complex structures on SC surfaces~\cite{Liebhaber2022quantum} and other systems. Thus, studying of this class of effective models remains physically relevant.

Within this scope, we expect several qualitative conclusions obtained here to remain robust in more realistic settings. Weakly coupled dots should still favor local SC singlets, whereas more densely coupled arrays can support competing magnetic and SC correlations that are sensitive to variations of parameters such as the inter-dot distance. Near half filling and for suppressed direct hopping, the regime considered here is particularly relevant for molecular structures on SC surfaces and engineered quantum-dot arrays. The appearance of stable triplet phases in two-dimensional clusters therefore provides a plausible consequence of the interplay between superconductivity and strong local interactions that may persist beyond the strict atomic limit.

A more realistic description can be obtained by explicitly including electronic degrees of freedom in the SC leads, for example within recently proposed chain-expansion schemes~\cite{Bobok2025che}. Such extensions naturally incorporate finite-gap effects and higher-order tunneling processes and may stabilize additional correlated phases, potentially including larger-spin ground states. At the same time, the corresponding Hilbert spaces grow rapidly with system size, making controlled numerical simulations increasingly demanding. In this sense, the present work also provides a methodological starting point: SC nanostructures can be treated accurately using relatively simple fermionic NQS, opening a route toward simulations of larger and more realistic hybrid SC systems.
    
\section*{Acknowledgements}
\paragraph{
M\v{Z} and  VP acknowledge the support from Czech Science Foundation through project No.~23-05263K. 
MF, JK and TP gratefully acknowledge financial support from the Czech Academy of Sciences (Praemium Academiae awarded to MF) and the Strategy AV21 (in particular the program "AI: Artificial Intelligence for Science and Society"). 
This work was also supported by the Ministry of Education, Youth and Sports of the Czech Republic through the e-INFRA CZ(ID:90254) at the IT4Innovations National Supercomputing Centre and 
the Czech Republic-Germany Mobility programme (ID:8J25DE001). 
VP acknowledges the support from the EU COST action CA21144 SUPERQUMAP. 
We thank Artur Slobodeniuk and Wolfgang Belzig for many productive discussions.}
\clearpage
\pagebreak
\begin{appendix}
\section{Superconducting atomic limit
\label{app:SCAL}}
We derive the SC atomic limit Hamiltonian by using the example of a double quantum dot (DQD) on an SC surface as the generalization to larger clusters is straightforward. Our starting point is the SC Anderson impurity model for a DQD,
\begin{align}
    \mathcal{H}^{\mathrm{AIM}} 
    &= \sum_{j=1}^2\sum_\sigma\epsilon_j n_{j\sigma} +
    \sum_{j=1}^{2}U_j n_{j\uparrow}n_{j\downarrow}-\sum_{\sigma}t_{12} \left(d_{2\sigma}^\dagger d_{1\sigma}^\pdag + d_{1\sigma}^\dagger d_{2\sigma}^\pdag\right) \label{eq:AIMa} \\
    &+\sum_{\mathbf{k}\sigma} \epsilon^{\pdag}_{\mathbf{k}} c^{\dagger}_{\mathbf{k} \sigma} c^{\pdag}_{\mathbf{k}\sigma} - \sum_{\mathbf{k}} \left(\bar{\Delta} c^{\dagger}_{\mathbf{k} \uparrow}
    c^{\dagger}_{-\mathbf{k} \downarrow} + \text{H.c.}\right)
    +\sum_{j=1}^2\sum_{\mathbf{k}\sigma} 
    \left(V_{j\mathbf{k}}  c^{\dagger}_{\mathbf{k} \sigma} d^{\pdag}_{j\sigma} + \text{H.c.}\right)\label{eq:AIMb}.
\end{align}
The first part, Eq.~\eqref{eq:AIMa}, describes the quantum dots, as in the main text. The second part, Eq.~\eqref{eq:AIMb}, describes the SC surface through a standard BCS Hamiltonian and the coupling between the dots and the superconductor. Here $c_{\mathbf{k}\sigma}^\dagger$ ($c_{\mathbf{k}\sigma}^\pdag$) creates (annihilates) an electron in the lead with spin $\sigma$ and energy $\epsilon^{\pdag}_{\mathbf{k}}$, and $\bar{\Delta}=\Delta e^{-i\varphi}$ is the complex SC order parameter, assumed here to be real as we are dealing with systems with only one SC lead. 

\paragraph{Non-interacting Green's function of a DQD\label{app:double dot}:}
Consider the Hamiltonian of a parallel DQD connected to one SC lead. We can write the Matsubara Green's function in the Nambu basis
$\mathbf{d}^T=(d^\pdag_{1\uparrow},d^\dag_{1\downarrow},d^\pdag_{2\uparrow},d^\dag_{2\downarrow},
c^\pdag_{\mathbf{k}},c^\dag_{-\mathbf{k}})$,

\begin{equation}
    G_0(i\omega_n)^{-1} =
    \begin{pmatrix}
    i\omega_n - \epsilon_{1} & 0 & -t_{12} & 0 & -V_{1\mathbf{k}} & 0 \\
    0 & i\omega_n + \epsilon_{1} & 0 & t_{12} & 0 & V_{1\mathbf{k}} \\
    -t_{12} & 0 & i\omega_n - \epsilon_{2} & 0 & -V_{2\mathbf{k}} & 0 \\
    0 & t_{12} & 0 & i\omega_n + \epsilon_{2} & 0 & V_{2\mathbf{k}} \\
    -V_{1\mathbf{k}} & 0 & -V_{2\mathbf{k}} & 0 & i\omega_n - \epsilon_{\mathbf{k}} & \Delta \\
    0 & V_{1\mathbf{k}} & 0 & V_{2\mathbf{k}} & \Delta & i\omega_n + \epsilon_{\mathbf{k}}
    \end{pmatrix},
\end{equation}
where we assume real $V_{j\mathbf{k}}$.
We are interested in the impurity Green's function $G^{\text{d}}_0$. Following a standard block-matrix inversion scheme~\cite{Bobok2025che} we obtain a 4$\times$4 matrix,

\begin{equation}
\begin{aligned}
    &[G^{\text{d}}_0(i\omega_n)]^{-1}  =
    \begin{pmatrix}
    i\omega_n - \epsilon_{1} & 0 & -t_{12} & 0 \\
    0 & i\omega_n + \epsilon_{1} & 0 & t_{12}  \\
    -t_{12} & 0 & i\omega_n - \epsilon_{2} & 0  \\
    0 & t_{12} & 0 & i\omega_n + \epsilon_{2} &  \\
    \end{pmatrix} \\
    &-\sum_\mathbf{k}\frac{1}{\omega_n^2+\epsilon_{\mathbf{k}}^2+\Delta^2} \\
    &\times\begin{pmatrix}
    (i\omega_n+\epsilon_{\mathbf{k}})V_{1\mathbf{k}}^2 
     & \Delta V_{1\mathbf{k}}^2 
     & (i\omega_n+\epsilon_{\mathbf{k}}) V_{1\mathbf{k}} V_{2\mathbf{k}} 
     & \Delta V_{1\mathbf{k}} V_{2\mathbf{k}} \\
    \Delta V_{1\mathbf{k}}^{2} 
     & (i\omega_n-\epsilon_{\mathbf{k}})V_{1\mathbf{k}}^2
     & \Delta V_{1\mathbf{k}}V_{2\mathbf{k}} 
     & (i\omega_n-\epsilon_{\mathbf{k}})V_{1\mathbf{k}}V_{2\mathbf{k}} \\
    (i\omega_n+\epsilon_{\mathbf{k}})V_{1\mathbf{k}}V_{2\mathbf{k}}
     & \Delta V_{1\mathbf{k}}V_{2\mathbf{k}}
     & (i\omega_n+\epsilon_{\mathbf{k}})V_{2\mathbf{k}}^2 
     & \Delta V_{2\mathbf{k}}^2 \\
    \Delta V_{1\mathbf{k}}V_{2\mathbf{k}} 
     & (i\omega_n-\epsilon_{\mathbf{k}})V_{1\mathbf{k}}V_{2\mathbf{k}}
     & \Delta V_{2\mathbf{k}}^2 
     & (i\omega_n-\epsilon_{\mathbf{k}})V_{2\mathbf{k}}^2
    \end{pmatrix}.
\end{aligned}
\end{equation}

Assuming a constant DOS with half-bandwidth $D$,
$
\rho(\epsilon)=\Theta(D-|\epsilon|)/2D,
$
we rewrite the momentum summations as integrals,
\begin{equation}
\sum_k F(\epsilon_k)
=
\frac{1}{2D}\int_{-D}^{D} d\epsilon\,F(\epsilon).
\end{equation}
This allows us to define real tunneling rates,
\begin{equation}
    \Gamma_{ij}(\epsilon) = \pi\sum_{\mathbf{k}} V_{i\mathbf{k}} V_{j\mathbf{k}}
    \delta(\epsilon-\epsilon_\mathbf{k}) = \Gamma_{ij}\Theta(D - |\epsilon|),
    \label{eq:WBL}
\end{equation}
which are constant in the wide-band ($D\rightarrow\infty$) limit, $\Gamma_{ij}(\epsilon)=\Gamma_{ij}$ for $D\rightarrow\infty$. We denote
\begin{equation}
\begin{aligned}
    \bar{\Gamma}_{ij}(i\omega_n)=\frac{\Gamma_{ij}}{\sqrt{\omega_n^2+\Delta^2}}A(i\omega_n), \\
    \label{eq:GammaFun}
\end{aligned}
\end{equation}
where 
\begin{equation}
    A(i\omega_n)=\frac{2}{\pi}\arctan\left[\frac{D}{\sqrt{\omega_n^2+\Delta^2}}\right]
\end{equation}
is a correction to finite bandwidth that approaches unity in the wide-band limit.
One can notice that the Green's function elements contain two types of summations. For zero magnetic field, the spin degeneracy should hold; thus, we can set $\epsilon_{\textbf{k}\uparrow} = \epsilon_{\textbf{k}\downarrow} = \epsilon_\textbf{k}$. Using these assumptions, we will get the following two different integrals 
\begin{align}
    \frac{\Gamma_{ij}}{\pi}&\int_{-\infty}^{\infty} d\epsilon \frac{(\epsilon \pm i\omega)}{\epsilon^2 + \omega^2 + \Delta^2} = \pm i\omega_n \tilde{\Gamma}_{ij},\\
    \frac{\Gamma_{ij}}{\pi}&\int_{-\infty}^{\infty} d\epsilon \frac{\Delta}{\epsilon^2 + \omega^2 + \Delta^2} = \Delta \tilde{\Gamma}_{ij},
\end{align}
where $i,j=1,2$ denote the dot index. 

The inverse impurity Green's function then reads

\begin{equation}
\begin{aligned}
    &[G^{\text{d}}_0(i\omega_n)]^{-1} = \\
    &\begin{pmatrix}
    i\omega_n[1+\bar{\Gamma}_{11}(i\omega_n)] -\epsilon_{1} 
    & -\Delta \bar{\Gamma}_{11}(i\omega_n) 
    & i\omega_n \bar{\Gamma}_{21}(i\omega_n) - t_{12}
    & -\Delta \bar{\Gamma}_{21}(i\omega_n)
    \\
    -\Delta \bar{\Gamma}_{11}(i\omega_n) & i\omega_n[1+\bar{\Gamma}_{11}(i\omega_n)] + \epsilon_{1}
    & -\Delta \bar{\Gamma}_{21}(i\omega_n)
    & i\omega_n \bar{\Gamma}_{21}(i\omega_n) + t_{12}
    \\
    i\omega_n \bar{\Gamma}_{12}(i\omega_n) - t_{12}
    & -\Delta \bar{\Gamma}_{12}(i\omega_n)
    & i\omega_n[1+\bar{\Gamma}_{22}(i\omega_n)] - \epsilon_{2}
    & -\Delta \bar{\Gamma}_{22}(i\omega_n)
    \\
    -\Delta \bar{\Gamma}_{12}(i\omega_n)
    & i\omega_n \bar{\Gamma}_{12}(i\omega_n) + t_{12}
    & -\Delta \bar{\Gamma}_{22}(i\omega_n)
    & i\omega_n[1+\bar{\Gamma}_{22}(i\omega_n)] + \epsilon_{2}
    \end{pmatrix}.
\end{aligned}
\end{equation}
\paragraph{Superconducting atomic limit:} Assuming first the wide-band limit $D \rightarrow \infty$ and then the infinite-gap limit $\Delta \rightarrow \infty$, the inverse impurity Green's function reduces to
\begin{equation}
    [G^{\text{d}}_0(i\omega_n)]^{-1} = \\
    \begin{pmatrix}
    i\omega_n -\epsilon_{1} 
    & -\Gamma_{11} 
    & - t_{12}
    & -\Gamma_{21}
    \\
    -\Gamma_{11}& i\omega_n + \epsilon_{1}
    & -\Gamma_{21}
    & t_{12}
    \\
    - t_{12}
    & -\Gamma_{12}
    & i\omega_n- \epsilon_{2}
    & -\Gamma_{22}
    \\
    -\Gamma_{12}
    & t_{12}
    & -\Gamma_{22}
    & i\omega_n + \epsilon_{2}
    \end{pmatrix}.
\end{equation}
This corresponds to the noninteracting Green's function of a simplified model, known as the superconducting atomic limit (SC-AL) model,
\begin{equation}
\begin{aligned}
\mathcal{H}^{\text{AL,NI}}_\text{DQD} = &\sum_{j=1}^{2} \sum_{\sigma} \epsilon_j n_{j \sigma} -\sum_{j=1}^{2} \Gamma_{jj}\left( d_{j\uparrow}^\dagger d_{j\downarrow}^\dagger + d_{j\downarrow} d_{j\uparrow} \right)-\sum_{\sigma}t_{12} \left(d_{2\sigma}^\dagger d_{1\sigma}^\pdag + d_{1\sigma}^\dagger d_{2\sigma}^\pdag\right)\\
 &- \Gamma_{21}  \left(d_{1\uparrow}^\dagger d_{2\downarrow}^\dagger + d_{2\uparrow}^\pdag d_{1\downarrow}^\pdag\right)
 - \Gamma_{12}  \left(d_{2\uparrow}^\dagger d_{1\downarrow}^\dagger + d_{1\uparrow}^\pdag d_{2\downarrow}^\pdag\right),
 \end{aligned}
\end{equation}
where $\Gamma_{jj}\equiv\Gamma_j$ represent the induced SC gaps on the dots, and the cross terms $\Gamma_{i \neq j}$ describe processes where a Cooper pair splits into two different QDs.

The cross terms have a simple microscopic
interpretation. They measure the overlap between the combinations of superconducting bath states coupled to dots $i$ and $j$. From
Eq.~\eqref{eq:WBL}, the Cauchy--Schwarz inequality implies
$
|\Gamma_{ij}|^2\leq \Gamma_{i}\Gamma_{j},
$
so that for equivalent dots, $\Gamma_{i}=\Gamma_{j}=\Gamma$, one has
$|\zeta|\leq1$ and $\Gamma_{ij}=\zeta\Gamma$.

The interacting version of the model includes repulsive Coulomb term,
\begin{equation}
    \mathcal{H}^{\text{AL}}_{\text{DQD}} 
    = \mathcal{H}^{\text{AL,NI}}_\text{DQD} + 
    \sum_{j=1}^{2}U_j n_{j\uparrow}n_{j\downarrow}.
\end{equation}
Generalization of this model to an arbitrary number of dots is straightforward. To simplify the analysis, we also assume identical dots with $\epsilon_j = \epsilon$, $U_j = U$, and $\Gamma_j = \Gamma$. The geometry of the dot assembly is then defined by cross terms $\Gamma_{i \neq j}$ and the hopping matrix $\mathbf{t}$. In the main text we restrict these to nearest neighbors, i.e., $t_{ij} = t$, and $\Gamma_{ij} = \zeta \Gamma$ for nearest neighbors and zero otherwise, with $\zeta \in [ -1,1]$. This leads to the general Hamiltonian used in the main text. The restriction to nearest-neighbor hopping and crossed pairing is a modeling assumption that simplifies the analysis. Naturally, longer-range terms can modify the quantitative phase boundaries and the stability of individual phases, but their systematic investigation is beyond the scope of the present work. 

The limit $\zeta=0$ corresponds to effectively uncorrelated (e.g., distant dots) 
whereas $|\zeta|=1$ corresponds to maximally correlated couplings (e.g., multilevel dot).
However, in a spatially resolved microscopic model, $\Gamma_{ij}$ generally
depends on the dot separation, contact geometry, tunneling amplitudes, Fermi-wavelength-scale interference, and coherence length.

In the effective description these spatial dependencies are not resolved explicitly, but are
parametrized by $\zeta$. Consequently, the formal $\Delta\rightarrow\infty$
limit should not be interpreted simultaneously as vanishing superconducting coherence length. Rather, the hybridization matrix $\Gamma_{ij}$ is retained as an effective low-energy parameter while the dynamical degrees of freedom of the
superconducting bath are eliminated.

\section{Single QD in the transformed particle conserving basis
\label{app:SQD}}
The Hamiltonian of a single QD in the SC-AL ($\Delta\rightarrow\infty$) in transformed basis~\eqref{eq:trans} reads as follows:
\begin{equation}
\begin{aligned}
\tilde{\mathcal{H}} &= \, \varepsilon \left( \tilde{n}_{\uparrow} -  \tilde{n}_{ \downarrow}\right) 
+ \frac{U}{2} \left( \tilde{n}_{\uparrow} - \tilde{n}_{\downarrow} \right)^2
- \Gamma \left( \tilde{d}_{\uparrow}^\dagger \tilde{d}_{\downarrow}^\pdag + \tilde{d}_{\downarrow}^\dagger \tilde{d}_{\uparrow}^\pdag \right).
\end{aligned}
\end{equation}
We can represent it in basis ${\ket{0},\ket{\uparrow},\ket{\downarrow},\ket{\uparrow\downarrow}}$ as
\begin{equation}
\tilde{\mathcal{H}} = \begin{pmatrix}
0 & 0 & 0 & 0\\
0 & \varepsilon+\frac{U}{2} & -\Gamma & 0\\
0 & -\Gamma & -\varepsilon+\frac{U}{2} & 0\\
0 & 0 & 0 & 0
\end{pmatrix}.
\end{equation}
This Hamiltonian has two trivial eigenvalues $E=0$ for eigenvectors $\ket{0}$ ($\tilde{N}_f=0$) and $\ket{\uparrow\downarrow}$ ($\tilde{N}_f=2$), and
\begin{equation}
E_{\pm}=\frac{U}{2}\pm\sqrt{\varepsilon^2+\Gamma^2}
\end{equation}
($\tilde{N}_f=1$) with the corresponding eigenvectors
\begin{equation}
\begin{aligned}
\ket{\psi_\pm} &= \frac{1}{\sqrt{\Gamma^2+(\varepsilon+\frac{U}{2}-E_{\pm})^2}}\begin{pmatrix}
0\\
\Gamma\\
\varepsilon+\frac{U}{2}-E_{\pm}\\
0
\end{pmatrix}.
\label{eq:1QDeig}
\end{aligned}
\end{equation}

\section{Exact solution for $L=2$ at $\varepsilon=0$ and $t=0$ in transformed basis}
\label{app:L2}
We consider the Hamiltonian Eq.~(\ref{eq:Ham3}) for $L=2$ where we omit the constant energy shift $-U$ and set the trivial HSC $\varepsilon=0$, $t=0$ and assume $\zeta\geq 0$ for simplicity. The Hamiltonian conserves the total particle number $\tilde{N}_f$, and for $\varepsilon=0$ it additionally commutes with the global spin-flip operator $\bar{\mathcal S}$. Therefore, it decomposes into independent blocks labeled by $(\tilde{N}_f,\bar{\mathcal S})$. For clarity we use here basis ordered as
$\ket{n_{1\uparrow},n_{1\downarrow},n_{2\uparrow},n_{2\downarrow}}$.

\subsection*{Sectors $\tilde{N}_f=0$ and $\tilde{N}_f=4$}
These sectors are trivial, each consisting of a single state with energy
$E_0 = E_4 = 0$ and respective eigenvectors 
\begin{equation}
\ket{\psi_0}=\ket{0,0,0,0},\quad\ket{\psi_4}=\ket{1,1,1,1}.
\end{equation}

\subsection*{Sector $\tilde{N}_f=1$}
In the basis
\begin{equation}
\{\ket{1,0,0,0},~\ket{0,1,0,0},~\ket{0,0,1,0},~\ket{0,0,0,1}\}
\end{equation}
the Hamiltonian reads
\begin{equation}
H_{1}=
\begin{pmatrix}
\frac U2 & -\Gamma & 0 & -\zeta\Gamma\\
-\Gamma & \frac U2 & -\zeta\Gamma & 0\\
0 & -\zeta\Gamma & \frac U2 & -\Gamma\\
-\zeta\Gamma & 0 & -\Gamma & \frac U2
\end{pmatrix}.
\end{equation}
By introducing spin-flip states
\begin{equation}
\begin{aligned}
|1,\pm\rangle &= \frac{1}{\sqrt2}(\ket{1,0,0,0}\pm\ket{0,1,0,0}),\\
|2,\pm\rangle &= \frac{1}{\sqrt2}(\ket{0,0,1,0}\pm\ket{0,0,0,1}),
\end{aligned}
\end{equation}
this Hamiltonian decomposes into
\begin{equation}
H_{1}^{(+)}=
\begin{pmatrix}
\frac U2-\Gamma & -\zeta\Gamma\\
-\zeta\Gamma & \frac U2-\Gamma
\end{pmatrix},\quad
H_{1}^{(-)}=
\begin{pmatrix}
\frac U2+\Gamma & \zeta\Gamma\\
\zeta\Gamma & \frac U2+\Gamma
\end{pmatrix}
\end{equation}
with eigenvalues
\begin{equation}
E_{1,+}=\frac U2 - \Gamma \pm \zeta\Gamma,\quad
E_{1,-}=\frac U2 + \Gamma \pm \zeta\Gamma
\end{equation}
and corresponding eigenvectors 
\begin{equation}
\ket{\psi_{1,+,\pm}}=\frac{1}{\sqrt{2}}(\ket{1,+}\pm\ket{2,+}),\quad
\ket{\psi_{1,-,\pm}}=\frac{1}{\sqrt{2}}(\ket{1,-}\pm\ket{2,-}).
\end{equation}
Assuming $\Gamma>0$ and $\zeta>0$ the minimal energy and the respective eigenvector read
\begin{equation}
\begin{aligned}
E_1(\zeta>0) &= \frac{U}{2}-\Gamma -\zeta\Gamma,\\
\ket{\psi_{1}}&=\frac{1}{\sqrt{2}}(\ket{1,+}+\ket{2,+})\\
&=\frac{1}{2}(\ket{1,0,0,0}+\ket{0,1,0,0} + \ket{0,0,1,0}+\ket{0,0,0,1}).
\end{aligned}
\end{equation}

\subsection*{Sector $\tilde{N}_f=3$}
This sector is connected to $\tilde{N}_f=1$ through particle-hole transformation. Therefore, it has an identical spectrum and the ground state:
\begin{equation}
\begin{aligned}
E_3(\zeta>0) &= \frac{U}{2}-\Gamma -\zeta\Gamma,\\
\ket{\psi_{3}}&=
\frac{1}{2}(\ket{0,1,1,1}+\ket{1,0,1,1} + \ket{1,1,0,1}+\ket{1,1,1,0}).
\end{aligned}
\end{equation}

\subsection*{Sector $\tilde{N}_f=2$}
In the basis
\begin{equation}
\{|1,1,0,0\rangle,\ |1,0,1,0\rangle,\ |0,1,1,0\rangle,\ |1,0,0,1\rangle,\ |0,1,0,1\rangle,\ |0,0,1,1\rangle\},
\end{equation}
the Hamiltonian reads
\begin{equation}
H_{2}=
\begin{pmatrix}
0 & -\zeta\Gamma & 0 & 0 & \zeta\Gamma & 0\\
-\zeta\Gamma & U & -\Gamma & -\Gamma & 0 & \zeta\Gamma\\
0 & -\Gamma & U & 0 & -\Gamma & 0\\
0 & -\Gamma & 0 & U & -\Gamma & 0\\
\zeta\Gamma & 0 & -\Gamma & -\Gamma & U & -\zeta\Gamma\\
0 & \zeta\Gamma & 0 & 0 & -\zeta\Gamma & 0
\end{pmatrix}.
\end{equation}
As before, using the spin-flip symmetry, we define
\begin{equation}
\begin{aligned}
\ket{\uparrow\uparrow_{\pm}} &= \tfrac{1}{\sqrt2}(\ket{1,0,1,0} \pm \ket{0,1,0,1}),\\
\ket{\uparrow\downarrow_{\pm}} &= \tfrac{1}{\sqrt2}(\ket{1,0,0,1} \pm \ket{0,1,1,0}),\\
D_1 &= \ket{1,1,0,0},\\
D_2 &= \ket{0,0,1,1},
\end{aligned}
\end{equation}
where $\ket{\uparrow\uparrow_{+}}$, $\ket{\uparrow\downarrow_{+}}$ belong into spin-flip even sector and $D_1$, $D_2$, $\ket{\uparrow\uparrow_{-}}$, $\ket{\uparrow\downarrow_{-}}$ belong into spin-flip odd sector. 

\paragraph{Spin-flip even sector} is described by
\begin{equation}
H_{2}^{(+)}=
\begin{pmatrix}
U & -2\Gamma\\
-2\Gamma & U
\end{pmatrix}
\end{equation}
with eigenvalues:
\begin{equation}
E^{(+)}_{2,\pm}=U\mp 2\Gamma
\end{equation}
and the respective eigenvectors
\begin{equation}
\frac{1}{\sqrt{2}}(\ket{\uparrow\uparrow_{+}}\pm \ket{\uparrow\downarrow_{+}}).
\end{equation}
The lower energy combination is
\begin{equation}
\begin{aligned}
E^{(+)}_{2}&=U-2\Gamma,\\
\ket{\psi_{2}^{(+)}}&=\frac{1}{\sqrt{2}}(\ket{\uparrow\uparrow_{+}}+\ket{\uparrow\downarrow_{+}})\\
&=\tfrac{1}{2}(\ket{1,0,1,0} + \ket{0,1,0,1}+\ket{1,0,0,1}+ \ket{0,1,1,0}).
\end{aligned}
\end{equation}

\paragraph{Spin-flip odd sector} is described by
\begin{equation}
H_{2}^{(-)}=
\begin{pmatrix}
0 & 0 & -\sqrt2\,\zeta\Gamma & 0\\
0 & 0 & \sqrt2\,\zeta\Gamma & 0\\
-\sqrt2\,\zeta\Gamma & \sqrt2\,\zeta\Gamma & U & 0\\
0 & 0 & 0 & U
\end{pmatrix}.
\end{equation}
This can be easily diagonalized as state $\ket{\uparrow\downarrow_{-}}$ with energy $U$ is decoupled. The rest of the eigenenergies read
\begin{equation}
0,\qquad \frac{U}{2} \pm\sqrt{\left(\frac{U}{2}\right)^2+\left(2\zeta\Gamma\right)^2}\equiv\frac{U}{2}\pm\Upsilon,
\end{equation}
with respective eigenvectors
\begin{equation}
\frac{1}{\sqrt{2}}(D_1+D_2),\qquad
\frac{1}{\sqrt{1+\left(\frac{U\pm2\Upsilon}{4\zeta\Gamma}\right)^2}}\left(\frac{1}{\sqrt{2}}(D_1-D_2)-\frac{U\pm 2\Upsilon}{4\zeta\Gamma}\ket{\uparrow\uparrow_{-}}\right).
\end{equation}
The state with the lowest energy for $\zeta>0$ in this sector is
\begin{equation}
\begin{aligned}
E^{(-)}_{2}&=\frac{U}{2}-\sqrt{\left(\frac{U}{2}\right)^2+\left(2\zeta\Gamma\right)^2},\\
\ket{\psi_{2}^{(-)}}&=\frac{1}{\sqrt{2}\sqrt{1+\left(\frac{U-2\Upsilon}{4\zeta\Gamma}\right)^2}}\left((\ket{1,1,0,0}-\ket{0,0,1,1})-\frac{U-2\Upsilon}{4\zeta\Gamma}(\ket{1,0,1,0}-\ket{0,1,0,1})\right).
\end{aligned}
\end{equation}
To better understand the nature of this, it is useful to explore the limit of small non-zero $\zeta$ and $U>2\Gamma$ for which $(U-2\Upsilon)/(4\zeta\Gamma)\approx 2\zeta\Gamma/U \ll 1$. We get
\begin{equation}
\begin{aligned}
    E^{(-)}_{2}(\zeta\ll 1)&\approx-\frac{4\zeta^2\Gamma^2}{U},\\
\ket{\psi_{2}^{(-)}}&\approx\frac{1}{\sqrt{2}}\left(\ket{1,1,0,0}-\ket{0,0,1,1}+\frac{2\zeta\Gamma}{U}(\ket{1,0,1,0}-\ket{0,1,0,1})\right).
\end{aligned}
\label{eq:psi2minapp}
\end{equation}
This can be interpreted as the antisymmetric doublon state hybridized with the intermediate states due to pair hopping processes that flip the spin, which lowers its energy with respect to the symmetric doublon state by $\approx (2\zeta\Gamma)^2/U$.
Note that the case $\zeta=0$ should be addressed separately. In this limit, we get a combination of independent QDs, solved in previous section. Consequently, for $\zeta=0$ the antisymmetric doublon state is no longer hybridized with $\ket{\uparrow\uparrow_{-}}$ and has the same energy as the symmetric one.  

Putting all sectors together allows us to construct the ground state phase diagram.
Using $\Gamma$ as the energy unit, the ground state is determined by competition between the lowest states in the $\tilde{N}_f=1,3$ and $\tilde{N}_f=2$ sectors. The first phase boundary arises from the crossing between the lowest even $\tilde{N}_f=2$ state and the lowest $\tilde{N}_f=1$ state:
\begin{equation}
U - 2\Gamma = \frac{U}{2} - \Gamma(1+\zeta)
\;\;\Rightarrow\;\;
U_{c1}=2\Gamma(1-\zeta).
\label{eq:phb1}
\end{equation}
The second boundary follows from the crossing between the lowest $\tilde{N}_f=1$ state and the lowest odd $\tilde{N}_f=2$ branch:
\begin{equation}
\frac{U}{2} - \Gamma(1+\zeta)
=
\frac{U - \sqrt{U^2+16\zeta^2\Gamma^2}}{2}
\;\;\Rightarrow\;\;
U_{c2}=2\Gamma\sqrt{(1-\zeta)(1+3\zeta)}.
\label{eq:phb2}
\end{equation}
With increasing $U$ (or $\zeta$), these boundaries separate the sequence of ground states:
\begin{equation}
\begin{aligned}
N_f=2;\;&\frac{1}{2}\left(\tilde{d}^\dagger_{1\uparrow}+\tilde{d}^\dagger_{1\downarrow}\right)\left(\tilde{d}^\dagger_{2\uparrow}+\tilde{d}^\dagger_{2\downarrow}\right)\ket{0}\quad(\text{trivial BCS singlet}),\\
&\downarrow\\
\tilde{N}_f=1;\;&\frac{1}{2}\left(\tilde{d}^\dagger_{1\uparrow}+\tilde{d}^\dagger_{1\downarrow}+\tilde{d}^\dagger_{2\uparrow}+\tilde{d}^\dagger_{2\downarrow}\right)\ket{0}\quad(\text{delocalized doublet}),\\
\tilde{N}_f=3;\;&\frac{1}{2}\left(\tilde{d}^\dagger_{1\downarrow}\tilde{d}^\dagger_{2\uparrow}\tilde{d}^\dagger_{2\downarrow} +
\tilde{d}^\dagger_{1\uparrow}\tilde{d}^\dagger_{2\uparrow}\tilde{d}^\dagger_{2\downarrow}+
\tilde{d}^\dagger_{1\uparrow}\tilde{d}^\dagger_{1\downarrow}\tilde{d}^\dagger_{2\downarrow}+
\tilde{d}^\dagger_{1\uparrow}\tilde{d}^\dagger_{1\downarrow}\tilde{d}^\dagger_{2\uparrow}
\right)\ket{0},\\
&\downarrow\\
\tilde{N}_f=2;\;&\frac{1}{\sqrt{2+2\left(\frac{U-2\Upsilon}{4\zeta\Gamma}\right)^2}}\left(\tilde{d}^\dagger_{1\uparrow}\tilde{d}^\dagger_{1\downarrow}-\tilde{d}^\dagger_{2\uparrow}\tilde{d}^\dagger_{2\downarrow}-\frac{U-2\Upsilon}{4\zeta\Gamma}(\tilde{d}^\dagger_{1\uparrow}\tilde{d}^\dagger_{2\uparrow}-\tilde{d}^\dagger_{1\downarrow}\tilde{d}^\dagger_{2\downarrow})\right)\ket{0}\quad
(\text{molecular singlet}),
\end{aligned}
\end{equation}
consistent with Fig.~\ref{fig:1DUzetaEO}(a1).  

\section{DMRG analysis of the three regimes in the one-dimensional case}
\label{app:MPS1D}

Here we provide a more detailed MPS analysis of the three regimes identified in the one-dimensional phase diagram. We study representative parameter points as a function of the chain length. For finite chains, detailed parameter scans become increasingly difficult because the number of level crossings grows with system size, including crossings between states within the same $\tilde{N}_f$ sector. We therefore concentrate on selected points characteristic of the three regimes and analyze their finite-size behavior through the entanglement entropy and excitation gaps. Even in the third regime, where the entanglement entropy increases with
system size, the DMRG calculations remain converged for the chain lengths
considered here, allowing us to reach $L=200$.

\begin{figure}[h]
    \centering
    \includegraphics[width=1.0\linewidth]{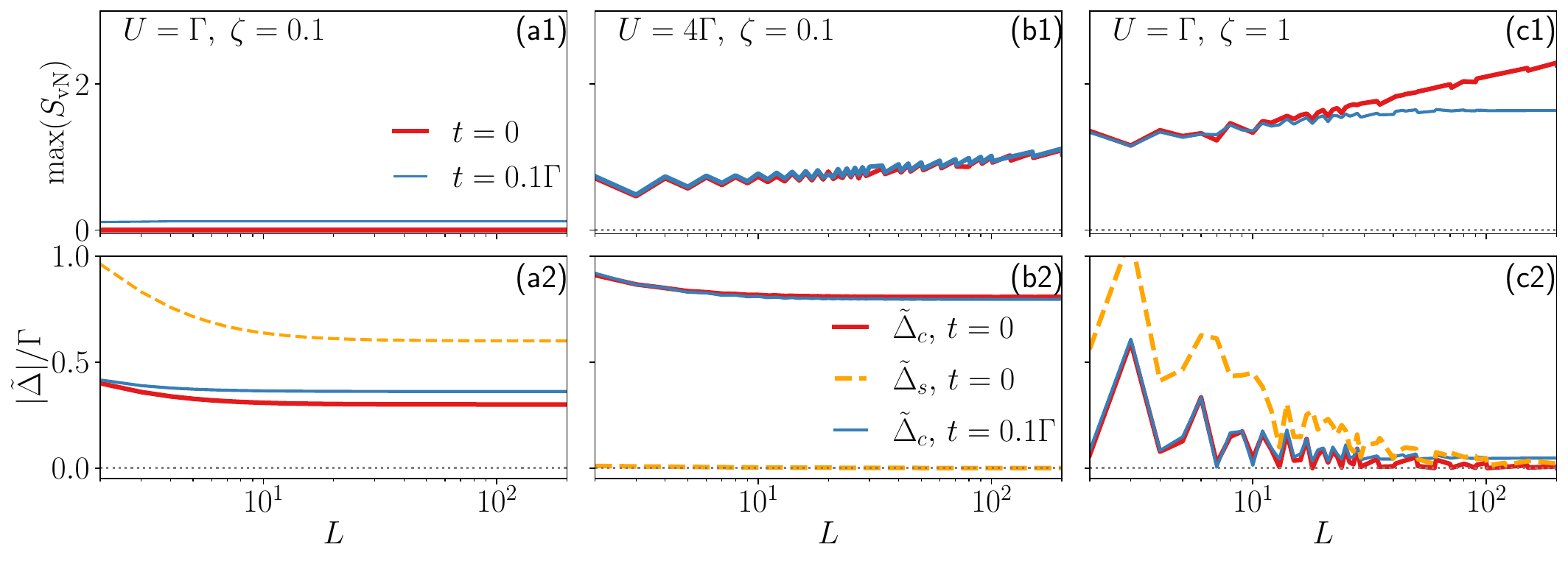}
    \caption{Finite-size scaling of the entanglement entropy (first row), charge gap (solid lines in the second row), and sector gap (dashed lines in the second row) for parameters representing the first (a), second (b), and third (c) regimes discussed in the text. The red and orange lines correspond to the HSC point for $\varepsilon=0$ and $t=0$, while the blue lines show results for $t=0.1\Gamma$.}
    \label{fig:Ldep}
\end{figure}

As shown in Fig.~\ref{fig:Ldep}, in the first regime the entanglement entropy remains zero for $t=0$, see panel (a1), reflecting the product-state character of the ground state. At the same time, increasing the chain length stabilizes a finite charge gap, shown by the red lines in panel (a2),
\begin{equation}
\tilde{\Delta}_c =
\abs{
E(\tilde{N}_f+1)
+
E(\tilde{N}_f-1)
-
2E(\tilde{N}_f)
}/2.
\end{equation}
This quantity is equivalent to the parity gap $\Delta_p$ in the original superconducting formulation, i.e., the gap between the even- and odd-parity sectors. We define it as one half of the conventional many-body charge gap in order to compare it directly with the excitation gap within a fixed $\tilde{N}_f$ sector,
\begin{equation}
\tilde{\Delta}_s(\tilde{N}_f)
=
E(\tilde{N}_f)_1
-
E(\tilde{N}_f)_0,
\end{equation}
shown by the orange dashed lines. The latter follows the same qualitative behavior but is generally larger. The first regime is therefore fully gapped.

In the second regime, the entanglement entropy increases with system size but remains below that of the third regime over the entire range of sizes shown in Fig.~\ref{fig:Ldep}. The charge gap stays finite within the accessible system sizes, whereas the sector gap rapidly becomes negligible. This behavior is consistent with the effective low-energy description in terms of coherent doublon hopping generated by second-order virtual single-particle processes, discussed in detail in Appendix~\ref{app:2nd}.

The third regime shows qualitatively different finite-size behavior. Both the charge and sector gaps decrease rapidly with increasing chain length, while the entanglement entropy grows approximately as $\log L$. These results are consistent with a critical, gapless phase in the thermodynamic limit.

It is important to stress that this critical behavior is specific to HSC. When HSC is broken, for example by introducing a finite hopping $t=0.1\Gamma$, a nonzero entanglement entropy appears already in the first regime for all studied system sizes, see the blue lines in panel (a1). More importantly, the third regime becomes gapped: the entanglement entropy saturates with increasing $L$, see panel (c1), while the charge gap approaches a finite value, see panel (c2).

\section{Strong-coupling expansion for $U>2\Gamma$ and $\zeta\ll1$}
\label{app:2nd}
We now use the single-dot and two-dot results derived above to construct an effective low-energy theory in the regime
$U>2\Gamma, \zeta\ll 1$. The unperturbed problem is taken to be the chain with $\zeta=0$, while the inter-dot spin-flip hopping proportional to $\zeta\Gamma$ is treated perturbatively.

For $\zeta=0$, each dot is independent. From Appendix~\ref{app:SQD}, the local eigenstates are
\begin{equation}
\ket{0}_j,\quad
\ket{D}_j
=
\tilde d^\dagger_{j\uparrow}
\tilde d^\dagger_{j\downarrow}
\ket{0}_j,
\end{equation}
with energy zero, and the singly occupied states
\begin{equation}
\ket{j,+}
=
\frac{1}{\sqrt2}
\left(
\tilde d^\dagger_{j\uparrow}
+
\tilde d^\dagger_{j\downarrow}
\right)\ket{0}_j,
\qquad
\ket{j,-}
=
\frac{1}{\sqrt2}
\left(
\tilde d^\dagger_{j\uparrow}
-
\tilde d^\dagger_{j\downarrow}
\right)\ket{0}_j,
\end{equation}
with energies
\begin{equation}
E_+=\frac{U}{2}-\Gamma,
\qquad
E_-=\frac{U}{2}+\Gamma.
\end{equation}
Therefore, for $U>2\Gamma,$
the singly occupied sector is separated from the empty/doubly occupied sector by a finite gap
$\frac{U}{2}-\Gamma$. 
Consequently, for sufficiently small $\zeta$, the low-energy Hilbert space is spanned predominantly by configurations containing only empty sites $\ket{0}_j$ and local doublons $\ket{D}_j$. The perturbation
\begin{equation}
V=-\zeta\Gamma
\sum_{\langle i,j\rangle}
\left(\tilde d^\dagger_{j\uparrow}\tilde d_{i\downarrow}+
\tilde d^\dagger_{i\uparrow}\tilde d_{j\downarrow}+
\tilde d^\dagger_{j\downarrow}\tilde d_{i\uparrow}+
\tilde d^\dagger_{i\downarrow}\tilde d_{j\uparrow}\right),
\end{equation}
changes the position of one quasiparticle while simultaneously flipping its spin. Consequently, it couples the low-energy doublon sector to virtual states with one quasiparticle on each of two neighboring sites.

Let us first consider a single bond $\langle i,j\rangle$. The relevant low-energy states are
\begin{equation}
\ket{D_i,0_j},
\qquad
\ket{0_i,D_j}.
\end{equation}
The perturbation connects these states to the normalized virtual state
\begin{equation}
\ket{T_{ij}}
=
\frac{1}{\sqrt2}
\left(
\ket{\uparrow_i,\uparrow_j}
-
\ket{\downarrow_i,\downarrow_j}
\right).
\end{equation}
Note that this state should not be identified with two spin-flip-even quasiparticles. Indeed, using
\begin{equation}
\ket{\uparrow_j}=\frac{1}{\sqrt2}\left(\ket{j,+}+\ket{j,-}\right),
\quad
\ket{\downarrow_j}=\frac{1}{\sqrt2}\left(\ket{j,+}-\ket{j,-}\right),
\end{equation}
one obtains
\begin{equation}
\ket{T_{ij}}
=
\frac{1}{\sqrt2}
\left(
\ket{i,+}\ket{j,-}
+
\ket{i,-}\ket{j,+}
\right).
\end{equation}
Therefore, the energy of the intermediate state relative to the low-energy doublon manifold is
\begin{equation}
\Delta_T = E_+ + E_-
= \left(\frac{U}{2}-\Gamma\right) + \left(\frac{U}{2}+\Gamma\right) = U.
\end{equation}
Explicitly, with the same fermionic ordering convention as in Appendix~\ref{app:L2},
\begin{equation}
V\ket{D_i,0_j} = -\sqrt2\,\zeta\Gamma\,\ket{T_{ij}},
\end{equation}
whereas
\begin{equation}
V\ket{0_i,D_j} = +\sqrt2\,\zeta\Gamma\,\ket{T_{ij}}.
\end{equation}

The effective Hamiltonian to the second order in $V$ reads
\begin{equation}
H_{\rm eff}^{(2)} = -PVQ\frac{1}{QH_0Q-E_0}QVP,
\end{equation}
where $P$ is a projector onto the low-energy subspace spanned by empty and doubly occupied sites, while $Q=1-P$ projects onto the virtual singly occupied states. Keeping the virtual state $\ket{T_{ij}}$, we obtain
\begin{equation}
H_{\rm eff}^{(2)} \simeq -\frac{1}{U}PVVP.
\end{equation}
The diagonal process
\begin{equation}
\ket{D_i,0_j}
\rightarrow
\ket{T_{ij}}
\rightarrow
\ket{D_i,0_j}
\end{equation}
gives the energy correction
\begin{equation}
\Delta E_{\rm diag}
=
-\frac{
|\langle T_{ij}|V|D_i,0_j\rangle|^2
}{U}
=
-\frac{2\zeta^2\Gamma^2}{U}.
\end{equation}
The same correction is obtained for $\ket{0_i,D_j}$.

The off-diagonal process
\begin{equation}
\ket{D_i,0_j}
\rightarrow
\ket{T_{ij}}
\rightarrow
\ket{0_i,D_j}
\end{equation}
generates coherent doublon hopping,
\begin{equation}
\langle 0_i,D_j|H_{\rm eff}^{(2)}|D_i,0_j\rangle
=
-\frac{
\langle 0_i,D_j|V|T_{ij}\rangle
\langle T_{ij}|V|D_i,0_j\rangle
}{U}
=
+\frac{2\zeta^2\Gamma^2}{U}.
\end{equation}
Therefore, in the basis
\begin{equation}
\{\ket{D_i,0_j},~\ket{0_i,D_j}\},
\end{equation}
the effective bond Hamiltonian becomes
\begin{equation}
H_{ij}^{\rm eff}
=
-\frac{2\zeta^2\Gamma^2}{U}
\begin{pmatrix}
1 & -1\\
-1 & 1
\end{pmatrix}.
\end{equation}
Its eigenstates are
\begin{equation}
\frac{1}{\sqrt2}
\left(
\ket{D_i,0_j}
-
\ket{0_i,D_j}
\right),
\qquad
E=-\frac{4\zeta^2\Gamma^2}{U}
\end{equation}
and
\begin{equation}
\frac{1}{\sqrt2}
\left(
\ket{D_i,0_j}
+
\ket{0_i,D_j}
\right),
\qquad
E=0.
\end{equation}
This reproduces the small-$\zeta$ limit of the exact two-site molecular singlet,
\begin{equation}
E_2^{(-)}
\simeq
-\frac{4\zeta^2\Gamma^2}{U}.
\end{equation}

To write the result for a chain, we introduce hard-core bosonic operators representing local doublons,
\begin{equation}
b_j^\dagger\ket{0}_j=\ket{D}_j,
\qquad
n_j=b_j^\dagger b_j.
\end{equation}

A given bond contributes only when one site is empty and the other doubly occupied. The diagonal contribution is therefore
\begin{equation}
-\frac{2\zeta^2\Gamma^2}{U}
\left[
n_i(1-n_j)+n_j(1-n_i)
\right],
\end{equation}
while the off-diagonal process corresponds to coherent doublon motion,
\begin{equation}
+\frac{2\zeta^2\Gamma^2}{U}
\left(
b_i^\dagger b_j+b_j^\dagger b_i
\right).
\end{equation}
The effective Hamiltonian can thus be written by summing over nearest-neighbor bonds
\begin{equation}
H_{\rm eff}
=
-J
\sum_{\langle i,j\rangle}
\left(n_i+n_j-2n_in_j\right)
+
J\sum_{\langle i,j\rangle}
\left(b_i^\dagger b_j+b_j^\dagger b_i\right),
\qquad
J=\frac{2\zeta^2\Gamma^2}{U}.
\label{eq:Heff}
\end{equation}
This model describes mobile hard-core doublons generated by virtual spin-flip fluctuations. The first term represents the second-order energy gain associated with virtual triplet-like fluctuations on bonds connecting an empty site and a doublon. The second term describes coherent doublon hopping generated by the same virtual process.

Importantly, the diagonal term energetically favors configurations maximizing the number of bonds connecting empty and doubly occupied sites. Note that, 
\begin{equation}
n_i+n_j-2n_in_j=
\begin{cases}
1,& (n_i,n_j)=(1,0)\ \text{or}\ (0,1),\\
0,& (n_i,n_j)=(0,0)\ \text{or}\ (1,1).
\end{cases}
\end{equation}
Therefore, each bond connecting a doublon and an empty site gains energy $-J$, whereas neighboring doublons do not. Consequently, the effective interaction between neighboring doublons is repulsive.

At or near half filling, the energy is minimized by maximizing the number of alternating bonds, naturally leading to a checkerboard configuration
$\ket{D,0,D,0,\dots}$
or
$\ket{0,D,0,D,\dots}$. 
For an even number of sites, the ideal checkerboard occurs at half filling. For an odd number of sites, a perfect alternating pattern is obtained in the sectors with one extra empty site or one extra doublon. However, the checkerboard product state is not an exact eigenstate of the effective Hamiltonian because the hopping term produces quantum fluctuations.
\begin{figure}[h]
    \centering \includegraphics[width=1.0\linewidth]{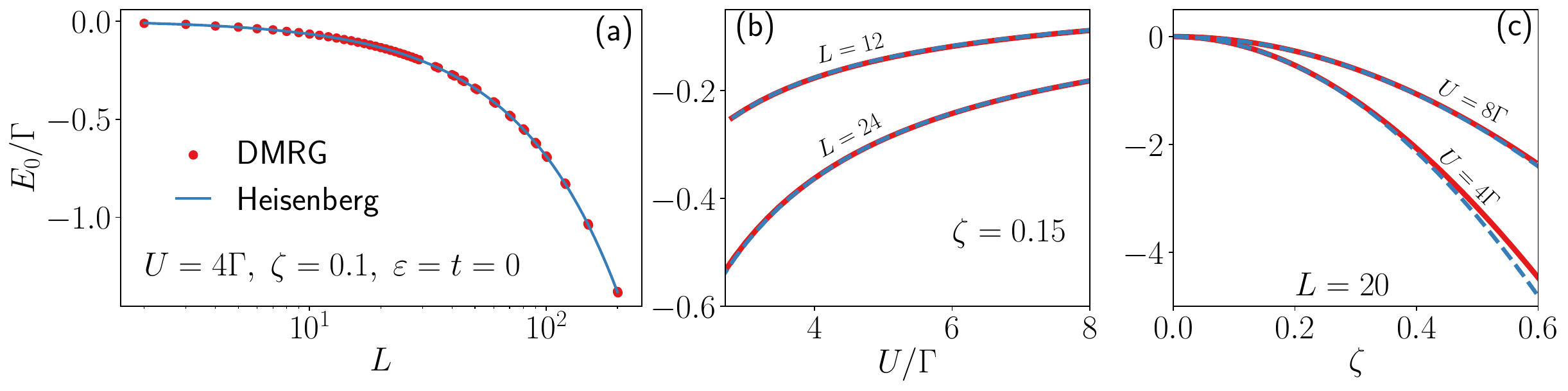}
    \caption{Comparison of DMRG results (red) with the effective Heisenberg-model approximation from Eq.~\ref{eq:1DH} (blue) for the corresponding one-dimensional regime. (a) Finite-size scaling. (b) Dependence on $U/\Gamma$ for small $\zeta=0.15$. (c) Evolution of the agreement between the two approaches with increasing $\zeta$ for different values of $U$, illustrating the range of validity of the effective Heisenberg description.}
    \label{fig:app:reg2}
\end{figure}

At half filling, Eq.~(\ref{eq:Heff}) maps onto the spin-$1/2$ antiferromagnetic Heisenberg chain. Using
\begin{equation}
n_j=\frac{1+\sigma_j^z}{2},
\qquad
b_j^\dagger=\sigma_j^+,
\qquad
b_j=\sigma_j^-,
\end{equation}
one obtains
\begin{equation}
H_{\rm eff}
=
2J\sum_{\langle i,j\rangle}\mathbf S_i\cdot\mathbf S_j
-\frac{J}{2}N_b,
\label{eq:Heisenberg}
\end{equation}
where $\mathbf S_j=\boldsymbol{\sigma}_j/2$ and $N_b$ is the number of nearest-neighbor bonds. The Bethe-ansatz ground-state energy of the spin-$1/2$ antiferromagnetic Heisenberg chain then gives~\cite{bethe1931theorie,lieb1961two,karbach1998introduction}
\begin{equation}
E_{1D}^{p} = -2J L \ln2 = -\frac{4\zeta^2\Gamma^2}{U}L \ln2
\end{equation}
for periodic boundary conditions and 
\begin{equation}
E_{1D}^{o} = E_{1D}^{p} + \frac{\zeta^2 \Gamma^2}{U}
\label{eq:1DH}
\end{equation}
for open boundary conditions. In Figure~\ref{fig:app:reg2}, we compare this approximation with the DMRG result for an open chain in the relevant regime. There is clear agreement that improves with both increasing lattice size and increasing $U$. As shown in panel (c), for $U \geq 4\Gamma$, the energies agree well up to $\zeta \approx 0.3$.

Using QMC results for the two-dimensional Heisenberg model~\cite{sandvik1997finite} we can generalize this result to two-dimensional clusters with periodic,
\begin{equation}
E_{2D}^{p}\simeq
-4.6777\frac{\zeta^2\Gamma^2}{U}L
\end{equation}
and open boundary conditions,
\begin{equation}
E_{2D}^{o}\simeq
E_{2D}^{p} + \frac{\zeta^2 \Gamma^2}{U}(L_x+L_y).
\end{equation}
This case is discussed in the main text, see Fig.~\ref{fig:L2D}.

\section{Neural quantum states benchmarks}
\label{app:NQS}
We consider a $3 \times 3$ cluster as a controlled benchmark to compare different NQS architectures against ED results. 
Despite its small size, this system exhibits several distinct ground-state phases, providing a nontrivial testbed for assessing both the expressivity and trainability of variational ans\"atze.

Benchmarks across different parameter points, marked by crosses in Fig.~\ref{fig:1DUzetat0}(a3), including a broad set of NQS architectures are summarized in Table~\ref{tab01:benchmarks}. To quantify the accuracy, we define the relative energy error as
\begin{equation}
   \mathrm{Err} = \min_{i} \left|
   \frac{E_{20}^{(i)} - E_{\mathrm{ex}}}{E_{\mathrm{ex}}}
   \right|,
   \label{eq:Err}
\end{equation}
where $E_{20}^{(i)}$ denotes the average variational energy over the final 20 iterations of the $i$-th training run. The averaging window is taken after the variational energy has reached a plateau and serves to reduce residual Monte Carlo and optimization fluctuations. We verified that using comparable windows of 10 or 40 iterations does not qualitatively change the reported errors or the conclusions. 

The minimum is taken over six independent runs initialized with different random parameters, and $E_{\mathrm{ex}}$ is the exact ground-state energy. The use of several independent initializations reduces the sensitivity of the comparison to an unfavorable individual training run. The resulting quantity $\mathrm{Err}$ should therefore be interpreted as the best relative energy accuracy obtained among these runs under the specified optimization protocol. It does not represent the ultimate accuracy achievable by a given architecture after architecture-specific hyperparameter optimization, nor is it intended as a measure of run-to-run stability.

There are some clear trends. In the trivial singlet phase ($A_{3\times3}$), all considered ans\"atze accurately reproduce the ground state, reflecting the product-state nature of the wave function.  In contrast, for the triplet phase ($B_{3\times3}$), purely correlational architectures such as Jastrow or RBM fail to converge to the correct energy.  For the triplet point $B_{3\times3}$, the Slater determinant reduces the relative error to $7\times10^{-3}$, compared to approximately $10^{-1}$ for the pure Jastrow and RBM states. Adding correlations to the determinant further reduces the error, to the $10^{-3}$ level for Jastrow--Slater and to approximately $2\times10^{-5}$ for neural BF. 

Furthermore, determinant-based ans\"atze capture the ground state reliably in weakly interacting regimes. 
In particular, for small interaction strength ($U = 0.1\Gamma$ in the $\tilde{C}_{3\times3}$ column), the pure Slater ansatz already yields an accurate description, indicating that both amplitude and sign structure are well approximated at this level. 
In contrast, purely correlational architectures such as Jastrow or small RBM fail even in this regime indicating that learning the required fermionic sign structure implicitly is difficult under the present optimization protocol.

As the interaction strength increases within the same phase (e.g.\ $U = 2\Gamma$ in the $C_{3\times3}$ column), the limitations of the pure Slater ansatz become apparent, as it lacks the flexibility to capture enhanced correlations. 
At the same time, purely correlational ans\"atze remain insufficient, while Jastrow--Slater and neural BF constructions provide significantly improved results. 
This highlights the necessity of combining an accurate antisymmetric structure with a flexible treatment of correlations.

A second observation concerns the strongly interacting regime ($U = 4\Gamma$, $D_{3\times3}$ column), where all considered architectures exhibit increased optimization difficulty. 
In this case, the performance of the neural BF ansatz can be substantially improved by modifying the training protocol. 
Specifically, in the $D^{\mathrm{TL}}_{3\times3}$ column we employ a simple transfer-learning strategy, in which training is initialized at $U = 4\Gamma$, followed by an adiabatic sweep to larger interaction strengths ($U \sim 7$--$8\Gamma$) and a subsequent return to the target value. 
This procedure reduces the error by approximately an order of magnitude, indicating that the BF ansatz is expressive enough to represent the ground state, and that optimization is an important limitation here.

For the other architectures considered, similar transfer-learning strategies did not lead to comparable improvements, as also reflected in Table~\ref{tab01:benchmarks}. 
While for specific parameter points Jastrow--Slater or RBM-based ans\"atze may achieve competitive or even better accuracy, the neural BF ansatz consistently provides the most consistently accurate performance.
For these reasons, the majority of NQS results presented in the main text are obtained using the neural BF ansatz.

To ensure a fair comparison, we employed an (almost) identical training protocol in all cases.

Each optimization begins with 400 iterations using the Adam optimizer,\\   (\texttt{opt = optax.chain(
        optax.clip\_by\_global\_norm(0.3),\\
        optax.adam(
            learning\_rate=lr\_schedule,
            b1=0.8,
            b2=0.99,
            eps=1e-6
            )))}\\
with an exponentially decaying learning rate, starting from \texttt{lr0} (typically \texttt{lr0 = 0.05}) and decreasing to \texttt{lr1} (typically \texttt{lr1 = 0.004}).

This initial phase is followed by 600 iterations of stochastic gradient descent (SGD) with a fixed learning rate \texttt{lr2 = 0.002}. Finally, we perform an additional 50 iterations using SGD combined with stochastic reconfiguration (SR), employing a reduced learning rate \texttt{lr2 = 0.0001} and a diagonal shift $diag\_shift = 10^{-3}$.

To maintain consistency with simulations of larger systems, sampling is performed using the \texttt{MetropolisFermionHop(hi, graph=graph, d\_max=1)} update rule throughout. Note that only learning rates \texttt{lr0}, \texttt{lr1} and the initial distribution of architecture parameters have been changed for some architectures for stability reasons.
\addtolength{\tabcolsep}{-2pt} 
\begin{table}[th]
\centering
\begin{tabular}{ll|cccccc}
\toprule
\multicolumn{6}{c}{Relative error: $L=3\times 3$, $\varepsilon=0$, $t=0$} \\
\midrule
architecture & params 
& $A_{3\times 3}$ 
& $B_{3 \times 3}$ 
& $C_{3 \times 3}$ 
& $\tilde{C}_{3 \times 3}$ 
& $D_{3 \times 3}$ 
& $D^{\mathrm{TL}}_{3 \times 3}$ \\
\midrule
Jastrow               & 153 & $1\!\times\!10^{-3}$ & $1\!\times\!10^{-1}$ & $4\!\times\!10^{-1}$ & $2\!\times\!10^{-1}$ & $4\!\times\!10^{-1}$ & $4\!\times\!10^{-1}$\\
RBM ($\alpha = 1$)    & 360 & $2\!\times\!10^{-5}$ & $1\!\times\!10^{-1}$ & $2\!\times\!10^{-1}$ & $2\!\times\!10^{-1}$ & $4\!\times\!10^{-1}$ & $4\!\times\!10^{-1}$\\
RBM ($\alpha = 4$)    & 1386 & $3\!\times\!10^{-4}$ & $1\!\times\!10^{-1}$ & $4\!\times\!10^{-1}$ & $9\!\times\!10^{-2}$ & $8\!\times\!10^{-1}$ & $7\!\times\!10^{-1}$\\
Slater   & 162 & $2\!\times\!10^{-5}$ & $7\!\times\!10^{-3}$ & $1\!\times\!10^{-1}$ & $5\!\times\!10^{-7}$ & $4\!\times\!10^{-1}$ & $4\!\times\!10^{-1}$\\
Jastrow-Slater ($h = L$)   & 333 & $2\!\times\!10^{-5}$ & $1\!\times\!10^{-3}$ & $1\!\times\!10^{-2}$ & $7\!\times\!10^{-8}$ & $8\!\times\!10^{-2}$ & $6\!\times\!10^{-2}$\\
Jastrow-Slater ($h = 2L$)   & 504 & $3\!\times\!10^{-5}$ & $7\!\times\!10^{-4}$ & $9\!\times\!10^{-3}$ & $1\!\times\!10^{-7}$ & $3\!\times\!10^{-2}$ & $2\!\times\!10^{-3}$\\
RBM-Slater ($\alpha = 1$)   & 522 & $3\!\times\!10^{-7}$ & $4\!\times\!10^{-4}$ & $8\!\times\!10^{-3}$ & $3\!\times\!10^{-7}$ & $3\!\times\!10^{-2}$ & $2\!\times\!10^{-2}$\\
RBM-Slater ($\alpha = 4$)   & 868 & $5\!\times\!10^{-6}$ & $4\!\times\!10^{-4}$ & $8\!\times\!10^{-3}$ & $4\!\times\!10^{-7}$ & $6\!\times\!10^{-2}$ & $2\!\times\!10^{-2}$\\
RBM-Slater ($\alpha = 10$)   & 3582 & $6\!\times\!10^{-6}$ & $3\!\times\!10^{-4}$ & $7\!\times\!10^{-3}$ & $3\!\times\!10^{-6}$ & $5\!\times\!10^{-2}$ & $1\!\times\!10^{-2}$\\
Multi-Slater ($ns=4$)   & 648 & $7\!\times\!10^{-4}$ & $7\!\times\!10^{-4}$ & $9\!\times\!10^{-2}$ &  $8\!\times\!10^{-5}$ & $3\!\times\!10^{-1}$ & $2\!\times\!10^{-1}$\\
Multi-Slater ($ns=8$)   & 1296& $5\!\times\!10^{-4}$ & $4\!\times\!10^{-4}$ & $8\!\times\!10^{-2}$ & $8\!\times\!10^{-5}$ & $3\!\times\!10^{-1}$ & $2\!\times\!10^{-1}$\\
Neural-Backflow ($h=L$)   & 1953 & $1\!\times\!10^{-5}$ & $2\!\times\!10^{-5}$ & $1\!\times\!10^{-3}$ & $2\!\times\!10^{-7}$ & $5\!\times\!10^{-2}$ & $5\!\times\!10^{-3}$\\
Neural-Backflow ($h=2L$)   & 3582 & $3\!\times\!10^{-5}$ & $2\!\times\!10^{-5}$ & $1\!\times\!10^{-3}$ & $9\!\times\!10^{-8}$ & $4\!\times\!10^{-2}$ & $4\!\times\!10^{-3}$\\
DMRG-MPS   &  & $0$ & $0$ & $0$ & $0$ & $0$ & $ $\\
\bottomrule
\end{tabular}
\caption{Comparison of the accuracy (lower is better) of NQS variational results on lattices $L=3\times 3$. The listed values of relative error follow Eq.~\eqref{eq:Err} (note that the constant shift $-UL/2$ is avoided to not underestimate the relative error at large $U$). The results are at $\varepsilon=0$, $t=0$ and $A_{3\times 3}:$ $U=0.5\Gamma$, $\zeta=0.1$, $\tilde{N}_f=9$;
$B_{3\times 3}:$ $U=0.9\Gamma$, $\zeta=0.5$, $\tilde{N}_f=7$; $C_{3\times 3}:$ $U=2\Gamma$, $\zeta=0.5$, $\tilde{N}_f=9$; $\tilde{C}_{3\times 3}:$ $U=0.1\Gamma$, $\zeta=0.85$, $\tilde{N}_f=9$ (the same ground state phase as $C_{3\times 3}$); $D_{3\times 3}$ and $D^\mathrm{TL}_{3\times 3}:$ $U=4\Gamma$, $\zeta=0.5$, $\tilde{N}_f=8$ where $D^\mathrm{TL}_{3\times 3}$ shows the transfer learning result.   Note that the number of parameters for each architecture is stated for the $N_f=L$ case. Away from half filling the number might change.}
\label{tab01:benchmarks}
\end{table}

The parameter set $A_{3\times 3}$: $U=0.5\Gamma$, $\zeta=0.1$, $\tilde{N}_f=9$ lies within the trivial BCS singlet phase, see Eq.~\eqref{eq:BCS}. All considered NQS architectures successfully reproduce this state. The remaining small variations in the relative error originate from stochastic Monte Carlo sampling and differences in the optimization dynamics; they do not reflect limitations in the expressive power of the respective architectures.  
\end{appendix}
\clearpage


\begin{thebibliography}{10}
\providecommand{\url}[1]{\texttt{#1}}
\providecommand{\urlprefix}{URL }
\expandafter\ifx\csname urlstyle\endcsname\relax
  \providecommand{\doi}[1]{doi:\discretionary{}{}{}#1}\else
  \providecommand{\doi}{doi:\discretionary{}{}{}\begingroup
  \urlstyle{rm}\Url}\fi
\providecommand{\eprint}[2][]{\url{#2}}

\bibitem{Wernsdorfer-2010}
S.~De~Franceschi, L.~Kouwenhoven, C.~Sch{\"o}nenberger and W.~Wernsdorfer,
\newblock \emph{Hybrid superconductor-quantum dot devices},
\newblock Nat. Nanotechnol. \textbf{5}, 703 (2010),
\newblock \doi{10.1038/nnano.2010.173}.

\bibitem{Heinrich-2018}
B.~W. Heinrich, J.~I. Pascual and K.~J. Franke,
\newblock \emph{Single magnetic adsorbates on s-wave superconductors},
\newblock Prog. Surf. Sci. \textbf{93}, 1 (2018),
\newblock \doi{10.1016/j.progsurf.2018.01.001}.

\bibitem{Balatsky2006impurity}
A.~V. Balatsky, I.~Vekhter and J.-X. Zhu,
\newblock \emph{Impurity-induced states in conventional and unconventional
  superconductors},
\newblock Rev. Mod. Phys. \textbf{78}, 373 (2006),
\newblock \doi{10.1103/RevModPhys.78.373}.

\bibitem{Krantz-2019-qubits}
P.~Krantz, M.~Kjaergaard, F.~Yan, T.~P. Orlando, S.~Gustavsson and W.~D.
  Oliver,
\newblock \emph{A quantum engineer's guide to superconducting qubits},
\newblock Appl. Phys. Rev. \textbf{6}, 021318 (2019),
\newblock \doi{10.1063/1.5089550}.

\bibitem{Natarajan-2012-detectors}
C.~M. Natarajan, M.~G. Tanner and R.~H. Hadfield,
\newblock \emph{Superconducting nanowire single-photon detectors: physics and
  applications},
\newblock Supercond. Sci. Technol. \textbf{25}, 063001 (2012),
\newblock \doi{10.1088/0953-2048/25/6/063001}.

\bibitem{Persky-2022-squid}
E.~Persky, I.~Sochnikov and B.~Kalisky,
\newblock \emph{Studying quantum materials with scanning {SQUID} microscopy},
\newblock Annu. Rev. Condens. Matter Phys. \textbf{13}, 385 (2022),
\newblock \doi{10.1146/annurev-conmatphys-031620-104226}.

\bibitem{Li2025individual}
C.~Li, V.~Pokorn\'y, M.~\v{Z}onda, J.-C. Liu, P.~Zhou, O.~Chahib, T.~Glatzel,
  R.~H{\"a}ner, S.~Decurtins, S.-X. Liu, R.~Pawlak and E.~Meyer,
\newblock \emph{Individual assembly of radical molecules on superconductors:
  Demonstrating quantum spin behavior and bistable charge rearrangement},
\newblock ACS Nano \textbf{19}, 3403 (2025),
\newblock \doi{10.1021/acsnano.4c12387}.

\bibitem{Li2025negative}
C.~Li, V.~Pokorn{\'y}, P.~Hapala, M.~{\v{Z}}onda, T.~Glatzel, P.~Zhou,
  S.~Decurtins, S.-X. Liu, F.~Song, E.~Meyer and R.~Pawlak,
\newblock \emph{Non-equilibrium correlated electron dynamics in triangular
  molecular assemblies},
\newblock Nat. Commun. \textbf{17}(1), 8406 (2026),
\newblock \doi{10.1038/s41467-026-75051-3}.

\bibitem{dvir2023realization}
T.~Dvir, G.~Wang, N.~van Loo, C.-X. Liu, G.~P. Mazur, A.~Bordin, S.~L.
  Ten~Haaf, J.-Y. Wang, D.~van Driel, F.~Zatelli \emph{et~al.},
\newblock \emph{Realization of a minimal kitaev chain in coupled quantum dots},
\newblock Nature \textbf{614}(7948), 445 (2023).

\bibitem{bordin2025enhanced}
A.~Bordin, C.-X. Liu, T.~Dvir, F.~Zatelli, S.~L. Ten~Haaf, D.~Van~Driel,
  G.~Wang, N.~Van~Loo, Y.~Zhang, J.~C. Wolff \emph{et~al.},
\newblock \emph{Enhanced majorana stability in a three-site kitaev chain},
\newblock Nature Nanotechnology \textbf{20}(6), 726 (2025).

\bibitem{SeoaneSouto-2024}
R.~Seoane~Souto and R.~Aguado,
\newblock \emph{Subgap States in Semiconductor-Superconductor Devices for
  Quantum Technologies: {Andreev} Qubits and Minimal {Majorana} Chains}, pp.
  133--223,
\newblock {Springer} {Nature} {Switzerland}, Cham,
\newblock ISBN 978-3-031-55657-9,
\newblock \doi{10.1007/978-3-031-55657-9_3} (2024).

\bibitem{Martin-Rodero-2012}
A.~Martín-Rodero and A.~Levy~Yeyati,
\newblock \emph{The {Andreev} states of a superconducting quantum dot: mean
  field versus exact numerical results},
\newblock J. Phys.: Condens. Matter \textbf{24}, 385303 (2012),
\newblock \doi{10.1088/0953-8984/24/38/385303}.

\bibitem{Zonda2015perturbation}
M.~\v{Z}onda, V.~Pokorn\'y, V.~Jani\v{s} and T.~Novotn\'y,
\newblock \emph{Perturbation theory of a superconducting $0-\pi$ impurity
  quantum phase transition},
\newblock Sci. Rep. \textbf{5}, 8821 (2015),
\newblock \doi{10.1038/srep08821}.

\bibitem{Pillet2013tunneling}
J.-D. Pillet, P.~Joyez, R.~\v{Z}itko and M.~F. Goffman,
\newblock \emph{Tunneling spectroscopy of a single quantum dot coupled to a
  superconductor: From {Kondo} ridge to {Andreev} bound states},
\newblock Phys. Rev. B \textbf{88}, 045101 (2013),
\newblock \doi{10.1103/PhysRevB.88.045101}.

\bibitem{Karrasch-2008-fRG}
C.~Karrasch, A.~Oguri and V.~Meden,
\newblock \emph{Josephson current through a single {Anderson} impurity coupled
  to {BCS} leads},
\newblock Phys. Rev. B \textbf{77}, 024517 (2008),
\newblock \doi{10.1103/PhysRevB.77.024517}.

\bibitem{Luitz2010weak}
D.~J. Luitz and F.~F. Assaad,
\newblock \emph{Weak-coupling continuous-time quantum {Monte Carlo} study of
  the single impurity and periodic {Anderson} models with $s$-wave
  superconducting baths},
\newblock Phys. Rev. B \textbf{81}, 024509 (2010),
\newblock \doi{10.1103/PhysRevB.81.024509}.

\bibitem{Luitz-2012}
D.~J. Luitz, F.~F. Assaad, T.~Novotn{\'y}, C.~Karrasch and V.~Meden,
\newblock \emph{Understanding the {Josephson} current through a
  {Kondo}-correlated quantum dot},
\newblock Phys. Rev. Lett. \textbf{108}, 227001 (2012),
\newblock \doi{10.1103/PhysRevLett.108.227001}.

\bibitem{Pokorny-2018}
V.~Pokorn\'y and M.~\v{Z}onda,
\newblock \emph{Correlation effects in superconducting quantum dot systems},
\newblock Physica B: Condens. Matter \textbf{536}, 488 (2018),
\newblock \doi{10.1016/j.physb.2017.08.059}.

\bibitem{Vecino-2003-ZBW}
E.~Vecino, A.~Mart\'{\i}n-Rodero and A.~L. Yeyati,
\newblock \emph{Josephson current through a correlated quantum level: {Andreev}
  states and $\ensuremath{\pi}$ junction behavior},
\newblock Phys. Rev. B \textbf{68}, 035105 (2003),
\newblock \doi{10.1103/PhysRevB.68.035105}.

\bibitem{Meng-2009self}
T.~Meng, S.~Florens and P.~Simon,
\newblock \emph{Self-consistent description of {Andreev} bound states in
  {Josephson} quantum dot devices},
\newblock Phys. Rev. B \textbf{79}, 224521 (2009),
\newblock \doi{10.1103/PhysRevB.79.224521}.

\bibitem{Pokorny-2023}
V.~Pokorn\'y and M.~\v{Z}onda,
\newblock \emph{Effective low-energy models for superconducting impurity
  systems},
\newblock Phys. Rev. B \textbf{107}, 155111 (2023),
\newblock \doi{10.1103/PhysRevB.107.155111}.

\bibitem{Bobok2025che}
D.~Bobok, L.~Frk, V.~Pokorn\'y and M.~\v{Z}onda,
\newblock \emph{Scalable effective models for superconducting nanostructures:
  Applications to double, triple, and quadruple quantum dots},
\newblock Phys. Rev. B \textbf{112}, 205418 (2025),
\newblock \doi{10.1103/mxsl-fc96}.

\bibitem{Meden2019the}
V.~Meden,
\newblock \emph{The {Anderson}{\textendash}{Josephson} quantum
  dot{\textemdash}a theory perspective},
\newblock J. Phys. Condens. Matter \textbf{31}, 163001 (2019),
\newblock \doi{10.1088/1361-648x/aafd6a}.

\bibitem{White1992DMRG}
S.~R. White,
\newblock \emph{Density matrix formulation for quantum renormalization groups},
\newblock Phys. Rev. Lett. \textbf{69}, 2863 (1992),
\newblock \doi{10.1103/PhysRevLett.69.2863}.

\bibitem{White1993DMRG}
S.~R. White,
\newblock \emph{Density-matrix algorithms for quantum renormalization groups},
\newblock Phys. Rev. B \textbf{48}, 10345 (1993),
\newblock \doi{10.1103/PhysRevB.48.10345}.

\bibitem{Catarina2023DMRG}
G.~Catarina and B.~Murta,
\newblock \emph{Density-matrix renormalization group: a pedagogical
  introduction},
\newblock Eur. Phys. J. B \textbf{96}, 111 (2023),
\newblock \doi{10.1140/epjb/s10051-023-00575-2}.

\bibitem{Gubernatis2016QMC}
J.~Gubernatis, N.~Kawashima and P.~Werner,
\newblock \emph{Quantum {Monte} {Carlo} methods},
\newblock Cambridge University Press, Cambridge, England,
\newblock \doi{10.1017/CBO9780511902581} (2016).

\bibitem{Bacsi2023Exchange}
{\'A}.~B{\'a}csi, L.~Pave{\v{s}}i{\'c} and R.~{\v{Z}}itko,
\newblock \emph{Exchange interaction between two quantum dots coupled through a
  superconducting island},
\newblock Phys. Rev. B \textbf{108}, 115160 (2023),
\newblock \doi{10.1103/PhysRevB.108.115160}.

\bibitem{Cirac2021MPS}
J.~I. Cirac, D.~P\'erez-Garc\'{\i}a, N.~Schuch and F.~Verstraete,
\newblock \emph{Matrix product states and projected entangled pair states:
  Concepts, symmetries, theorems},
\newblock Rev. Mod. Phys. \textbf{93}, 045003 (2021),
\newblock \doi{10.1103/RevModPhys.93.045003}.

\bibitem{Banuls2023TN}
M.~C. Banuls,
\newblock \emph{Tensor network algorithms: A route map},
\newblock Annu. Rev. Condens. Matter Phys. \textbf{14}, 173 (2023),
\newblock \doi{10.1146/annurev-conmatphys-040721-022705}.

\bibitem{verstraete2004PEPS}
F.~Verstraete and J.~I. Cirac,
\newblock \emph{Renormalization algorithms for quantum-many body systems in two
  and higher dimensions},
\newblock \eprint{https://arxiv.org/abs/cond-mat/0407066}.

\bibitem{Orus2014TN}
R.~Or\'us,
\newblock \emph{A practical introduction to tensor networks: Matrix product
  states and projected entangled pair states},
\newblock Ann. Phys. \textbf{349}, 117 (2014),
\newblock \doi{10.1016/j.aop.2014.06.013}.

\bibitem{carleo2017solving}
G.~Carleo and M.~Troyer,
\newblock \emph{Solving the quantum many-body problem with artificial neural
  networks},
\newblock Science \textbf{355}, 602 (2017),
\newblock \doi{10.1126/science.aag2302}.

\bibitem{carleo2018_constructing}
G.~Carleo, Y.~Nomura and M.~Imada,
\newblock \emph{Constructing exact representations of quantum many-body systems
  with deep neural networks},
\newblock Nat. Commun. \textbf{9}, 5322 (2018),
\newblock \doi{10.1038/s41467-018-07520-3}.

\bibitem{Medvidovic2024NQS}
M.~Medvidovi{\'{c}} and J.~R. Moreno,
\newblock \emph{Neural-network quantum states for many-body physics},
\newblock Eur. Phys. J. Plus \textbf{139}, 631 (2024),
\newblock \doi{10.1140/epjp/s13360-024-05311-y}.

\bibitem{chooSymmetriesManyBodyExcitations2018}
K.~Choo, G.~Carleo, N.~Regnault and T.~Neupert,
\newblock \emph{Symmetries and {{Many-Body Excitations}} with {{Neural-Network
  Quantum States}}},
\newblock Phys. Rev. Lett. \textbf{121}, 167204 (2018),
\newblock \doi{10.1103/PhysRevLett.121.167204}.

\bibitem{dawidModernApplicationsMachine}
A.~Dawid, J.~Arnold, B.~Requena, A.~Gresch, M.~P{\l}odzien, K.~Donatella,
  K.~Nicoli, P.~Stornati, R.~Koch, M.~B{\"u}ttner, R.~Oku{\l}a, G.~Mu{\~n}oz
  \emph{et~al.},
\newblock \emph{Modern applications of machine learning in quantum sciences} p.
  268,
\newblock \doi{10.48550/arXiv.2204.04198}.

\bibitem{Padila2025autoregressive}
E.~Ibarra-Garc\'{\i}a-Padilla, H.~Lange, R.~G. Melko, R.~T. Scalettar,
  J.~Carrasquilla, A.~Bohrdt and E.~Khatami,
\newblock \emph{Autoregressive neural quantum states of {Fermi} {Hubbard}
  models},
\newblock Phys. Rev. Res. \textbf{7}, 013122 (2025),
\newblock \doi{10.1103/PhysRevResearch.7.013122}.

\bibitem{mezera2023NNQS}
M.~Mezera, J.~Men\v{s}\'ikov\'a, P.~Bal\'a\v{z} and M.~\v{Z}onda,
\newblock \emph{{Neural network quantum states analysis of the
  Shastry-Sutherland model}},
\newblock SciPost Phys. Core \textbf{6}, 088 (2023),
\newblock \doi{10.21468/SciPostPhysCore.6.4.088}.

\bibitem{gu2026solving}
Y.~Gu, W.~Li, H.~Lin, B.~Zhan, R.~Li, Y.~Huang, D.~He, Y.~Wu, T.~Xiang, M.~Qin,
  L.~Wang and D.~Lv,
\newblock \emph{Solving the hubbard model with neural quantum states},
\newblock Nature Communications \textbf{17}(1), 7838 (2026),
\newblock \doi{10.1038/s41467-026-74028-6}.

\bibitem{carrasquillaNeuralNetworksQuantum2021}
J.~Carrasquilla and G.~Torlai,
\newblock \emph{Neural networks in quantum many-body physics: A hands-on
  tutorial},
\newblock \eprint{http://arxiv.org/abs/2101.11099}.

\bibitem{kim2024ultra}
J.~Kim, G.~Pescia, B.~Fore, J.~Nys, G.~Carleo, S.~Gandolfi, M.~Hjorth-Jensen
  and A.~Lovato,
\newblock \emph{Neural-network quantum states for ultra-cold {Fermi} gases},
\newblock Commun. Phys. \textbf{7}, 148 (2024),
\newblock \doi{10.1038/s42005-024-01613-w}.

\bibitem{jordan1928paulische}
P.~Jordan and E.~Wigner,
\newblock \emph{{\"U}ber das {Paulische} {\"a}quivalenzverbot},
\newblock Z. Physik \textbf{47}, 631 (1928),
\newblock \doi{10.1007/BF01331938}.

\bibitem{Barrett2022autoregressive}
T.~D. Barrett, A.~Malyshev and A.~I. Lvovsky,
\newblock \emph{Autoregressive neural-network wavefunctions for ab initio
  quantum chemistry},
\newblock Nat. Mach. Intell. \textbf{4}, 351 (2022),
\newblock \doi{10.1038/s42256-022-00461-z}.

\bibitem{Bajdich2006pfaffian}
M.~Bajdich, L.~Mitas, G.~Drobn\'y, L.~K. Wagner and K.~E. Schmidt,
\newblock \emph{Pfaffian pairing wave functions in electronic-structure quantum
  monte carlo simulations},
\newblock Phys. Rev. Lett. \textbf{96}, 130201 (2006),
\newblock \doi{10.1103/PhysRevLett.96.130201}.

\bibitem{gao2023generalizing}
N.~Gao and S.~Günnemann,
\newblock \emph{Generalizing neural wave functions},
\newblock \eprint{http://arxiv.org/abs/2302.04168}.

\bibitem{gao2024neural}
N.~Gao and S.~Günnemann,
\newblock \emph{Neural {Pfaffians}: Solving many many-electron {Schr\"odinger}
  equations},
\newblock \eprint{http://arxiv.org/abs/2405.14762}.

\bibitem{Chen2024empowering}
A.~Chen and M.~Heyl,
\newblock \emph{Empowering deep neural quantum states through efficient
  optimization},
\newblock Nat. Phys. \textbf{20}, 1476 (2024),
\newblock \doi{10.1038/s41567-024-02566-1}.

\bibitem{chen2025neural}
A.~Chen, Z.-Q. Wan, A.~Sengupta, A.~Georges and C.~Roth,
\newblock \emph{Neural network-augmented {Pfaffian} wave-functions for scalable
  simulations of interacting fermions},
\newblock \eprint{http://arxiv.org/abs/2507.10705}.

\bibitem{Xu2022optimized}
R.~G. Xu, T.~Okubo, S.~Todo and M.~Imada,
\newblock \emph{Optimized implementation for calculation and fast-update of
  {Pfaffians} installed to the open-source fermionic variational solver
  {mVMC}},
\newblock Comput. Phys. Commun. \textbf{277}, 108375 (2022),
\newblock \doi{10.1016/j.cpc.2022.108375}.

\bibitem{netket2:2019}
G.~Carleo, K.~Choo, D.~Hofmann, J.~E. Smith, T.~Westerhout, F.~Alet, E.~J.
  Davis, S.~Efthymiou, I.~Glasser, S.-H. Lin, M.~Mauri, G.~Mazzola
  \emph{et~al.},
\newblock \emph{{NetKet}: A machine learning toolkit for many-body quantum
  systems},
\newblock SoftwareX p. 100311 (2019),
\newblock \doi{10.1016/j.softx.2019.100311}.

\bibitem{netket3:2022}
F.~Vicentini, D.~Hofmann, A.~Szab{\'o}, D.~Wu, C.~Roth, C.~Giuliani, G.~Pescia,
  J.~Nys, V.~Vargas-Calder\'{o}n, N.~Astrakhantsev and G.~Carleo,
\newblock \emph{{NetKet 3}: Machine learning toolbox for many-body quantum
  systems},
\newblock SciPost Phys. Codebases p.~7 (2022),
\newblock \doi{10.21468/SciPostPhysCodeb.7}.

\bibitem{Nomura2017restricted}
Y.~Nomura, A.~S. Darmawan, Y.~Yamaji and M.~Imada,
\newblock \emph{Restricted {Boltzmann} machine learning for solving strongly
  correlated quantum systems},
\newblock Phys. Rev. B \textbf{96}, 205152 (2017),
\newblock \doi{10.1103/PhysRevB.96.205152}.

\bibitem{stokes2020phases}
J.~Stokes, J.~R. Moreno, E.~A. Pnevmatikakis and G.~Carleo,
\newblock \emph{Phases of two-dimensional spinless lattice fermions with
  first-quantized deep neural-network quantum states},
\newblock Phys. Rev. B \textbf{102}, 205122 (2020),
\newblock \doi{10.1103/PhysRevB.102.205122}.

\bibitem{luo2019backflow}
D.~Luo and B.~K. Clark,
\newblock \emph{Backflow transformations via neural networks for quantum
  many-body wave functions},
\newblock Phys. Rev. Lett. \textbf{122}, 226401 (2019),
\newblock \doi{10.1103/PhysRevLett.122.226401}.

\bibitem{Liu2024Unifying}
Z.~Liu and B.~K. Clark,
\newblock \emph{Unifying view of fermionic neural network quantum states: From
  neural network backflow to hidden fermion determinant states},
\newblock Phys. Rev. B \textbf{110}, 115124 (2024),
\newblock \doi{10.1103/PhysRevB.110.115124}.

\bibitem{Antonelli2025exploring}
T.~Antonelli, M.~Coraiola, D.~C. Ohnmacht, A.~E. Svetogorov, D.~Sabonis, S.~C.
  ten Kate, E.~Cheah, F.~Krizek, R.~Schott, J.~C. Cuevas, W.~Belzig,
  W.~Wegscheider \emph{et~al.},
\newblock \emph{Exploring the energy spectrum of a four-terminal {Josephson}
  junction: Toward topological {Andreev} band structures},
\newblock Phys. Rev. X \textbf{15}, 031066 (2025),
\newblock \doi{10.1103/qd3y-f912}.

\bibitem{Ebert2025sextets}
M.~R. Ebert, D.~C. Ohnmacht, W.~Belzig and J.~C. Cuevas,
\newblock \emph{Sextets in four-terminal {Josephson} junctions},
\newblock Phys. Rev. B \textbf{112}, 195430 (2025),
\newblock \doi{10.1103/phvj-vfnx}.

\bibitem{tenpy2024}
J.~Hauschild, J.~Unfried, S.~Anand, B.~Andrews, M.~Bintz, U.~Borla, S.~Divic,
  M.~Drescher, J.~Geiger, M.~Hefel, K.~H\'emery, W.~Kadow \emph{et~al.},
\newblock \emph{{Tensor network Python {(TeNPy)} version 1}},
\newblock SciPost Phys. Codebases p.~41 (2024),
\newblock \doi{10.21468/SciPostPhysCodeb.41}.

\bibitem{Shiba-1972}
H.~Shiba,
\newblock \emph{Thermodynamic properties of the one-dimensional
  half-filled-band {Hubbard} model. {II}: Application of the grand canonical
  method},
\newblock Prog. Theor. Phys. \textbf{48}(6), 2171 (1972),
\newblock \doi{10.1143/PTP.48.2171}.

\bibitem{lieb-1989}
E.~H. Lieb,
\newblock \emph{Two theorems on the {Hubbard} model},
\newblock Phys. Rev. Lett. \textbf{62}, 1201 (1989),
\newblock \doi{10.1103/PhysRevLett.62.1201}.

\bibitem{Scalettar-1989}
R.~T. Scalettar, E.~Y. Loh, J.~E. Gubernatis, A.~Moreo, S.~R. White, D.~J.
  Scalapino, R.~L. Sugar and E.~Dagotto,
\newblock \emph{Phase diagram of the two-dimensional negative-{U} {Hubbard}
  model},
\newblock Phys. Rev. Lett. \textbf{62}, 1407 (1989),
\newblock \doi{10.1103/PhysRevLett.62.1407}.

\bibitem{yang-1990}
C.~N. YANG and S.~ZHANG,
\newblock \emph{So$_4$ symmetry in a {Hubbard} model},
\newblock Mod. Phys. Lett. B \textbf{04}(11), 759 (1990),
\newblock \doi{10.1142/S0217984990000933},
\newblock \eprint{https://doi.org/10.1142/S0217984990000933}.

\bibitem{essl-2024}
H.~E\ss{}l, M.~Reitner, G.~Sangiovanni and A.~Toschi,
\newblock \emph{General {Shiba} mapping for on-site four-point correlation
  functions},
\newblock Phys. Rev. Res. \textbf{6}, 033061 (2024),
\newblock \doi{10.1103/PhysRevResearch.6.033061}.

\bibitem{Pokorny2021footprints}
V.~Pokorn\'y and T.~Novotn\'y,
\newblock \emph{Footprints of impurity quantum phase transitions in quantum
  {Monte} {Carlo} statistics},
\newblock Phys. Rev. Research \textbf{3}, 023013 (2021),
\newblock \doi{10.1103/PhysRevResearch.3.023013}.

\bibitem{hornik1989multilayer}
K.~Hornik, M.~Stinchcombe and H.~White,
\newblock \emph{Multilayer feedforward networks are universal approximators},
\newblock Neural Netw. \textbf{2}, 359 (1989),
\newblock \doi{10.1016/0893-6080(89)90020-8}.

\bibitem{devore2021neural}
R.~DeVore, B.~Hanin and G.~Petrova,
\newblock \emph{Neural network approximation},
\newblock Acta Numer. \textbf{30}, 327 (2021),
\newblock \doi{10.1017/S0962492921000052}.

\bibitem{Carleo2017_notes}
G.~Carleo,
\newblock \emph{{Neural-Network Quantum States}. {A} lecture for the machine
  learning and many-body physics workshop},
\newblock
  \urlprefix\url{https://www.csrc.ac.cn/upload/file/20170703/1499072201537152.pdf}
  (2017).

\bibitem{lange2024architectures}
H.~Lange, A.~Van~de Walle, A.~Abedinnia and A.~Bohrdt,
\newblock \emph{From architectures to applications: a review of neural quantum
  states},
\newblock Quantum Sci. Technol. \textbf{9}, 040501 (2024),
\newblock \doi{10.1088/2058-9565/ad7168}.

\bibitem{gitlabrepo}
M.~{\v{Z}}onda,
\newblock \emph{Correlated states in quantum dot clusters coupled to a common
  superconductor},
\newblock \url{https://gitlab.mff.cuni.cz/zondam/sc_qd_clusters},
\newblock GitLab repository, accessed 2026-06-02 (2026).

\bibitem{LiebMattis1962}
E.~Lieb and D.~Mattis,
\newblock \emph{Theory of ferromagnetism and the ordering of electronic energy
  levels},
\newblock Physical Review \textbf{125}(1), 164 (1962),
\newblock \doi{10.1103/PhysRev.125.164}.

\bibitem{Tasaki1998}
H.~Tasaki,
\newblock \emph{The hubbard model---an introduction and selected rigorous
  results},
\newblock Journal of Physics: Condensed Matter \textbf{10}(20), 4353 (1998),
\newblock \doi{10.1088/0953-8984/10/20/004}.

\bibitem{Luo-2024}
J.-J. Luo, H.~Pu and X.-W. Guan,
\newblock \emph{Exact results of the one-dimensional repulsive {Hubbard}
  model},
\newblock Rep. Prog. Phys. \textbf{87}(11), 117601 (2024),
\newblock \doi{10.1088/1361-6633/ad7b70}.

\bibitem{bethe1931theorie}
H.~Bethe,
\newblock \emph{Zur theorie der metalle: I. eigenwerte und eigenfunktionen der
  linearen atomkette},
\newblock Z. Physik \textbf{71}, 205 (1931),
\newblock \doi{10.1007/BF01341708}.

\bibitem{sandvik1997finite}
A.~W. Sandvik,
\newblock \emph{Finite-size scaling of the ground-state parameters of the
  two-dimensional {Heisenberg} model},
\newblock Phys. Rev. B \textbf{56}, 11678 (1997),
\newblock \doi{10.1103/PhysRevB.56.11678}.

\bibitem{Zonda2023generalized}
M.~\v{Z}onda, P.~Zalom, T.~Novotn\'y, G.~Loukeris, J.~B\"atge and V.~Pokorn\'y,
\newblock \emph{Generalized atomic limit of a double quantum dot coupled to
  superconducting leads},
\newblock Phys. Rev. B \textbf{107}, 115407 (2023),
\newblock \doi{10.1103/PhysRevB.107.115407}.

\bibitem{zalom2024double}
P.~Zalom, K.~Wrze\'{s}niewski, T.~Novotn\'y and I.~Weymann,
\newblock \emph{Double quantum dot {Andreev} molecules: Phase diagrams and
  critical evaluation of effective models},
\newblock Phys. Rev. B \textbf{110}, 134506 (2024),
\newblock \doi{10.1103/PhysRevB.110.134506}.

\bibitem{annabi2026fermion}
S.~Annabi, H.~Riechert, K.~Watanabe, T.~Taniguchi, J.~Griesmar, E.~Arrighi,
  L.~Bretheau and J.-D. Pillet,
\newblock \emph{Fermion parity of an andreev molecule probed by nonlocal
  josephson effect},
\newblock arXiv preprint arXiv:2607.15786  (2026).

\bibitem{Liebhaber2022quantum}
E.~Liebhaber, L.~M. R{\"u}tten, G.~Reecht, J.~F. Steiner, S.~Rohlf,
  K.~Rossnagel, F.~von Oppen and K.~J. Franke,
\newblock \emph{Quantum spins and hybridization in artificially-constructed
  chains of magnetic adatoms on a superconductor},
\newblock Nat. Commun. \textbf{13}, 2160 (2022),
\newblock \doi{10.1038/s41467-022-29879-0}.

\bibitem{lieb1961two}
E.~Lieb, T.~Schultz and D.~Mattis,
\newblock \emph{Two soluble models of an antiferromagnetic chain},
\newblock Ann. Phys. \textbf{16}, 407 (1961),
\newblock \doi{10.1016/0003-4916(61)90115-4}.

\bibitem{karbach1998introduction}
M.~Karbach, K.~Hu and G.~Muller,
\newblock \emph{Introduction to the {Bethe} ansatz {II}},
\newblock \eprint{https://arxiv.org/abs/cond-mat/9809163}.

\end{thebibliography}

\nolinenumbers

\end{document}